\documentclass[amsmath,amssymb,aps,twocolumn,pra,freq,floatfix, superscriptaddress]{revtex4-2}
\usepackage{graphicx}
\usepackage{dcolumn}
\usepackage{bm}
\usepackage{hyperref}
\usepackage{color}
\usepackage{subfigure}
\usepackage{flexisym}
\usepackage{float}
\usepackage{placeins}

\begin{document}

\title{On the Road with a Diamond Magnetometer}

\author{S. M. Graham}
\altaffiliation{s.graham.4@warwick.ac.uk}
\affiliation{Department of Physics University of Warwick Gibbet Hill Road Coventry CV4 7AL United Kingdom}
\affiliation{Diamond Science and Technology Centre for Doctoral Training University of Warwick Coventry CV4 7AL United  Kingdom}%

\author{A. J. Newman}
\affiliation{Department of Physics University of Warwick Gibbet Hill Road Coventry CV4 7AL United Kingdom}
\affiliation{Diamond Science and Technology Centre for Doctoral Training University of Warwick Coventry CV4 7AL United Kingdom}

\author{C. J. Stephen}
\affiliation{Department of Physics University of Warwick Gibbet Hill Road Coventry CV4 7AL United Kingdom}

\author{A. M. Edmonds}
\affiliation{Element Six Innovation Fermi  Avenue Harwell  Didcot  OX11 0QR  Oxfordshire  United  Kingdom}%

\author{D. J. Twitchen}
\affiliation{Element Six Innovation Fermi  Avenue Harwell  Didcot  OX11 0QR  Oxfordshire  United  Kingdom}%

\author{M. L. Markham}
\affiliation{Element Six Innovation Fermi  Avenue Harwell  Didcot  OX11 0QR  Oxfordshire  United  Kingdom}%

\author{G. W. Morley}
\altaffiliation{gavin.morley@warwick.ac.uk}
\affiliation{Department of Physics University of Warwick Gibbet Hill Road Coventry CV4 7AL United Kingdom}%
\affiliation{Diamond Science and Technology Centre for Doctoral Training University of Warwick Coventry CV4 7AL United Kingdom}%


%


\begin{abstract}

Nitrogen vacancy centres in diamond can be used for vector magnetometry. In this work we present a portable vector diamond magnetometer. Its vector capability, combined with feedback control and robust structure enables operation on moving platforms. While placed on a trolley, magnetic mapping of a room is demonstrated and the magnetometer is also shown to be operational in a moving van with the measured magnetic field shifts for the x, y, and z axes being tagged with GPS coordinates. These magnetic field measurements are in agreement with measurements taken simultaneously with a fluxgate magnetometer. 
\end{abstract}

\maketitle

\section{Introduction}

Portable, robust, and high sensitivity magnetometers may be employed for a wide range of industrial and geophysical applications; from condition monitoring and battery scanning to ore prospecting \cite{glenn2017micrometer, gooneratne2017downhole, ingleby2022digital, zhang2021battery, zhou2021imaging, hatano2021simultaneous, chatzidrosos2019eddy, graham2023fiber, newman2024tensor, kubota2023wide, hatano2022high}. The nitrogen vacancy centre (NVC) in diamond is a promising candidate for such magnetometry applications \cite{taylor2008high, acosta2009diamonds}. Contained within the diamond lattice, NVCs can sense the projection of external magnetic fields along their symmetry axis in extreme environments with temperatures and pressures of up to 600 K and 130 GPa respectively. NVC magnetometers are capable of operating without external heating, cryogenic cooling, magnetic shielding or vacuum environments \cite{taylor2008high, rondin2014magnetometry, barry2020sensitivity, zhou2021imaging, plakhotnik2014all, toyli2013fluorescence, toyli2012measurement, liu2019coherent, doherty2014electronic, ivady2014pressure, hsieh2019imaging, hilberer2023enabling}. Employing ensembles of NVCs allows sensitivity to be improved by the factor 1/$\sqrt{\textrm{N}}$, where N is the number of NVC centres, at the cost of reduced spatial resolution \cite{acosta2009diamonds, taylor2008high, rondin2014magnetometry, barry2020sensitivity}. For NVC ensembles, the NVCs may exist with their symmetry axis along four possible crystallographic orientations within the tetrahedral diamond lattice, providing an inherent vector capability, without the need for additional sensors, the rigid diamond lattice ensuring the axes are well-defined \cite{le2013optical, maertz2010vector, pham2011magnetic, clevenson2018robust, schloss2018simultaneous, newman2024tensor, zhang2018vector}. This minimises non-linearity and non-orthogonality issues and high spatial resolutions are possible using a single diamond sensor head. Feedback control can also be employed to greatly enhance dynamic range \cite{clevenson2018robust, schloss2018simultaneous, wang2023realization, newman2024tensor}. These features make NVC magnetometers suitable for deployment on moving platforms \cite{fu2020sensitive}. 

High-sensitivity NVC magnetometers are typically stationary devices that cannot be practically employed outside of the laboratory and on relatively small moving platforms such as cars or aircraft. High sensitivities may be retained using mobile fiber-coupled sensor heads \cite{rabeau2005diamond, liu2013fiber, ampem2009nano, mayer2016direct, ruan2015nanodiamond, henderson2011diamond, liu2013fiber, fedotov2014fiber, fedotov2014fiber2, dmitriev2016concept, patel2020subnanotesla, sturner2021integrated, chatzidrosos2021fiberized, graham2023fiber, xie2023microfabricated, xie2022microfabricated, zhao2023all}. Several integrated or partially integrated NVC magnetometers have been demonstrated with nT and sub-nT sensitivities including some employing flux concentrators \cite{song2023integrated, mao2023integrated, sturner2019compact, ran2023portable, wang2022portable, ran2023highly, wang2023hybrid} though not all with vector and feedback control capability. The highest achieved vector sensitivity is 50 pT/$\sqrt{\textrm{Hz}}$ in a table-top configuration \cite{schloss2018simultaneous}. While the calibration and testing of a vector NVC magnetometer on a boat has been performed \cite{fleig2018maritime, frontera2018shipboard}, none of these examples were demonstrated working on moving land vehicles or used for indoor magnetic field mapping.

Potential applications include magnetic mapping which is useful for both magnetic navigation and geophysical surveying \cite{canciani2016absolute, canciani2017airborne, li2014magnetic, gaffney2008detecting, wang2023quantum}. GPS is employed to aid inertial navigation systems, which rapidly accumulate error, however GPS is readily jammed and inaccessible underwater so there is an interest in a passive, un-jammable system of navigation \cite{grant2009gps, taraldsen2011underwater}. Magnetic navigation making use of map-matching techniques and magnetometers is such a system. These high-sensitivity magnetometers could also be employed for producing the high-quality maps of the Earth's magnetic anomaly field required for such navigation. The Earth's magnetic anomaly field is of the order of hundreds of nT and represents the scalar difference from the Earth's core field. The anomaly field varies over relatively short distances compared to the core field (1 km at an altitude of 1 km as opposed to a minimum of 4000 km) \cite{canciani2016absolute}. Other applications include non-destructive testing and space exploration, diamond being radiation hard \cite{acuna2002space, cochrane2016vectorized, fu2020sensitive}.

In this work, we demonstrate a portable vector NVC magnetometer capable of taking magnetic field measurements both indoors and outdoors whilst placed on a moving trolley platform, and it is used to map a laboratory room. It is shown to be operational outside of the laboratory even when placed in a moving vehicle, in this case a van. For a single projection, it has an unshielded mean sensitivity of approximately 0.5 nT/$\sqrt\textrm{Hz}$ in a (10-150)-Hz frequency range. Also unshielded and on the trolley platform with vector and feedback control it achieves mean sensitivities of approximately 180 nT/$\sqrt\textrm{Hz}$, 210 nT/$\sqrt\textrm{Hz}$, and 140 nT/$\sqrt\textrm{Hz}$ for the x, y, and z axes respectively in a (0.1-0.66)-Hz frequency range. 

\FloatBarrier
 
\section{Methods}

\begin{figure}[t]
\includegraphics[width=\columnwidth]{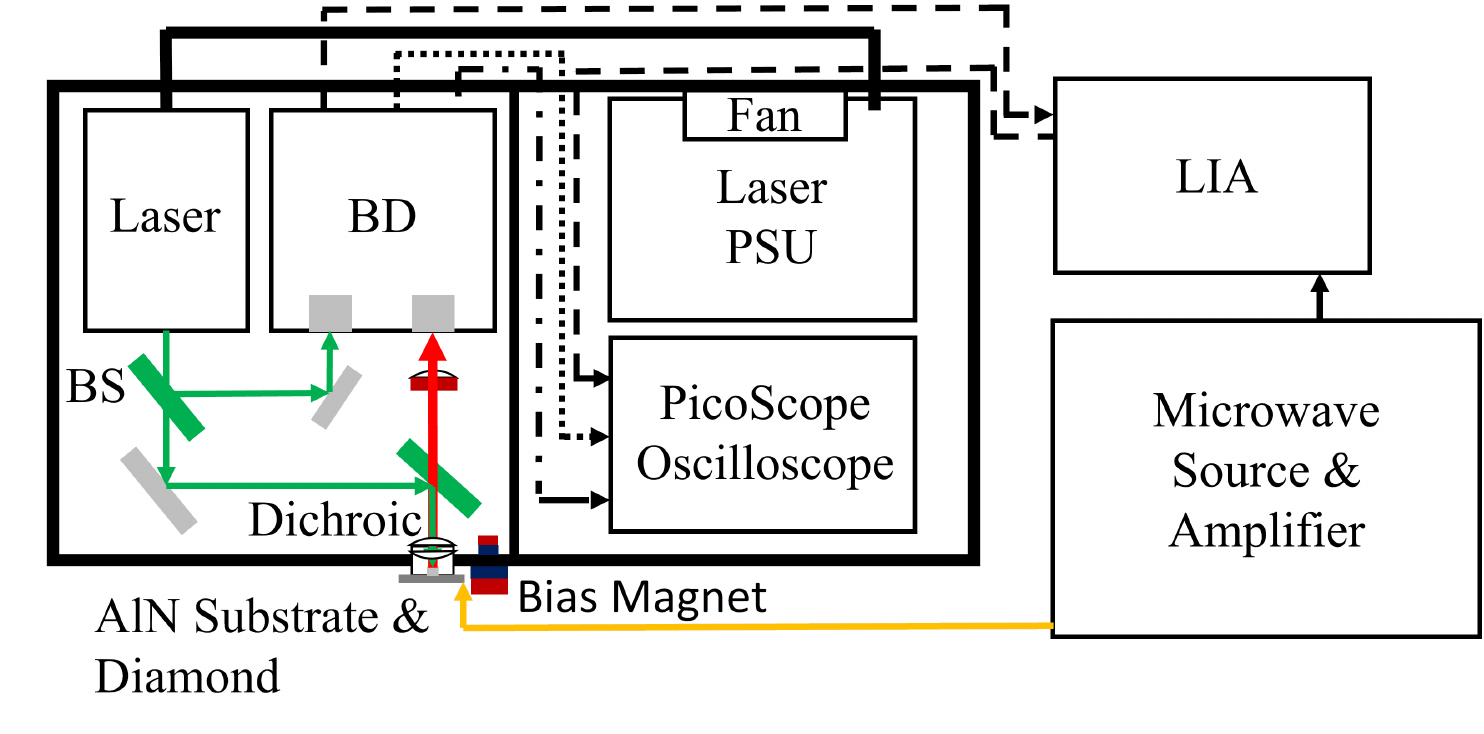} 
\caption{\small A schematic of the experimental setup: BS, 100:1 beam sampler; BD, balanced detector; PSU, power supply unit; LIA, lock-in amplifier.} 
\label{fig: ExperimentalSetup}
\end{figure}

The portable magnetometer consists of a (300 mm $\times$ 474 mm $\times$ 135 mm) box with a partition at the centre - optics and diamond being found on one side and electronics such as a PicoScope Oscilloscope (4424) and laser power supply unit (PSU) on the other. Figure \ref{fig: ExperimentalSetup} shows the experimental setup as a schematic. 
For the vector measurements an external microwave source (Agilent N5172B), 43-dB gain microwave amplifier (Mini-circuits ZHL-16W-43-S+) and lock-in amplifier (LIA) (Zurich MFLI-500 kHz) are employed. The box has a total weight of approximately 5 kg and under normal operating conditions with vector magnetometry has a power draw of approximately 403 W. The portable magnetometer unit is placed onto the bottom level of a two-level trolley, with the microwave source, microwave amplifier, and LIA being placed on the top level, helping to minimise magnetic noise at the sensor head. Further details on the trolley are provided in appendix B, and appendix K describes a more portable configuration, without vector capabilities, that does not require a trolley. 

A continuous-wave optically detected magnetic resonance (cw ODMR) magnetometry scheme is employed. A Laser Quantum 532-nm GEM laser with a maximum output power of 1 W is used to excite the NVC ensemble, however only 0.5 W is employed for this work - as the laser is found to be more unstable at higher powers and noise associated with temperature fluctuations in the diamond is increased. Approximately 1\% of the laser power is picked off with a beam sampler and sent to a Thorlabs PD450A balanced detector, onto which the red fluorescence from the diamond is also focused, to allow for the cancellation of common-mode laser intensity noise. The diamond is placed on an aluminium nitride (AlN) substrate antenna which is fixed to one of the walls of the box. 
Both the green and red light is collimated using a 0.25 NA aspheric lens, with a 0.7 NA aspheric lens being placed immediately prior to the diamond for focusing the laser light. The green-to-red photon conversion efficiency is found to be approximately 0.07\%.

Sine-wave frequency modulation of the microwaves is employed to allow lock-in detection. For the optimum sensitivity measurements, but not for the vector mapping, prior to the amplifier, the microwaves are mixed using a Mini-Circuits ZX05-U432H-S+ up-converter with a 2.158-MHz sine-wave generated with an RSPro AFG21005 arbitrary function generator, enabling hyperfine excitation. 
The microwaves are supplied through a transmission line to a 4.4-mm loop antenna. The transmission line and antenna are both made from copper (Cu) with a tin (Sn) coating. The AlN substrate is employed as a heat spreader. AlN has a thermal conductivity of up to 321 W/mK, approaching that of SiC and sharing its dielectric nature \cite{watari1993phonon}. A circular section of Cu is placed in the middle of the antenna loop below the diamond to improve the optical excitation and collection efficiency. The diamond is a CVD grown $^{12}\textrm{C}$ purified Element Six diamond with a high NVC concentration that has been electron irradiated and annealed \cite{edmonds2021characterisation}. 

A permanent Ne-Fe-B magnet is employed as a bias field, this being attached to the side of the portable magnetometer box. This splits out the NVC magnetic resonances into eight distinct peaks. The magnetometry signal is extracted from the noise using a LIA and a fixed microwave frequency is applied to the zero-crossing point of a given resonance, which is the steepest part. 
Additional external magnetic fields shift the resonance frequency via the Zeeman effect producing changes in the NVC fluorescence and thus shifts in the LIA voltage output. These shifts in the LIA output voltage are used to provide an error signal for a proportional linear feedback system with the microwave frequency being continually adjusted to remain at the zero-crossing points of the four $m_s$ = 0 to $m_s$ = +1 transition ODMR peaks, greatly improving our dynamic range. The dynamic range would otherwise be limited to the linear region of an ODMR peak \cite{clevenson2018robust, wang2022portable}. To turn this voltage error into a frequency for the microwave source it is first necessary to convert it into a frequency shift using the zero-crossing slope (ZCS) of the ODMR resonance in V/MHz. It can then be added to the initial frequency of the resonance set by the operator. Simultaneously, the microwave frequency is sequentially looped over the zero-crossing points of the four $m_s$ = 0 to $m_s$ = +1 transitions, sitting on each peak for around 70 ms, allowing the projection of an external field to be measured along all four NVC orientation axes from the frequency shift. From these frequency shifts the x, y, and z components of such a field can also be calculated. The resultant magnetic field components are shifts in the magnetic field relative to the initial environmental and bias field at the diamond, as opposed to the absolute magnetic field. The method used to convert the frequency shifts into the x, y, and z field components involves the use of a matrix of effective gyromagnetic ratios (referred to henceforth as the $\mathbf{A}$-matrix) and is expanded upon in Refs. \cite{schloss2018simultaneous, newman2024tensor} and appendix F. The sequential approach to vector measurements limit the sampling rate to approximately 1.3 Hz. This limits the maximum speed that a moving platform can travel at and collect data, and limits the spatial resolution of the measurements at a given speed. This sampling rate is an average as the data collection rate is not entirely constant. 

Parameter optimisation for the microwave power and frequency modulation are performed as in Refs. \cite{el2017optimised, graham2023fiber, newman2024tensor}. With a 150-Hz or 100-Hz LIA low-pass filter (LPF) 3-dB point, the optimum parameters are found to be 14.003 kHz, 400.23 kHz, and -4 dBm for the modulation frequency, modulation amplitude, and microwave power respectively. The microwave power stated is the value selected on the microwave source, rather than the measured power reaching the diamond. For the vector measurements the modulation amplitude is increased to 3.5 MHz. The parameter optimisation measurements are detailed in appendix C.


\section{Results}

Figure \ref{fig: OptimumSensitivityMeasurement} shows a sensitivity measurement taken with the magnetometer in the laboratory, using hyperfine excitation and with the optimum parameters identified above. The sensitivity for a single NVC projection is determined by applying a linear fit to a given ODMR resonance. This gives us the ZCS in V/MHz that can be converted into a calibration constant in V/nT using the gyromagnetic ratio of the NVC, approximately equal to 28 GHz/T \cite{doherty2013nitrogen}. The voltage noise floor is then measured by setting our microwave frequency to said ODMR resonance such that it is magnetically sensitive. 30 1-s time traces are then taken in the PicoScope software with a sampling rate of 20 kHz. The power-spectral-density (PSD) of each time trace is then calculated and the mean of these 30 PSDs is then taken. Taking the square root of the mean PSD then gives the amplitude-spectral-density (ASD) in $\textrm{V}$/$\sqrt{\textrm{Hz}}$. The ASD is converted into $\textrm{nT}$/$\sqrt{\textrm{Hz}}$ using the calibration constant. The mean of the flat noise floor is then taken, with the 50 Hz and its harmonics being masked. This is found to be (0.5 $\pm$ 0.1) nT/$\sqrt{\textrm{Hz}}$ in the (10-150)-Hz frequency range. This range is selected for comparison to works which employ a similar range \cite{wang2022portable, ran2023portable, patel2020subnanotesla, graham2023fiber} and is limited by the use of a 150 Hz 3-dB point on the LIA LPF with a filter slope of 48 dB/octave. Time traces are also taken when off-resonance (3.5 GHz) and thus magnetically insensitive as well as with the laser off to quantify electronic noise levels. As can be seen we are limited by magnetic noise from equipment on the trolley. Several of the peaks around 50 Hz can be attributed to the two fans on the trolley as they disappear when the fan is turned off. The photon-shot-noise-limited sensitivity is determined to be (0.06 $\pm$ 0.01) nT/$\sqrt{\textrm{Hz}}$. The average noise floors, in the same frequency range, are found to be (0.19 $\pm$ 0.04) nT/$\sqrt{\textrm{Hz}}$ and (0.11 $\pm$ 0.05) nT/$\sqrt{\textrm{Hz}}$ for the magnetically insensitive and electronic noise cases respectively. 
Sensitivity measurements using the vector settings as well as looking at lower frequency bands can be found in appendix D alongside Allan deviation (AD) measurements in appendix E. 

\begin{figure}[h!]
\includegraphics[width=\columnwidth, trim={1.5cm 1.5cm 1.5cm 1.5cm}]{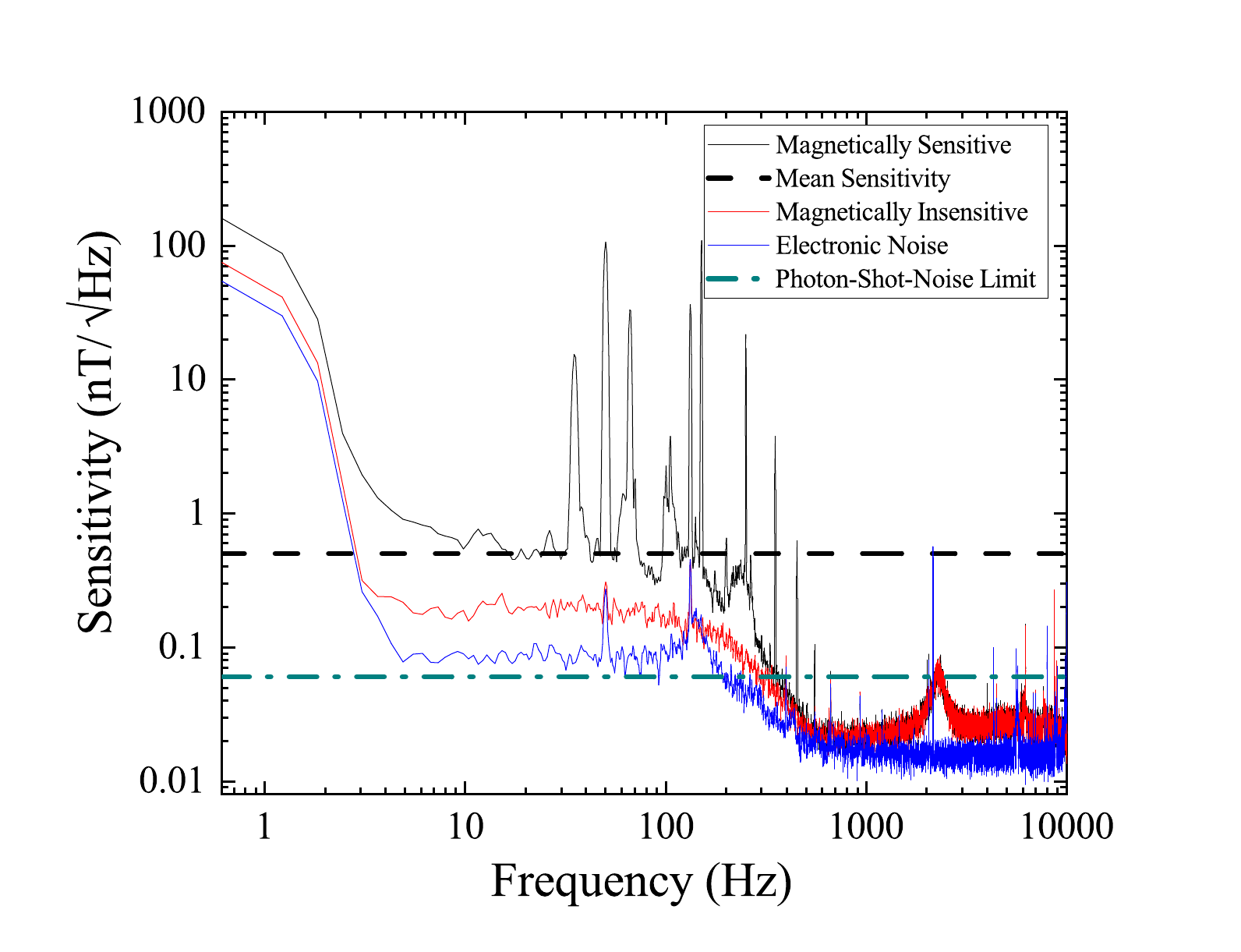} 
\caption{\small Sensitivity spectra, using optimum settings and hyperfine excitation with a mean sensitivity of $\break$ (0.5 $\pm$ 0.1) nT/$\sqrt{\textrm{Hz}}$ from (10-150)-Hz, indicated by the dashed line. The noise floor is also shown when magnetically insensitive (off-resonance at 3.5 GHz) and with no applied microwaves or laser (electronic noise). The photon-shot-noise-limited sensitivity is also shown. For these measurements a Blackman window is being used for the amplitude-spectral-densities.} 
\label{fig: OptimumSensitivityMeasurement}
\end{figure}

\begin{figure}[h!]
\includegraphics[width=\columnwidth, trim={1cm 1cm 0.6cm 1cm}]{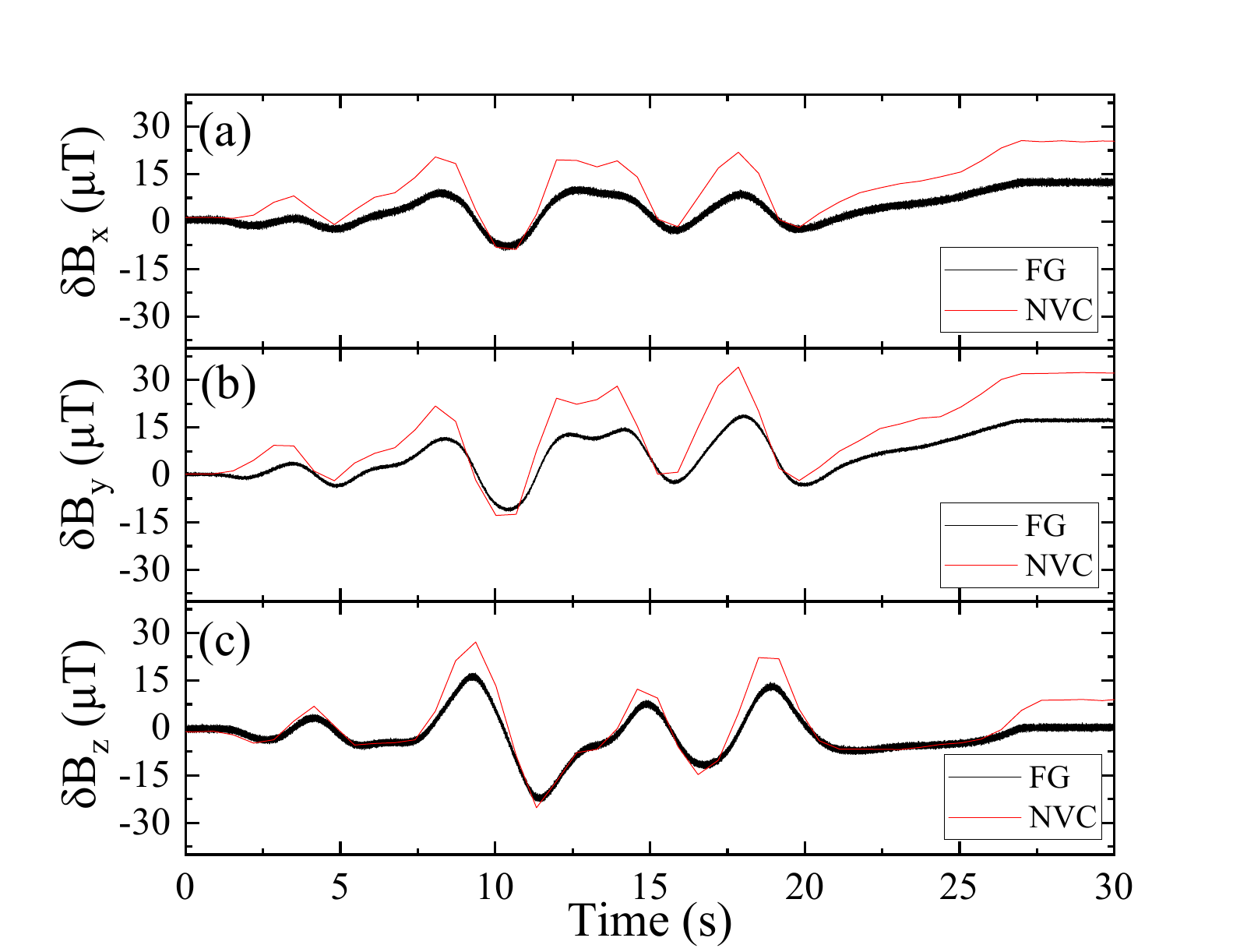} 
\caption{\small Straight line profiles comparing the magnetic field shifts; (a) $\delta\textit{B}_{\textrm{x}}$, (b) $\delta\textit{B}_{\textrm{y}}$, and (c) $\delta\textit{B}_{\textrm{z}}$, taken with the fluxgate (FG) and NVC magnetometers when the trolley is moved in the Y direction for approximately 8.25 m in the laboratory.} 
\label{fig: LabLineProfiles}
\end{figure}

With the laboratory empty of other equipment, magnetic maps of the room are taken with the trolley. For all of these mapping measurements the LIA scaling factor is set to $\times$250 and the 3-dB of the LPF is 100 Hz. Figure \ref{fig: LabLineProfiles} shows continuous magnetic-field shift ($\delta\textit{B}_{\textrm{x}}$, $\delta\textit{B}_{\textrm{y}}$, and $\delta\textit{B}_{\textrm{z}}$) line profiles taken within the laboratory with both our NVC magnetometer and a Bartington Mag-03MS100 fluxgate (FG). For these line profiles the trolley containing both the NVC and FG magnetometers is continuously pushed at an average speed of approximately 0.4 m/s (0.9 mi/h or 1.4 km/h), demonstrating its ability to take continuous vector measurements without losing feedback control. Additionally, at this velocity the sampling rate is sufficient to avoid serious aliasing of the magnetic signals, as can be seen when compared to the FG which had a sampling rate of 1000 Hz (limited by the PicoScope Oscilloscope settings). The signals seen may be attributed to steel rebars found under the laboratory floor. The FG is positioned such that its three x, y, and z triaxial coils are aligned with the x, y, and z axes defined for the NVC magnetometer. The diamond of the NVC magnetometer is located approximately 7 cm below and 16 cm behind the FG and this accounts for some of the difference in the magnitude of the magnetic signals observed.

\begin{figure}[h!]
\includegraphics[width=\columnwidth]{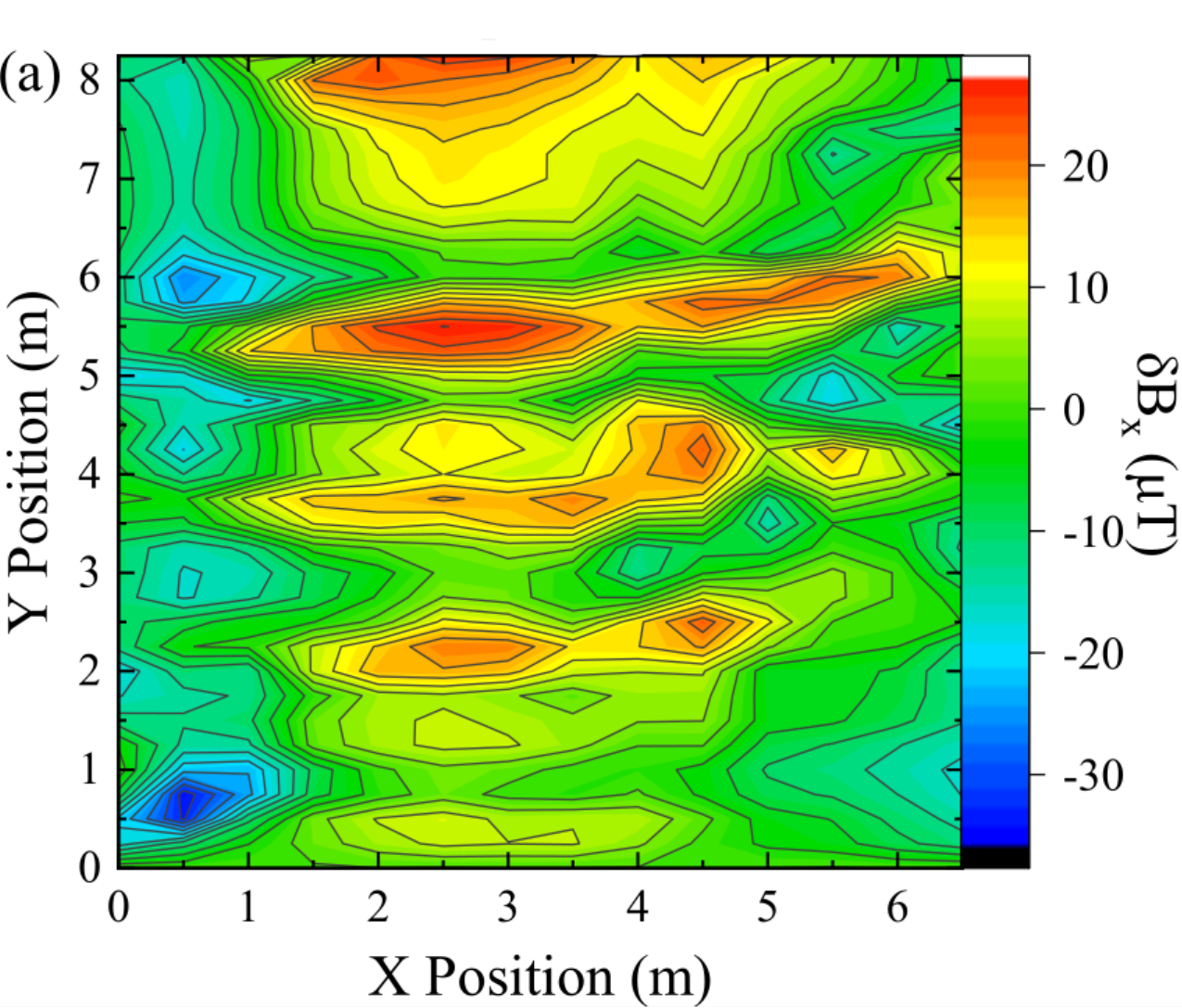} 
\includegraphics[width=\columnwidth]{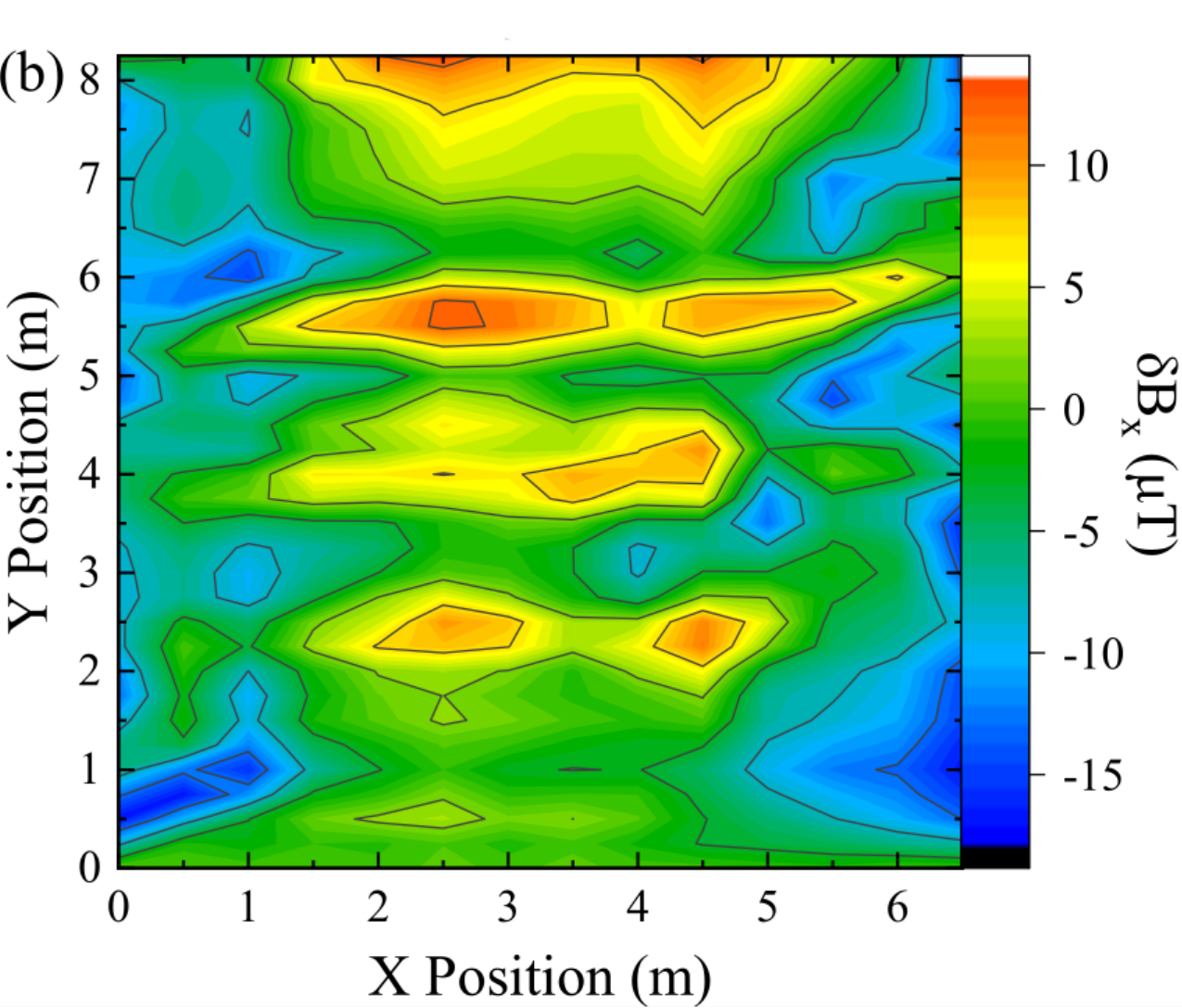}
\caption{\small Laboratory room 2D point maps of $\delta\textit{B}_{\textrm{x}}$ with 0.5 m and 0.25 m resolutions along the X and Y axes respectively taken with the NVC (a) and fluxgate (FG) (b) magnetometers.} 
\label{fig: Lab2DPointMap}
\end{figure}

Expanding upon these line profiles, 2D point maps of the entire laboratory room can be taken. The laboratory room is marked with a 6.5 $\times$ 8.25 m grid (X $\times$ Y) with an X and Y resolution of 0.5 m and 0.25 m respectively. A series of line profiles can then be taken, starting at X = 0 for each. Together these produced a map of the vector components of the magnetic field shift, as seen in Fig. \ref{fig: Lab2DPointMap} which shows the qualitative agreement between the FG and NVC measurements for the $\delta\textit{B}_{\textrm{x}}$ component. The rebar below the floor can be clearly seen in these maps. The X and Y axes of the grid do not align with the x and y axes of the magnetometers, which are rotated by approximately $45^{\circ}$ relative to the grid axes. Pushing the trolley on foot, it can also be taken and operated outside of the laboratory building.

\begin{figure}[h!]
\includegraphics[width=\columnwidth, trim={1cm 1cm 0.6cm 1cm}]{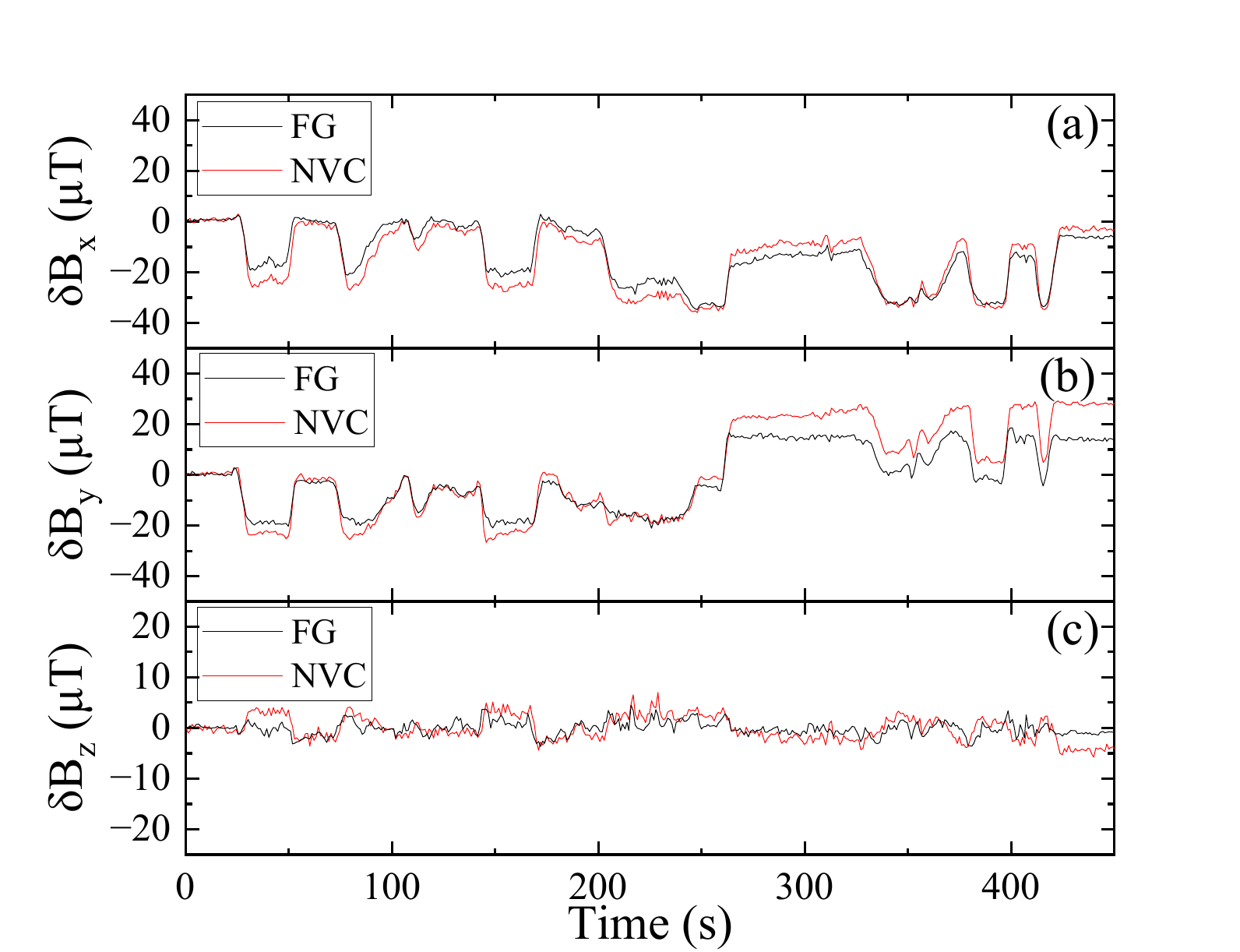} 
\caption{\small Line profiles comparing the magnetic field shifts; (a) $\delta\textit{B}_{\textrm{x}}$, (b) $\delta\textit{B}_{\textrm{y}}$, and (c) $\delta\textit{B}_{\textrm{z}}$, taken with the fluxgate (FG) and NVC magnetometers in the electric van.} 
\label{fig: VanLineProfiles}
\end{figure}

To demonstrate its ability to operate on a moving land vehicle, the trolley was secured via ratchet straps into the back of first a diesel (Renault Trafic SL30) then an electric van (Nissan e-NV200). These vans were driven around a nearby residential and commercial area, with roundabouts, gentle inclines and speedbumps, for approximately 20 minutes. These two sets of measurements were taken in early September 2023 and late October 2023 respectively with differing weather conditions - it being approximately $40^{\circ}$C and $10^{\circ}$C inside of the diesel and electric vans respectively. These temperatures were measured with a Multicomp Pro MP780618 temperature data-logger. No issues associated with these temperature differences were observed, despite changes in the laser head and PSU temperatures. The van was driven at an average speed of approximately 7 m/s (15.7 mi/h or 25.2 km/h), with a maximum speed of approximately 11 m/s (24.6 mi/h or 39.6 km/h). As with the laboratory room measurements, the FG was placed aligned with the NVC x, y, and z axes and in approximately the same position relative to the diamond. Figure \ref{fig: VanLineProfiles} shows time traces of $\delta\textit{B}_{\textrm{x}}$, $\delta\textit{B}_{\textrm{y}}$ and $\delta\textit{B}_{\textrm{z}}$ for both the NVC and FG magnetometer over the course of the drive in the electric van. 

\begin{figure}[h!]
\includegraphics[width=\columnwidth]{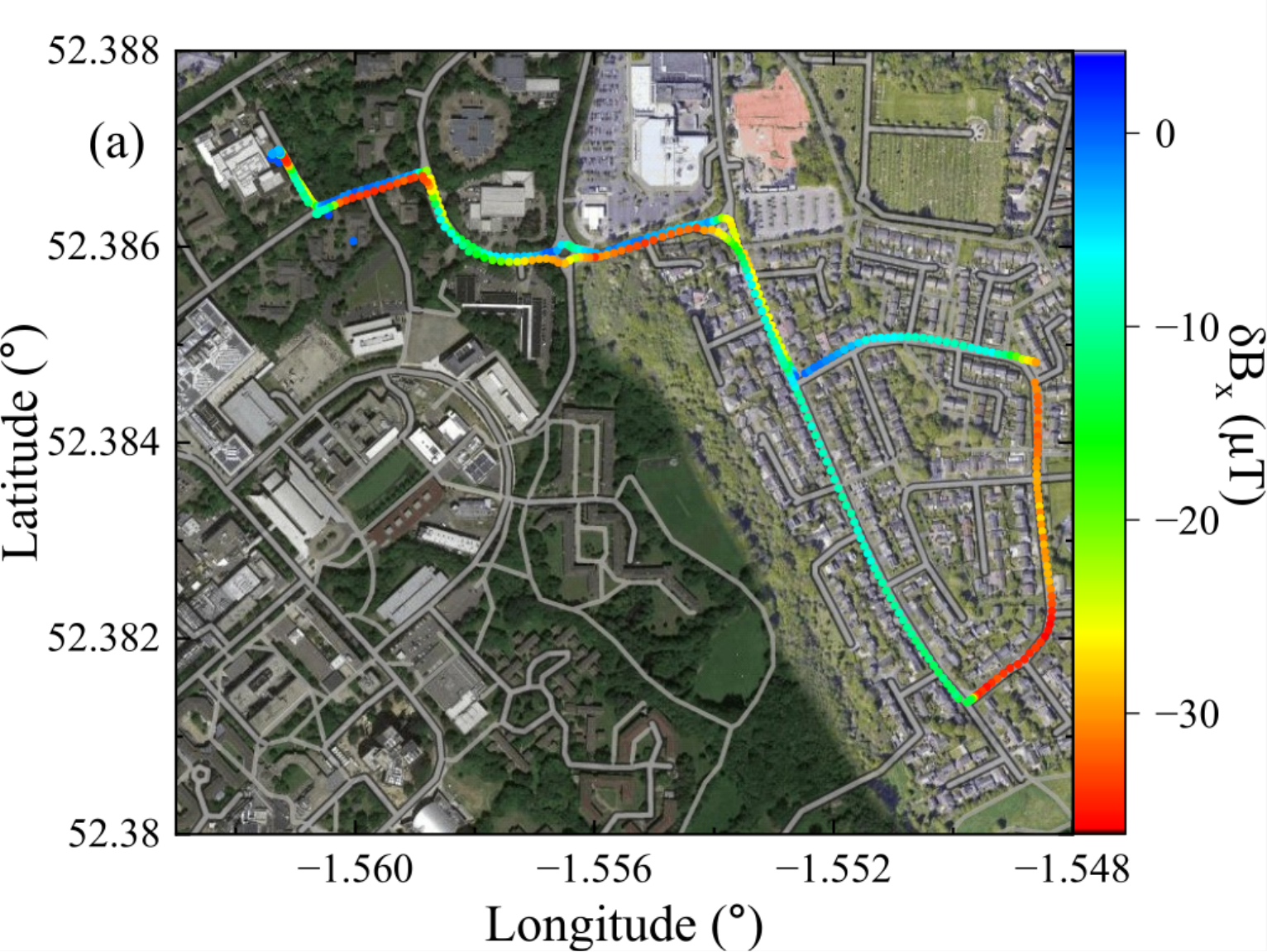} 
\includegraphics[width=\columnwidth]{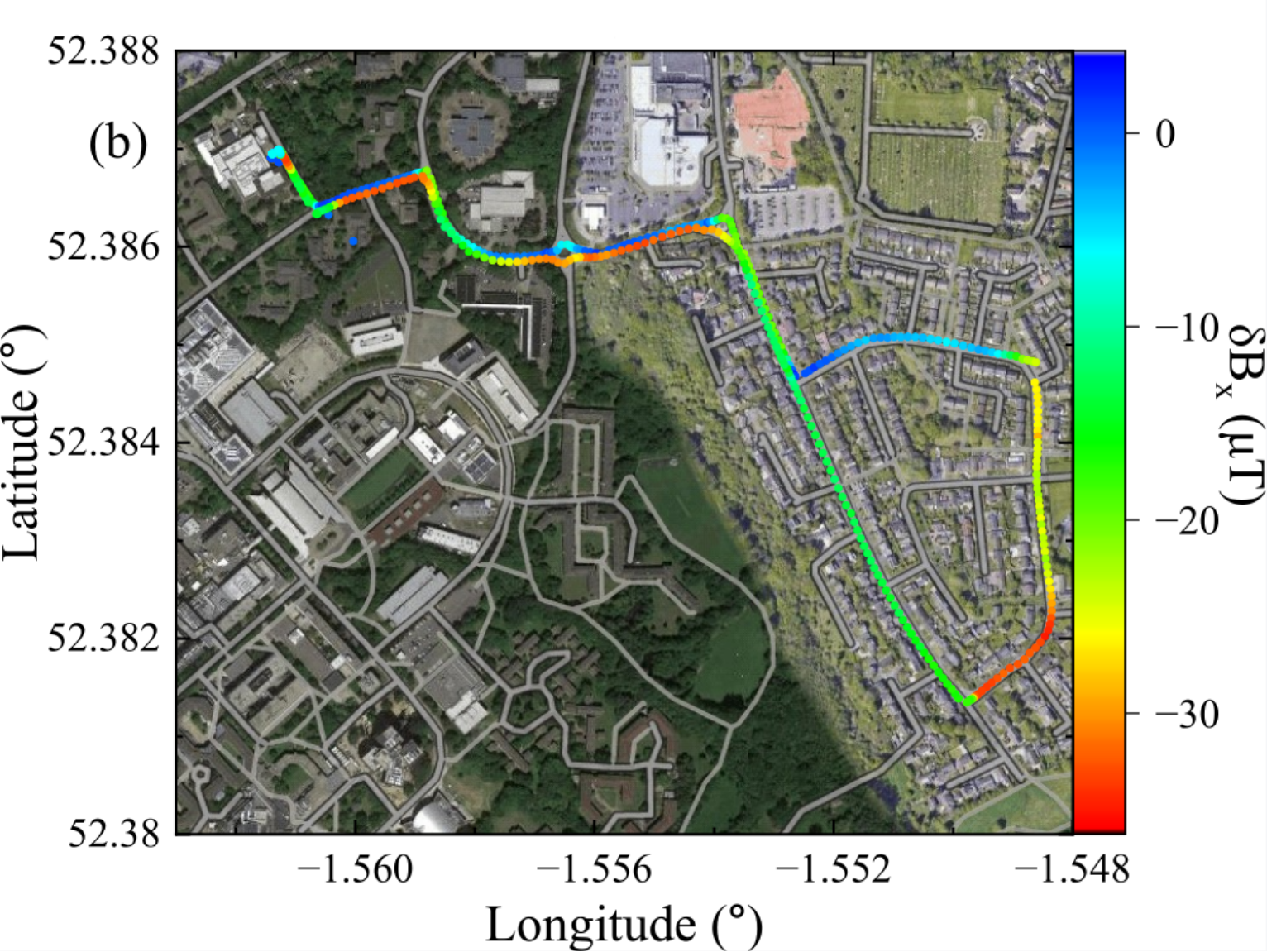} 
\caption{\small Magnetic field shifts of; (a) NVC magnetometer $\delta\textit{B}_{\textrm{x}}$, and (b) fluxgate (FG) magnetometer $\delta\textit{B}_{\textrm{x}}$, as a function of latitude and longitude plotted on a Google hybrid map. These measurements were taken in the electric van.} 
\label{fig: VanGPSMaps}
\end{figure}

A smartphone sensor logger is used to obtain the GPS coordinates of the van at each point in the drive path with a horizontal (X-Y) accuracy of approximately 5 m. Figure \ref{fig: VanGPSMaps} shows the path travelled by the electric van during the drive colour tagged with the magnetic field shifts (for the x component, with the y and z components found in appendix G) measured with the NVC and FG at each set of latitude and longitude coordinates. In the van, the magnetic field at the position of the diamond includes not only the Earth's and bias field, but also a disturbance field from the van and the trolley. These measurements have not been compensated to account for this disturbance field. The disturbance fields within the van have three principle components; permanent fields, induced fields and eddy current fields \cite{canciani2016absolute}. The NVC magnetometer measures the shift in the magnetic field along the x, y, and z axes. As the permanent field of the van is approximately constant in the NVC magnetometer frame of reference we do not observe significant shifts due to the changing vector addition of the van's permanent magnetic field with the Earth's core field. Changes in magnetic field from the initial value may thus be attributed to variations in the component of the Earth's core field along the x, y, and z axes as the van rotated and went up and down inclines. This is confirmed by the agreement between the measurements taken in the diesel and electric van, as well as on foot with a smartphone magnetometer as shown in appendix H. Some of the variation could be due to the induced and eddy current fields of the van, however, it is difficult to attribute specific features to these fields. We are not directly measuring the Earth's crustal anomaly field that would be required for magnetic map matching in these measurements. The FG and NVC magnetometer are in qualitative agreement. The differences in the measured magnetic field shift between the FG and NVC can likely be attributed to a combination of their differences in position within the van, and inaccuracies in the calibration of the NVC magnetometer (in terms of both the ODMR ZCSs and $\mathbf{A}$-matrix). It should be noted that the FG sampling rate was limited to the same rate as that of the NVC for these measurements for synchronisation purposes. Additionally, the PicoScope oscilloscope employed to measure its voltage outputs for x, y, and z limited its sensitivity to approximately 1-3 nT/$\sqrt\textrm{Hz}$ comparable to the NVC. This is due to the use of a $\pm$ 10 V voltage range to accommodate the full dynamic range of the FG which leads to a high level of quantisation noise given the 12-bit ADC resolution. With the appropriate voltage measurement equipment the Bartington Mag-03MS100 FG has a nominal sensitivity of approximately 6 pT/$\sqrt{\textrm{Hz}}$ at 1 Hz for each axis, though this would in practice be limited by magnetic noise from the platform \cite{canciani2016absolute}. All the magnetometry data is also further resampled to match the approximately 1 Hz sampling rate of the GPS data-logger. 

\FloatBarrier

\section{Discussion and Conclusions}

These measurements demonstrate that our NVC magnetometer is capable of taking vector measurements whilst being moved around on a trolley platform inside the laboratory, and also outside of the laboratory, with the trolley placed in a moving van driven on public roads. The vector measurements are in good qualitative agreement with those of the FG. Furthermore, this magnetometer is capable of obtaining sub-nT sensitivities with hyperfine excitation and would in principle be capable of such sensitivities combined with vector and feedback control given improvements in the robustness of our feedback control and a non-sequential vector approach. 

However, further work is required for it to be useful for magnetic mapping applications, especially for mapping of the Earth's anomaly field. Firstly, at present the sampling rate in vector mode is greatly limited from kHz to only 1.3 Hz by the need for sequential frequency hopping. This issue could be avoided by employing multiple microwave sources to simultaneously address the four resonances as in \cite{schloss2018simultaneous} or by mixing in a direct digital synthesiser signal. An additional issue is that the feedback method relies on the ZCSs of the ODMR resonances and this is susceptible to changes in the experimental parameters such as laser power \cite{clevenson2018robust}. Furthermore, in contrast to Ref. \cite{schloss2018simultaneous}, in which the diamond axes are used to define the x, y, and z axes and thus obtain the vector $\mathbf{A}$-matrix of effective gyromagnetic ratios, in our case we defined our own x, y, and z axes with a Helmholtz coil. These features of our setup negate some of the advantages of the NVC magnetometers, notably their potential for accurate magnetic field measurement just using well-known fundamental constants and for the latter point the potential for low non-orthogonality by exploiting the well-defined diamond axes. A feedback control method such as that employed in Ref. \cite{clevenson2018robust} would ensure our measurements were not dependent on experimental parameters. To obtain an $\mathbf{A}$-matrix based on the diamond axes we would need to fit the ODMR spectrum to obtain the bias field components. We would then use a numerical approach to determine the $\mathbf{A}$-matrix components (as opposed to just defining them according to the unit vectors of the four NVC symmetry axes in the diamond axes defined frame). This is due to the potentially significant transverse bias magnetic field along each NVC symmetry axis, which lead to additional shifts in the resonance frequencies \cite{schloss2018simultaneous, clevenson2018robust}. Furthermore, the fact that we are measuring the shift in, as opposed to the absolute, magnetic field along each of the axes at the diamond poses an issue. This means we cannot then determine the total magnetic intensity (TMI), which would be desirable for magnetic mapping measurements. 

To observe the magnetic anomaly field a higher sub-Hz vector sensitivity would also be required, especially if employed on an aircraft a substantial height from the surface of the Earth \cite{wang2023quantum}. With improvements to the robustness of our feedback control, this would be possible using techniques such as hyperfine excitation. Additional gains in sensitivity could also be achieved via improvements to the green-to-red photon conversion efficiency \cite{barry2020sensitivity, zhang2022high, zhang2017fluorescence, ma2018magnetometry, le2012efficient, duan2018enhancing, xu2019high}. Greater stability and sensitivity could also be achieved using dual-resonance magnetometry with all eight ODMR peaks being addressed and tracked simultaneously \cite{barry2020sensitivity, acosta2010temperature, clevenson2015broadband, schloss2018simultaneous, clevenson2018robust, fescenko2020diamond}. Various compensation techniques would also need to be employed for magnetic mapping measurements to account for the fields of the vehicle, especially as in many cases the magnetometer cannot be placed on a boom \cite{canciani2016absolute, canciani2017airborne, fleig2018maritime}. Compensation is discussed further in appendix B. More robust feedback control could be achieved via the use of a PID controller, though this would likely be slower than our current method \cite{newman2024tensor, clevenson2018robust}.

In conclusion, we have developed a portable vector diamond magnetometer that is capable of operation both inside and outside of the laboratory. It is shown to be operational whilst moving on a trolley platform and employed for mapping a laboratory room. Furthermore magnetic field shift measurements dominated by the Earth's core field are taken in a moving land vehicle, in this case a van, whilst driving on public roads. With further improvements in sensitivity and accuracy this could be useful for magnetic mapping applications such as magnetic navigation and geophysical surveying.

\FloatBarrier

\section{Acknowledgements}

We thank Jeanette Chattaway, Matty Mills, and Lance Fawcett of the Warwick Physics Mechanical Workshop, and Robert Day and David Greenshield of the Warwick Physics Electronics Workshop. The Ph.D. studentships of S. M. G. and A. J. N. are funded by the Defence Science and Technology Laboratory (DSTL) and the National Nuclear Laboratory (NNL) respectively. We would like to thank Owen Griffiths at DSTL for useful conversations about this work. This work is supported by the UK Hub NQIT (Networked Quantum Information Technologies) and the UK Hub in Quantum Computing and Simulation, part of the UK National Quantum Technologies Programme, with funding from UK Research and Innovation (UKRI) EPSRC Grants No. EP/M013243/1 and No. EP/T001062/1, respectively. This work is also supported by Innovate UK Grant No. 10003146, EPSRC Grant No. EP/V056778/1 and EPSRC Impact Acceleration Account (IAA) awards (Grants No. EP/R511808/1 and No. EP/X525844/1).

\FloatBarrier

\section*{Appendix A: NVC Physics}

\begin{figure}[h!]
\includegraphics[width=0.75\columnwidth]{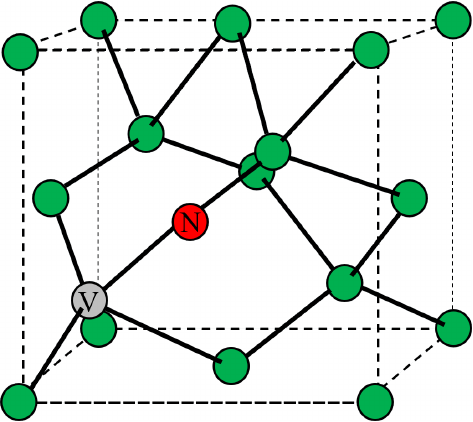} 
\caption{\small The physical structure of the NVC.} 
\label{fig: NVCentrePS}
\end{figure}

The NVC is a point defect found in diamond that consists of a substitutional nitrogen atom and an adjacent vacancy. The physical structure of the NVC is shown in Fig. \ref{fig: NVCentrePS}. It has trigonal symmetry, with the major symmetry axis being the line between the nitrogen and vacancy along $\langle$111$\rangle$. For an NVC ensemble four alignments are possible and are typically equally likely; [111], [$\overline{1}$$\overline{1}$1], [1$\overline{1}$$\overline{1}$], and [$\overline{1}$$1\overline{1}$]. NVCs measure the projection of external magnetic fields along their symmetry axis, and thus by exploiting the entire NVC population it is possible to perform vector measurements, by determining the projection of an external field along each of the four possible alignments, with the x, y, and z coordinate system potentially being defined by the geometry of the diamond crystal.

\begin{figure}[h!]
\includegraphics[width=\columnwidth]{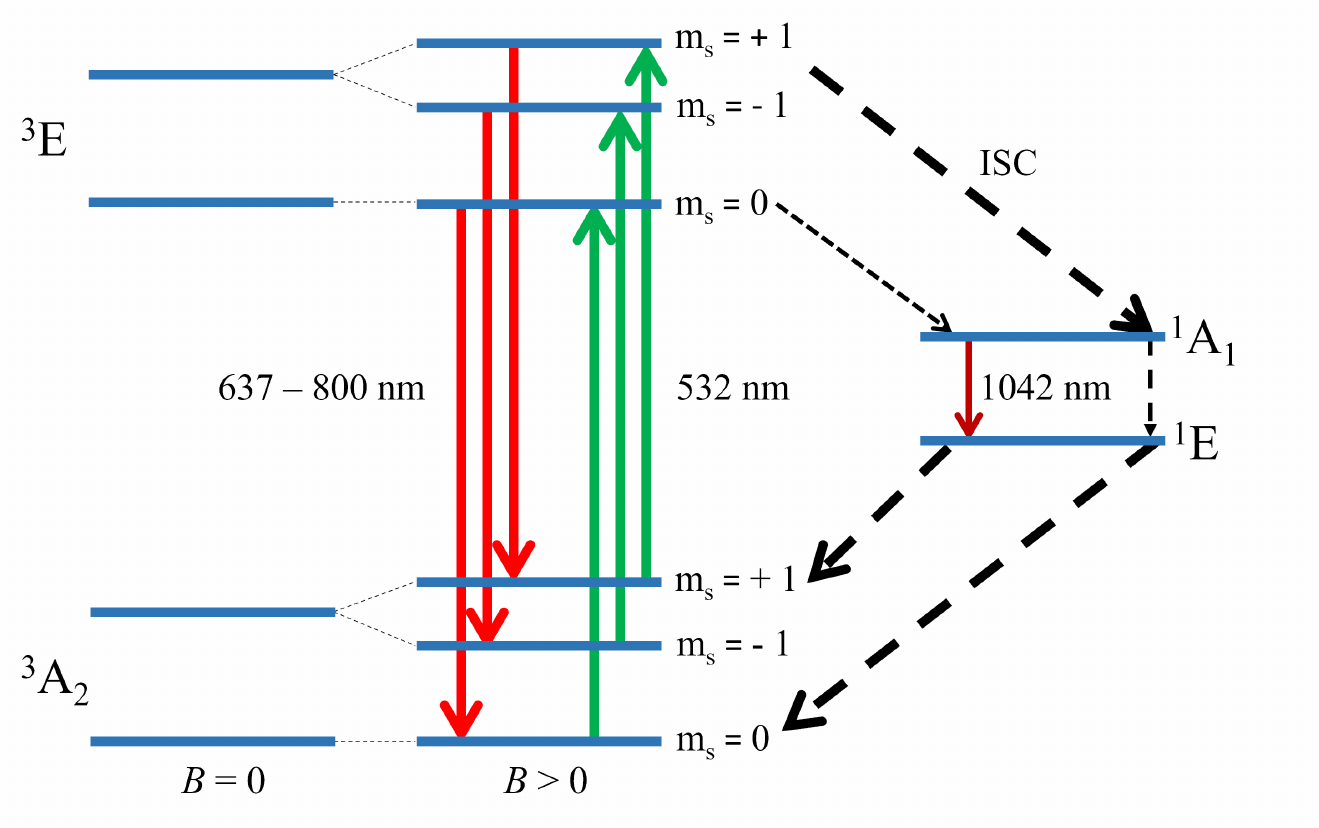} 
\caption{\small A schematic of the NVCs electronic structure, with the intersystem crossings (ISC) and electric dipole transitions indicated.} 
\label{fig: NVCentreES}
\end{figure}

The NVC can be found in a neutral, positive and negative charge state. It is the latter of these that is of interest for magnetometry applications. Its energy level structure is shown in Fig. \ref{fig: NVCentreES}.

The negative charge state is a spin-1 defect with a spin-triplet ground state. This state can be addressed using ODMR spectroscopy \cite{rondin2014magnetometry, barry2020sensitivity}. The NVC is optically initialized into the $m_s$ = 0 sublevel of the $^3A_2$ ground state via a 532-nm laser. This exploits the fact that if the defect is initially in the $\break$ $m_s$ = $\pm$ 1 sublevels prior to excitation, it has a higher probability of taking the singlet states pathway to the ground state and this requires non-spin-conserving transitions. This pathway is also nonradiative and thus measurement of the fluorescence intensity allows the spin state to be read out. Microwaves can be used to manipulate and drive the spin state from the $m_s$ = 0 to $m_s$ = $\pm$ 1 sublevels \cite{rondin2014magnetometry, barry2020sensitivity}.

There is a zero-field splitting (ZFS) of approximately 2.87 GHz for the $^3A_2$ ground state at room temperature. Applying an external magnetic field lifts the degeneracy of the $\break$ $m_s$ = $\pm$ 1 sublevels through the Zeeman effect. This produces resonances at two frequencies (for a given NVC alignment) split by \cite{rondin2014magnetometry, barry2020sensitivity}

\begin{equation}
\label{eq:ZeemanSplitting2}
\Delta f = 2\gamma B_{\textrm{NVC}},
\end{equation}

where $B_{\textrm{NVC}}$ an external magnetic field's projection along the NVC symmetry axis for a given alignment. There is a hyperfine interaction between the $\break$ S = 1 electron spin and the $\break$ I = 1 nuclear spin of the $^{14}\textrm{N}$ atom of the NVC. This leads to the splitting of each resonance into three, with separations of approximately 2.158 MHz \cite{doherty2013nitrogen, barry2016optical}. 

\FloatBarrier

\section*{Appendix B: Trolley} 

The two-level trolley has a width of 440 mm, a height of 850 mm and a length of 1000 mm. The trolley is light-tight with boards covering three sides and a blind over the remaining side to allow access for servicing. The magnetometer is powered using a 25 m extension cable within the laboratory, however, when operating within the vans the magnetometer is powered using an uninterruptible power supply (UPS) based on a battery. With 100\% charge this UPS is able to power the magnetometer for approximately 30 minutes. This limited how far we could travel with the magnetometer in the van. The trolley is entirely plastic, with the exception of its aluminium wheels which have stainless steel screws and bolts, to minimise any permanent disturbance field at the diamond. The principle magnetic noise sources on the trolley are the microwave source, microwave amplifier, and LIA. The fans are also a significant source of magnetic noise - one of which is located in the electronics section of the portable box and the second of which is located in the front panel of the trolley. This fan drew air in from outside the trolley and funnelled it into the fan of the electronics box - blowing cool air from outside the trolley over the laser PSU and preventing it from overheating. This did, however, mean that variable temperatures outside of the trolley could lead to laser power fluctuations as the laser PSU and head took longer to thermalise and is readily knocked out of equilibrium (see appendix J). The optics and electronics box is placed on the bottom level of the trolley, whereas the microwave source, amplifier and LIA are placed on the top level. The 25 m extension cable is located next to the optics and electronics box - this is only used when operating inside the laboratory or at short distances from the laboratory building. 

In principle the disturbance fields from the trolley and also van could be compensated for to allow the magnetic anomaly field of the Earth to be recovered from the data. The standard approach is to use the Tolles-Lawson model. In this model the disturbance field is made up of permanent, induced and eddy current components \cite{woloszyn2012analysis, canciani2016absolute}. The two principle contributions to the trolley and van's disturbance field come from the permanent (hard iron) and induced (soft iron) fields. The first of these originates from ferromagnetic objects, whereas the latter comes from paramagnetic and ferromagnetic objects, the magnetisation of which may vary with their orientation within the Earth's field. The eddy current contribution is likely to be minimal. For the FG the bias magnet of the NVC magnetometer would be a significant permanent magnetic field source. Typical compensation techniques such as the Tolles-Lawson model work best when the disturbance field is significantly smaller than the Earth's field to be measured \cite{canciani2016absolute, fleig2018maritime}. This is not the case for the NVC magnetometer in the van and trolley. Machine learning and neural network approaches have been shown to be effective in these more demanding cases \cite{williams1993aeromagnetic, wu2017aeromagnetic, yu2021improved}. As noted in the main text, as we are measuring the shift in magnetic field assuming the diamond is static relative to the trolley and van's magnetic objects much of the permanent field would have been removed automatically. This, however, would not account for induced and eddy current magnetic fields. Reducing the size of disturbance fields could be possible by ensuring the diamond is kept far from sources of noise, whether electronic equipment like the fans and microwave sources, or large static fields such as the walls and doors of the van. The use of a fiber-coupled sensor head could be useful for this, especially if multiple such sensor heads were employed in a tensor gradiometry configuration \cite{noriega2014aeromagnetic, yin2019compensation}. Space weather effects are unlikely to be a serious source of magnetic noise over the approximately 20 minute duration of the measurements \cite{canciani2016absolute}.

\FloatBarrier

\section*{Appendix C: Parameter Optimisation}

\begin{figure}[h!]
\includegraphics[width=\columnwidth, trim={1.5cm 1.5cm 1.5cm 1.5cm}]{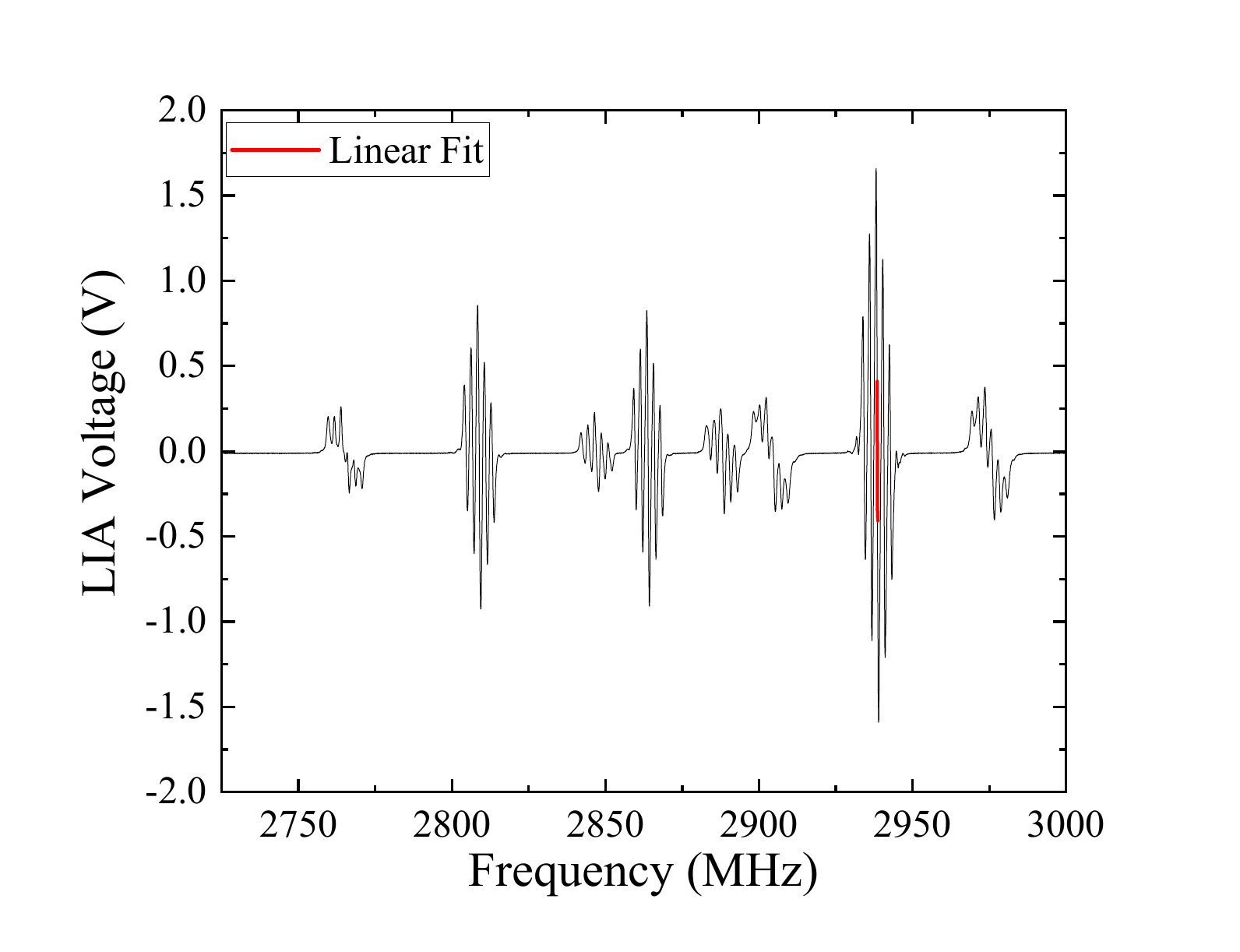} 
\caption{\small The demodulated optically detected magnetic resonance (ODMR) spectrum using the optimum microwave and frequency modulation settings with the resonance used for sensitivity measurements indicated by the red linear fit line.} 
\label{fig: OptimumSensitivityODMRSpectrum}
\end{figure}

Experimental parameters such as microwave power, modulation frequency, and modulation amplitude are optimised to maximise the sensitivity of the magnetometer. For these measurements ODMR spectra are taken with a bias field aligned such that eight ODMR resonances could be observed in the spectrum. 
Due to the frequency response of the microwave amplifier used the amount of microwave power delivered to the diamond varied significantly over the 2.7 to 3.05 GHz frequency range of the ODMR spectrum. This made it difficult to optimise the microwave power to obtain optimal performance for each peak and made it undesirable to use these optimum settings with hyperfine excitation for vector measurements exploiting multiple resonances simultaneously.  For the optimum sensitivity measurements the parameter optimisation is investigated using the single resonance labelled in Fig. \ref{fig: OptimumSensitivityODMRSpectrum}. The ZCS is taken by applying a linear fit to the derivative slope of this resonance. A higher ZCS, for a given scaling factor, is indicative of greater responsivity and thus sensitivity. However, in the case of the modulation frequency (and more generally in the magnetic noise limited case) increases in ZCS may be accompanied by an increase in technical noise and thus increases in ZCS do not always correspond to improvements in sensitivity. Accordingly, in these cases the sensitivity is also measured as a function of the given experimental parameter.

\begin{figure}[h!]
\includegraphics[width=\columnwidth, trim={1.5cm 1.5cm 1.5cm 1.5cm}]{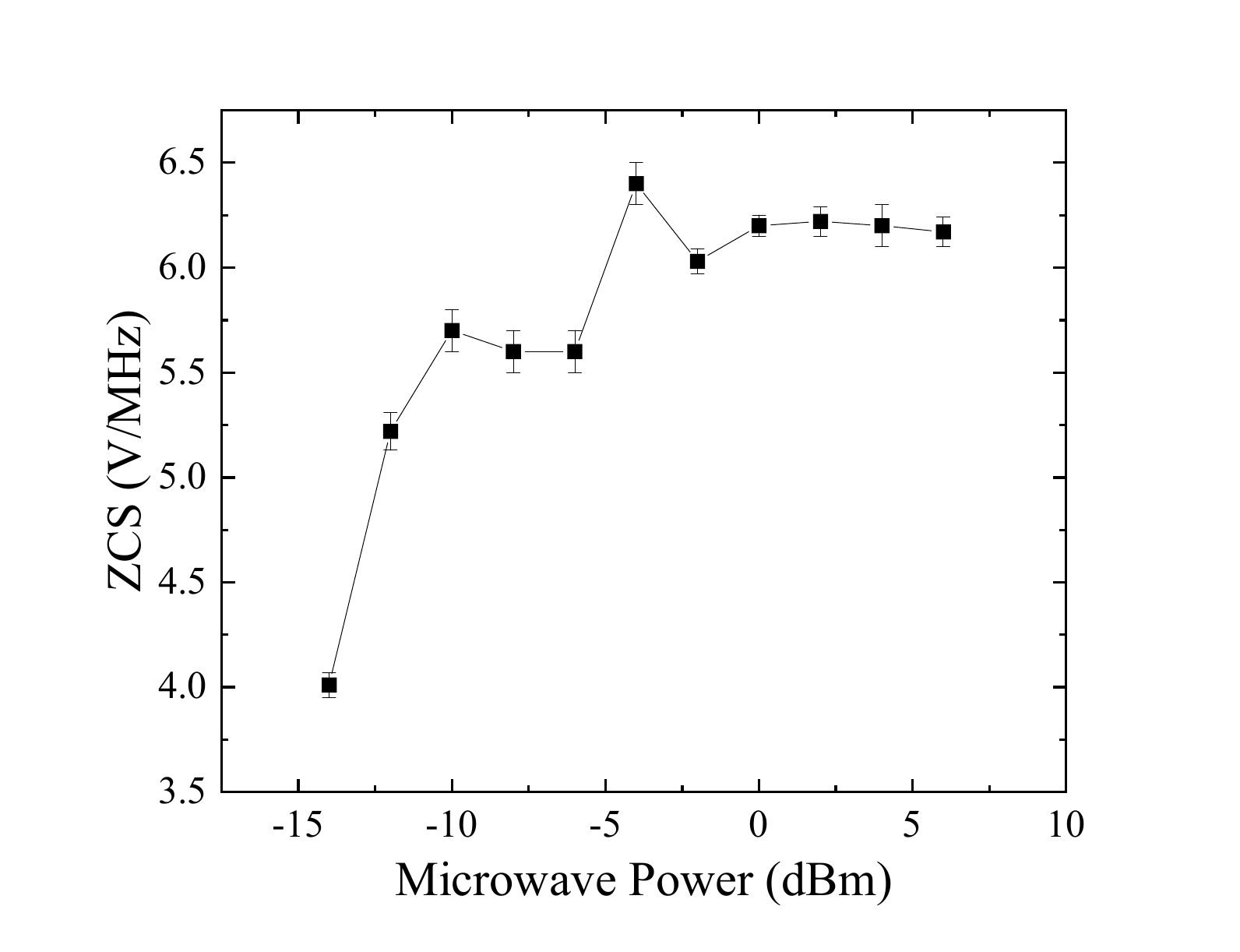} 
\caption{\small The zero-crossing slope (ZCS) as a function of microwave power in dBm. This is the power at the output of the microwave source which is proportional to the power at the diamond. All the measurements are taken with a modulation frequency of 4.003 kHz and a modulation amplitude of 400.23 kHz.} 
\label{fig: ZCSvsMWPowerMeasurements}
\end{figure}

Figure \ref{fig: ZCSvsMWPowerMeasurements} shows the zero-crossing slope (ZCS) as a function of microwave power. The stated microwave power is the value of the output from the Agilent N5172B microwave source in dBm. This power is amplified by a 43-dB microwave amplifier and there would be significant losses in the cables and transmission line. Additionally the IQ modulation of the microwave source is employed to mix the microwave frequency with a 2.158 MHz signal with a 0.3 Vpp (peak-to-peak voltage) from an external function generator - enabling hyperfine excitation. As can be seen the ZCS peaks at a value of -4 dBm. Increasing the microwave power beyond this does increase the contrast, but it causes broadening of the resonance linewidth. As noted previously, however, the amount of microwave power differs at the frequencies of the other ODMR resonances and thus the optimum output power of the microwave source differs for each resonance.

\begin{figure}[h!]
\includegraphics[width=\columnwidth, trim={1.5cm 1.5cm 1.5cm 1.5cm}]{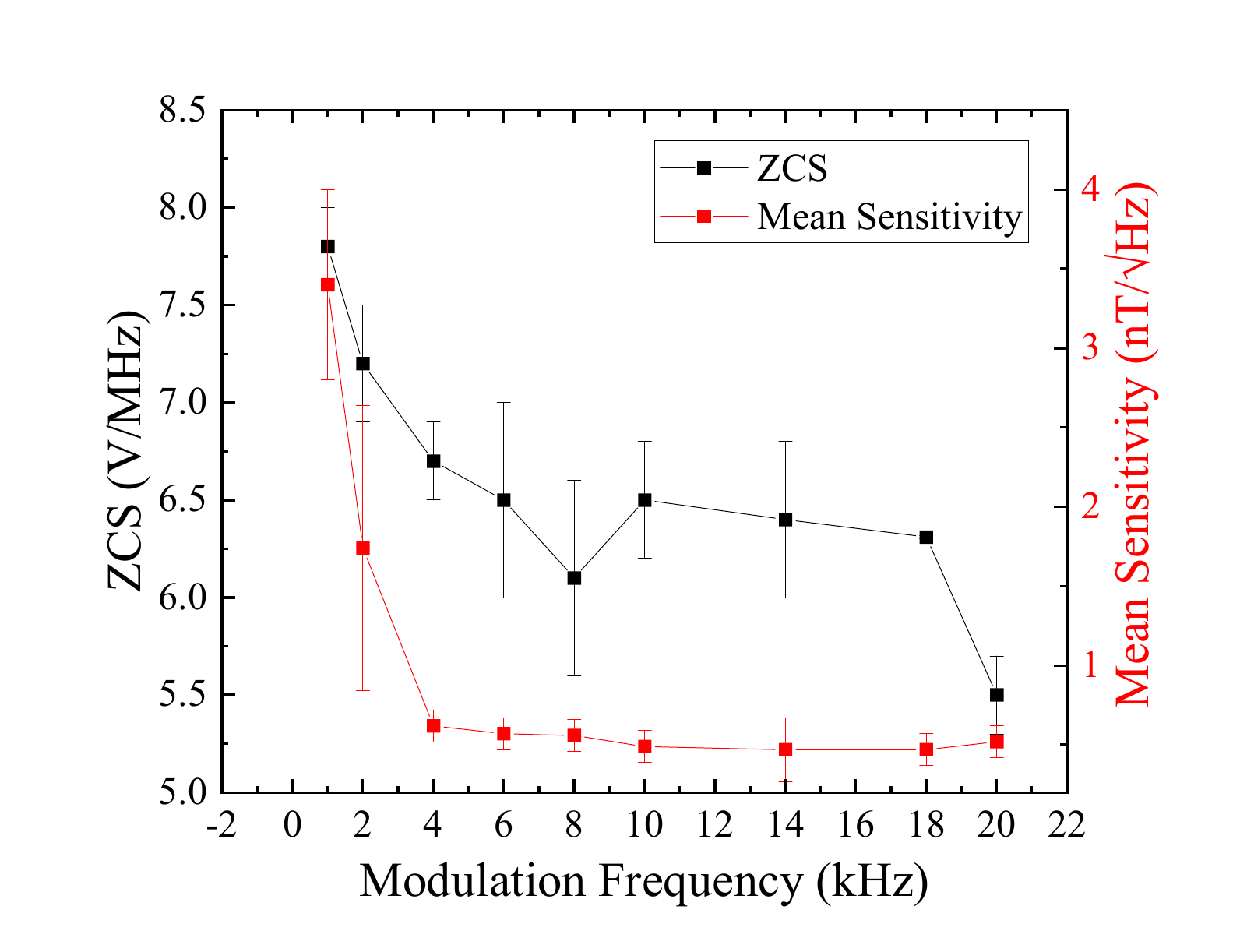} 
\caption{\small The zero-crossing slope (ZCS) and mean sensitivity, taken in a (5-150)-Hz frequency range as a function of modulation frequency. All the measurements are taken with a modulation amplitude of 400.23 kHz and with the microwave power set to -4 dBm at the source.} 
\label{fig: ZCSSensvsFMRate}
\end{figure}

The ZCS increases with decreasing modulation frequency as can be seen in Fig. \ref{fig: ZCSSensvsFMRate}. The change in ZCS with modulation frequency is relatively low \cite{SMGraham2023}. At higher modulation frequencies the effective contrast and thus ZCS decreases due to the finite repolarisation rate of the NVC \cite{wang2022portable, xie2020dissipative}. The optimum modulation frequency depends on the properties of the diamond and its NVC ensemble as well as the laser power. Increasing the laser power increases the effective bandwidth by increasing the repolarisation rate \cite{wang2022portable, zhang2021diamond}. Considering that the noise floors rise for lower modulation frequencies due to increasing difficulty in rejecting technical noise approaching dc, a modulation frequency of 14.003 kHz is selected for a laser power of 0.5 W. As can be seen in Fig. \ref{fig: ZCSSensvsFMRate}, a similar sensitivity is obtained for a wide range of modulation frequencies at this laser power and thus this decision is somewhat arbitrary. For these measurements only 10 1-s time traces were taken and averaged, as opposed to 30.

\begin{figure}[h!]
\includegraphics[width=\columnwidth, trim={1.5cm 1.5cm 1.5cm 1.5cm}]{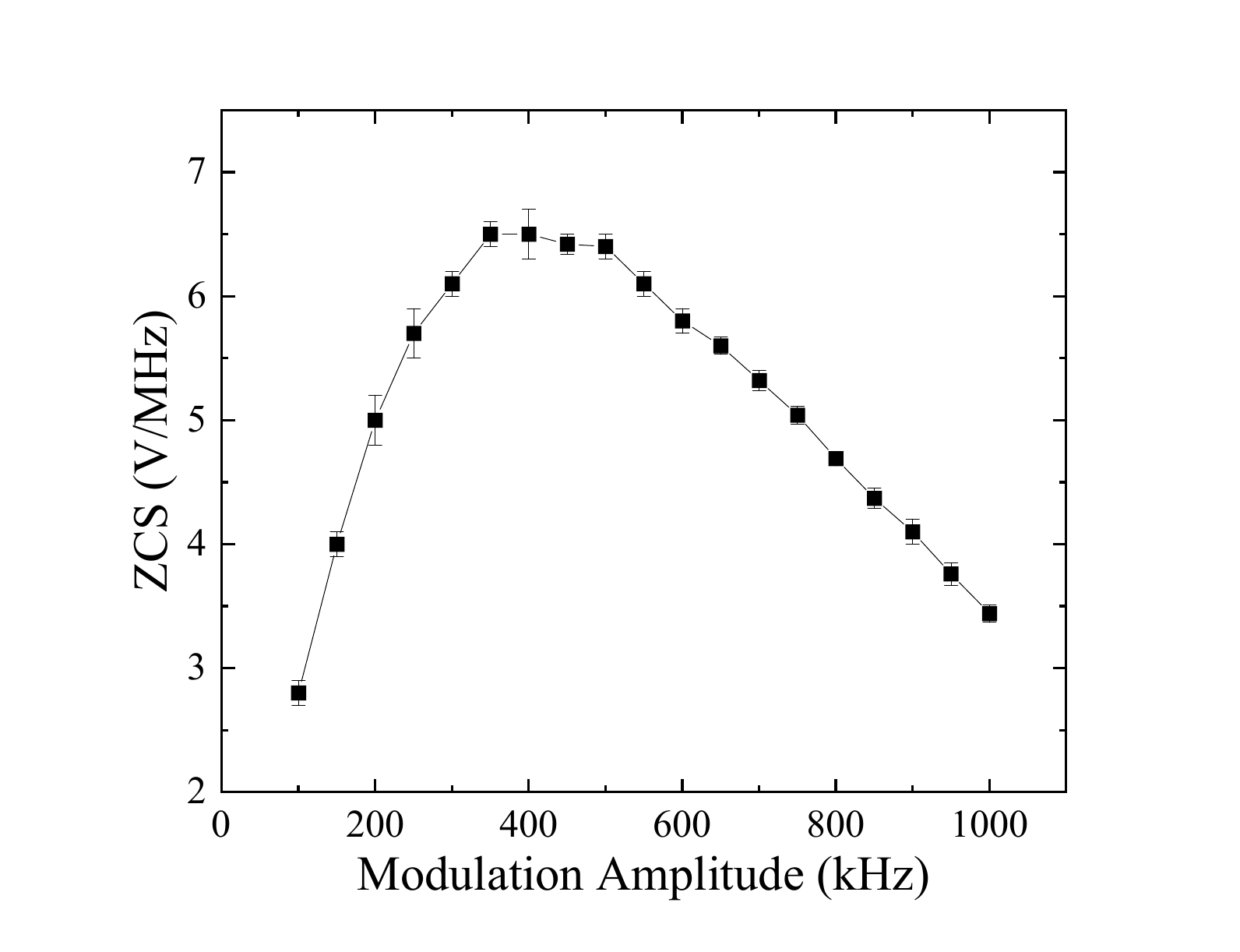} 
\caption{\small The zero-crossing slope (ZCS) as a function of the modulation amplitude. All the measurements are taken with a modulation frequency of 4.003 kHz and with the microwave power set to -4 dBm at the source.} 
\label{fig: ZCSSensvsFMDepth}
\end{figure}

Figure \ref{fig: ZCSSensvsFMDepth} shows the ZCS as a function of modulation amplitude. As we are magnetically noise limited the actual sensitivity would be relatively uniform over this range of modulation amplitudes as the noise floor would scale with the ZCS. A modulation amplitude of 400.23 kHz is found to provide the optimum performance. For the vector measurements, using hyperfine excitation is found to cause difficulties with feedback control for tracking. Accordingly hyperfine excitation is not used and the modulation amplitude is also increased to 3.5 MHz. This is larger than the linewidth of the ODMR resonances but ensured a more uniform ZCS for each of the four ODMR resonances used for vector measurements and increased the dynamic range (with and without feedback control). Ultimately we sacrificed some sensitivity for robustness with our present vector and feedback control implementation. Were we to use four microwave sources \cite{schloss2018simultaneous}, as opposed to the current frequency hopping approach, we would likely be able to make use of the high sensitivity settings in vector mode. 

\FloatBarrier

\section*{Appendix D: Sensitivity Measurements}

\begin{figure}[h!]
\includegraphics[width=\columnwidth, trim={1.5cm 1.5cm 1.5cm 1.5cm}]{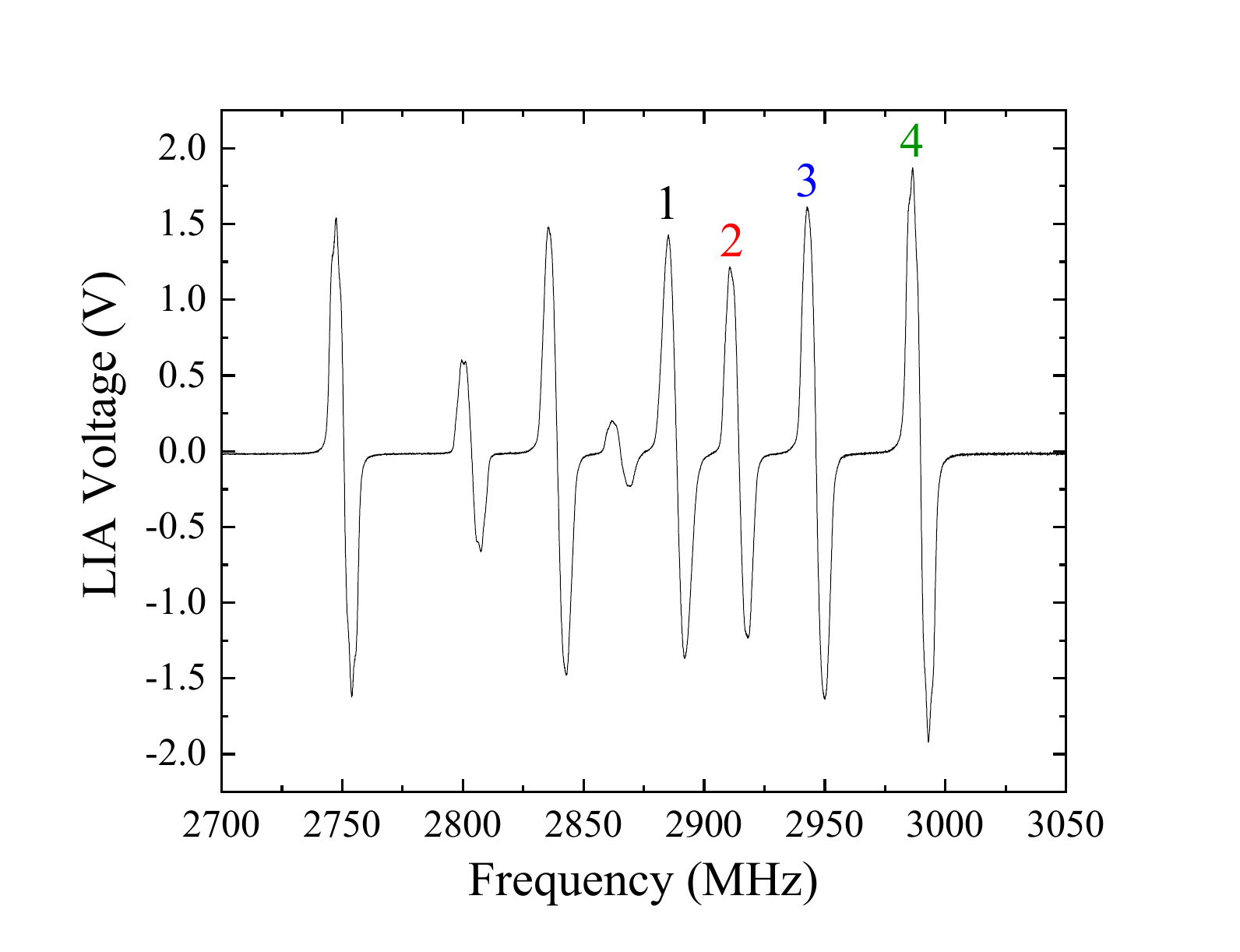} 
\caption{\small The demodulated optically detected magnetic resonance (ODMR) spectrum using the vector microwave and frequency modulation settings with the four $m_s$ = 0 to $m_s$ = + 1 resonances labelled 1 to 4.} 
\label{fig: ODMRVectorMeasurement}
\end{figure}

In the main paper the sensitivity measurements taken using the optimum parameters for high sensitivity are shown, with the mean being taken over a relatively high frequency noise band. As previously noted, we used slightly different parameters when taking vector measurements to improve dynamic range and robustness. Figure \ref{fig: ODMRVectorMeasurement} shows the ODMR spectrum with the vector settings: as can be seen the hyperfine features are no longer visible due to the high modulation amplitude. For both the ODMR and time-trace measurements the LIA scaling factor is set to $\times$250. 

\begin{figure}[h!]
\includegraphics[width=\columnwidth, trim={1.5cm 1.5cm 1.5cm 1.5cm}]{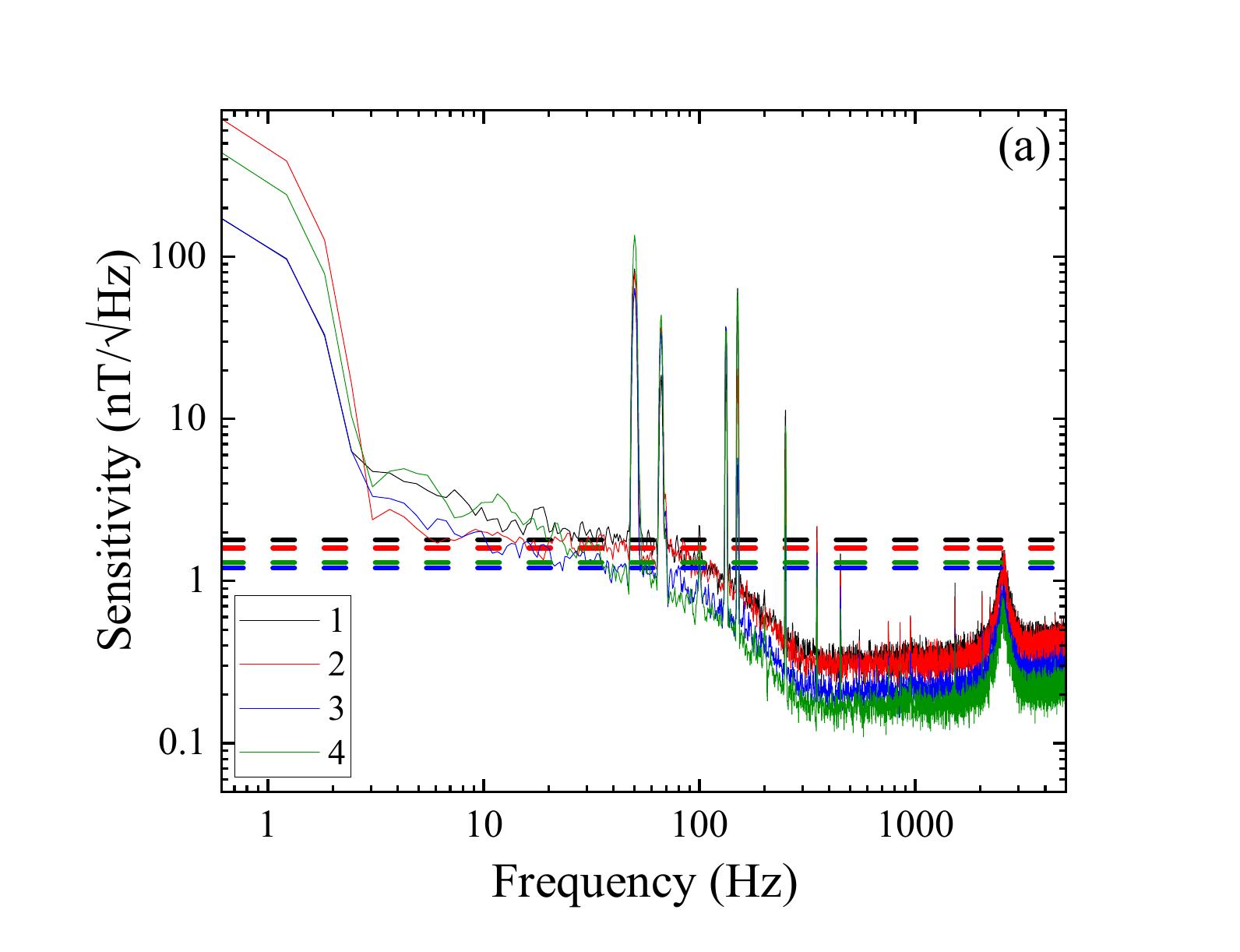}
\includegraphics[width=\columnwidth, trim={1.5cm 1.5cm 1.5cm 1.5cm}]{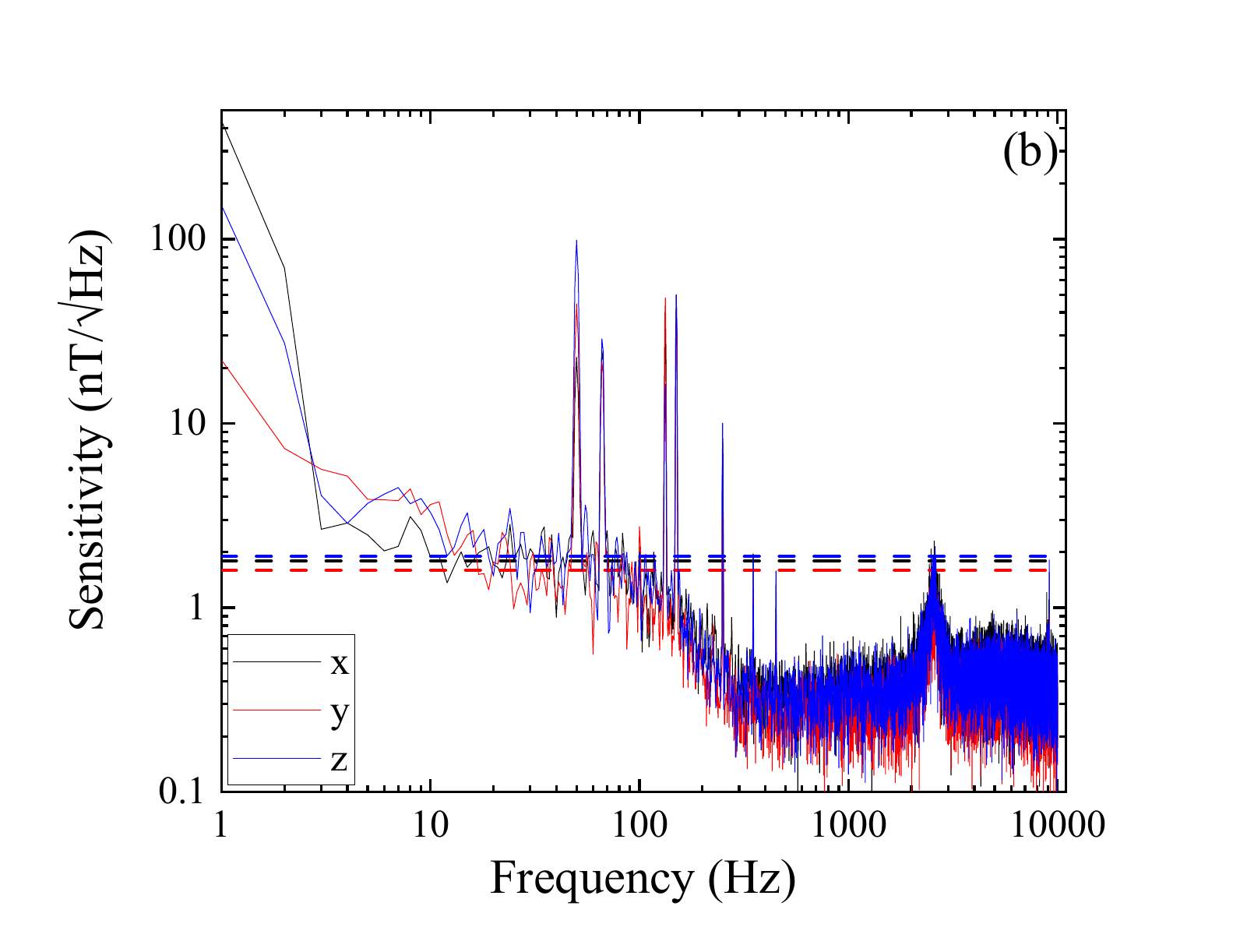} 
\caption{\small (a) Sensitivity spectra for the four NVC symmetry axes 1 to 4 (labelled in Fig. \ref{fig: ODMRVectorMeasurement}) the mean sensitivity value in the frequency range (10-100)-Hz indicated for each.  (b) Sensitivity spectra for the x, y, and z axes without feedback control or vector hopping with the mean sensitivity value in the frequency range (10-100)-Hz indicated for each. For all of these measurements the 3-dB point of the lock-in amplifier low-pass filter was set to 100 Hz. For these measurements a Blackman window is being used for the amplitude-spectral-densities.} 
\label{fig: 4ProjectionsSensitivityMeasurements+xyzVectorSensitivity}
\end{figure}

Figure \ref{fig: 4ProjectionsSensitivityMeasurements+xyzVectorSensitivity}(a) shows the sensitivity measurements taken when considering the four NVC symmetry axes (with associated ODMR resonances) separately in non-vector mode and Fig. \ref{fig: 4ProjectionsSensitivityMeasurements+xyzVectorSensitivity}(b) shows the sensitivity measurements for the x, y, and z axes we defined. The time traces for these measurements are taken with a sampling rate of 10 kHz. The reduction in sensitivity compared to the measurements shown in the main paper are principally due to the use of a modulation amplitude of 3.5 MHz in order to artificially broaden the linewidth and thus improve the dynamic range, and the lack of hyperfine excitation. For the x, y, and z vector measurements we take the four separate 1-s time traces for each of the four ODMR resonances and then convert these into frequency shifts using their respective ZCSs. The vector $\mathbf{A}$-matrix is then employed to convert these into magnetic field noise time traces along x, y, and z. Due to phase differences between the x, y, and z time traces the peak heights of the sinusoidal signals are likely incorrect, though this does not affect the white noise floor. These measurements do not account for additional noise produced by the feedback control method and frequency hopping for vector. For the projection sensitivity measurements the mean sensitivities are found to be (1.8 $\pm$ 0.4) nT/$\sqrt{\textrm{Hz}}$, (1.6 $\pm$ 0.3) nT/$\sqrt{\textrm{Hz}}$, (1.2 $\pm$ 0.3) nT/$\sqrt{\textrm{Hz}}$, and (1.3 $\pm$ 0.6) nT/$\sqrt{\textrm{Hz}}$ for ODMR peaks one, two, three, and four respectively for a (10-100)-Hz frequency range. With the same frequency range, for the vector sensitivity measurements the mean sensitivities are found to be (1.8 $\pm$ 0.4) nT/$\sqrt{\textrm{Hz}}$,(1.6 $\pm$ 0.5) nT/$\sqrt{\textrm{Hz}}$, and (1.9 $\pm$ 0.6) nT/$\sqrt{\textrm{Hz}}$ for the x, y, and z axes respectively. For these measurements 32 time traces were taken, as opposed to 30 for the optimum sensitivity measurements.

\begin{figure}[h!]
\includegraphics[width=\columnwidth, trim={1.5cm 1.5cm 1.5cm 1.5cm}]{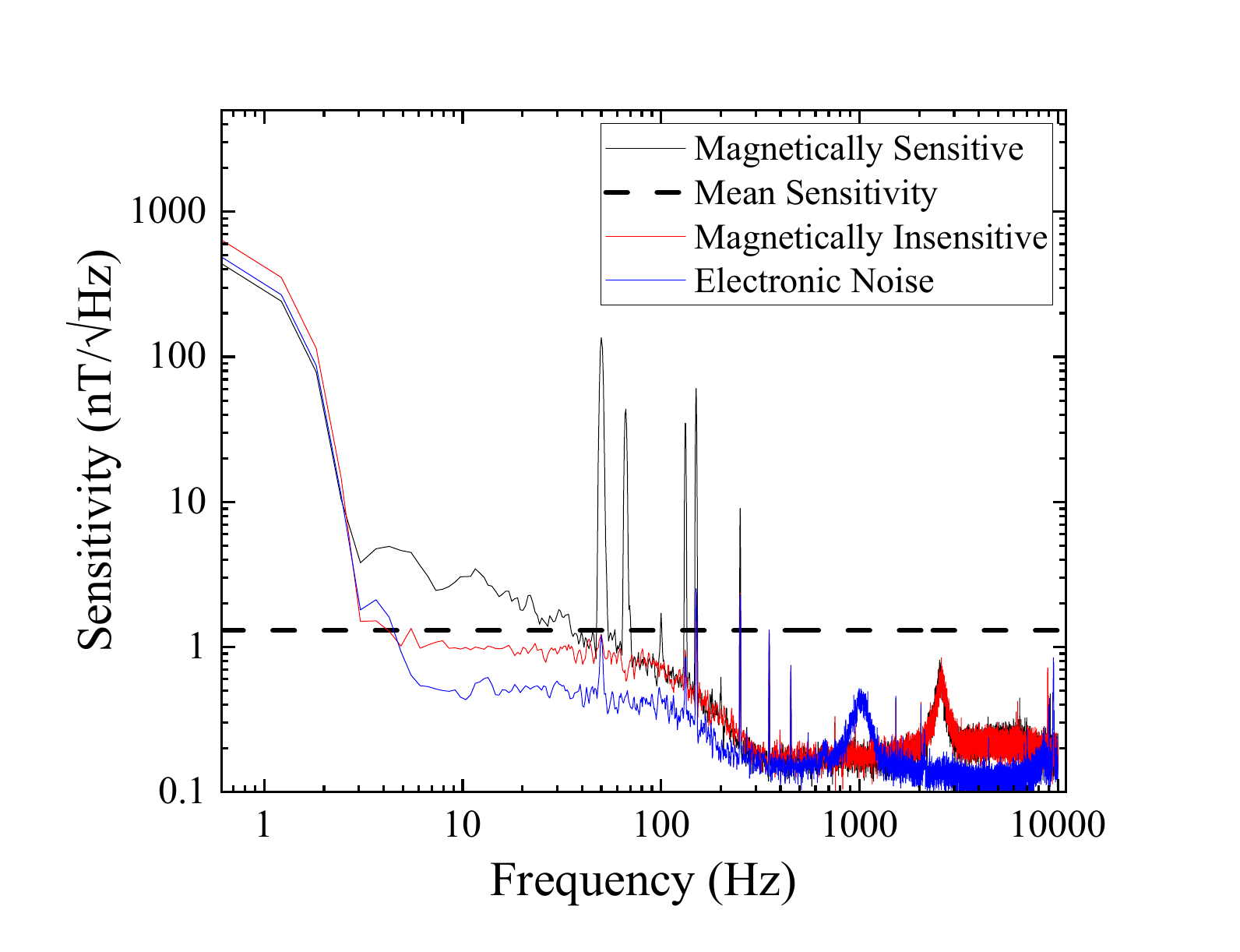} 
\caption{\small Sensitivity spectra of peak 4, using vector settings and without hyperfine excitation with a mean sensitivity of $\break$ (1.3 $\pm$ 0.6) nT/$\sqrt{\textrm{Hz}}$ from (10-100)-Hz. The noise floor is also shown when magnetically insensitive (off-resonance at 3.5 GHz) and with no applied laser or microwaves (electronic noise). For these measurements a Blackman window is being used for the amplitude-spectral-densities.} 
\label{fig: VectorSensitivityLNL}
\end{figure}

In contrast to the optimum sensitivity measurements these are laser noise limited around 50 Hz as is shown in Fig. \ref{fig: VectorSensitivityLNL}. The excess noise on the magnetically sensitive spectra below 40 Hz is not necessarily magnetic in origin, and could be due to temperature fluctuations, laser power fluctuations or even variations in the collection efficiency. Microwave phase noise is unlikely to be a limiting factor at present \cite{ibrahim2020high}.

\begin{figure}[h!]
\includegraphics[width=\columnwidth, trim={1.5cm 1.5cm 1.5cm 1.5cm}]{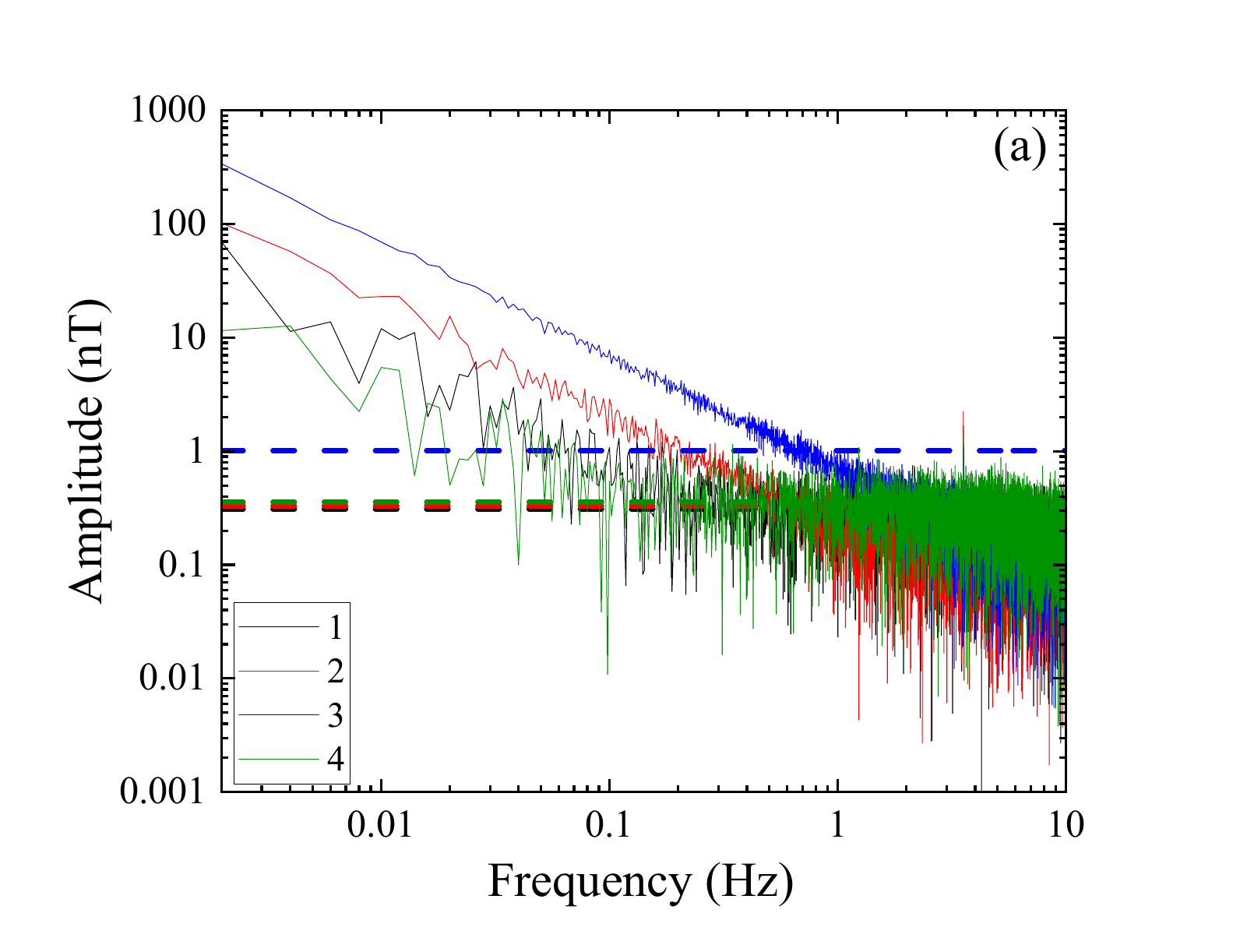}
\includegraphics[width=\columnwidth, trim={1.5cm 1.5cm 1.5cm 1.5cm}]{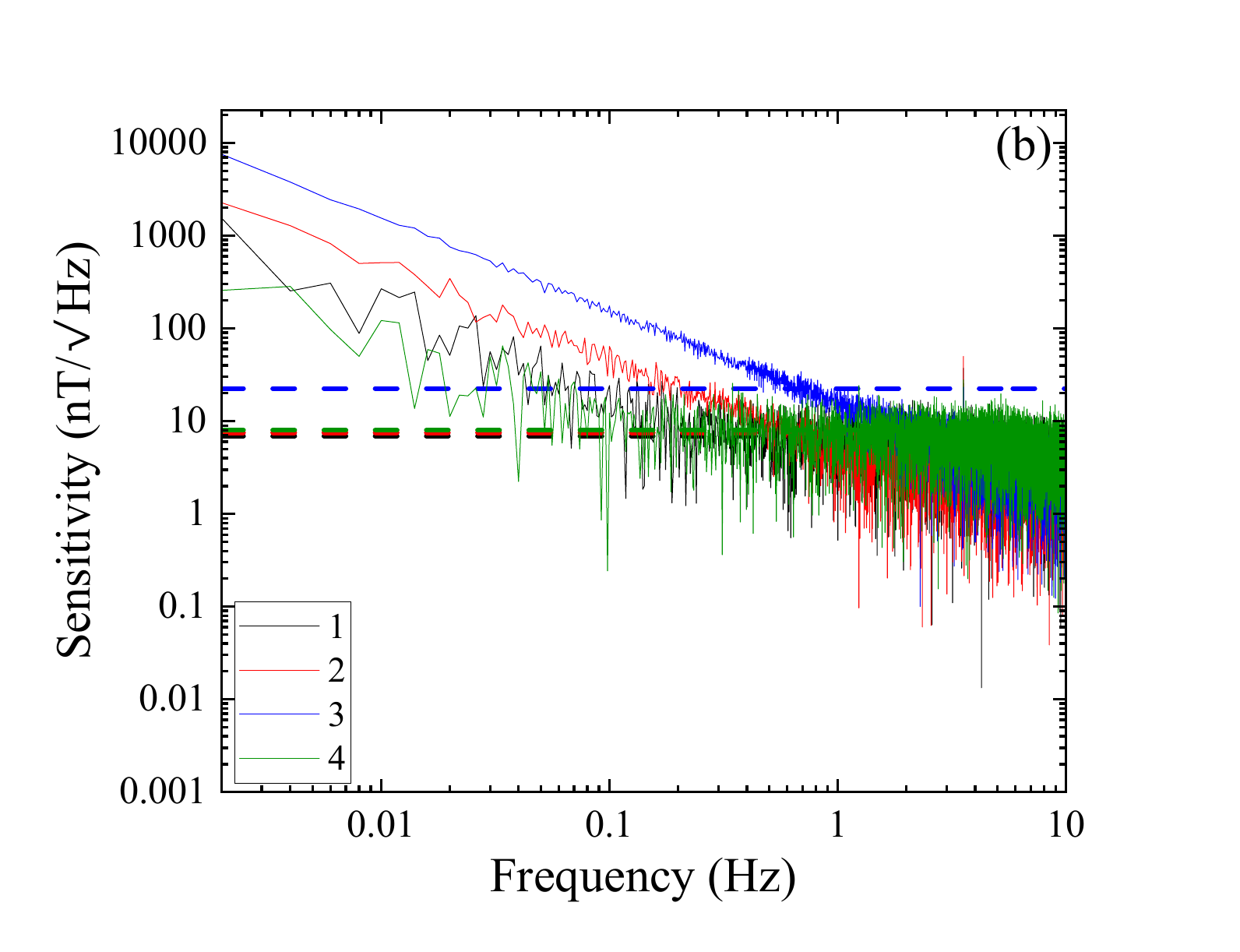} 
\caption{\small (a) Amplitude spectra of the four different optically detected magnetic resonance (ODMR) time traces, the mean values of the amplitude in a (0.1-5)-Hz frequency range are indicated by the dashed lines. The duration of the time traces are 500-s and the bin size is 0.002 Hz. (b) Sensitivity spectra of the four different optically detected magnetic resonance (ODMR) time traces, the mean values of the sensitivity in a (0.1-5)-Hz frequency range are indicated by the dashed lines. A 100 Hz 3-dB point lock-in amplifier low-pass filter is used for these measurements and for the amplitude-spectral-densities a rectangular window.} 
\label{fig: SubHz-SensitivityMeasurements}
\end{figure}

For the magnetic mapping measurements the frequency range of interest is sub-Hz. For a 1-s time trace with a bin size of 1 Hz, the statistics at 1 Hz are very poor. Accordingly longer time traces of 500-s duration are taken, with a sampling rate of 400 Hz, this allowing us to observe the noise level for sub-Hz frequencies through the improved frequency resolution (or bin size) as well as to see the benefits of averaging in the reduction of the white noise floor and subsequent improvement in the signal-to-noise. For these measurements only a single acquisition is taken. Figure \ref{fig: SubHz-SensitivityMeasurements}(a) shows the amplitude spectrum for each of the four ODMR resonance time traces with a bin size of 0.002 Hz, and \ref{fig: SubHz-SensitivityMeasurements}(b) shows the ASD normalised to 1 Hz bin size. 

As can be seen for longer time traces the white noise floor drops as the power is distributed over a greater number of bins, with the bin size $F_{res}$ = 1/$\textrm{T}_{d}$, where $\textrm{T}_{d}$ is the duration of the time trace. Sinusoidal noise peaks remain relatively constant in amplitude, and thus the signal-to-noise is improved though you are potentially more susceptible to 1/f and especially random walk noise at lower frequencies. For the amplitude spectra noise floors of (0.310 $\pm$ 0.008) nT, (0.33 $\pm$ 0.01) nT, (1.01 $\pm$ 0.04) nT, and (1.01 $\pm$ 0.04) nT for ODMR peaks one, two, three, and four respectively are determined. For the ASD spectra mean sensitivities of (6.9 $\pm$ 0.2) nT/$\sqrt{\textrm{Hz}}$, (7.5 $\pm$ 0.3) nT/$\sqrt{\textrm{Hz}}$, (22.5 $\pm$ 0.9) nT/$\sqrt{\textrm{Hz}}$, and (8 $\pm$ 0.2) nT/$\sqrt{\textrm{Hz}}$ are found all in a frequency range of (0.1-5)-Hz. The mean value for the second and especially third peaks is distorted by the high levels of drift in the time traces leading to significant 1/f noise (with a gradient of -1 in the ASD). Given a mean sensitivity of approximately 8 nT/$\sqrt{\textrm{Hz}}$ you would expect a noise level of 0.36 nT, assuming it is broadband white-noise, for 500-s which is consistent with the noise level measured in the amplitude spectrum between (0.1-5)-Hz.  

\begin{figure}[h!]
\includegraphics[width=\columnwidth, trim={1.5cm 1.5cm 1.5cm 1.5cm}]{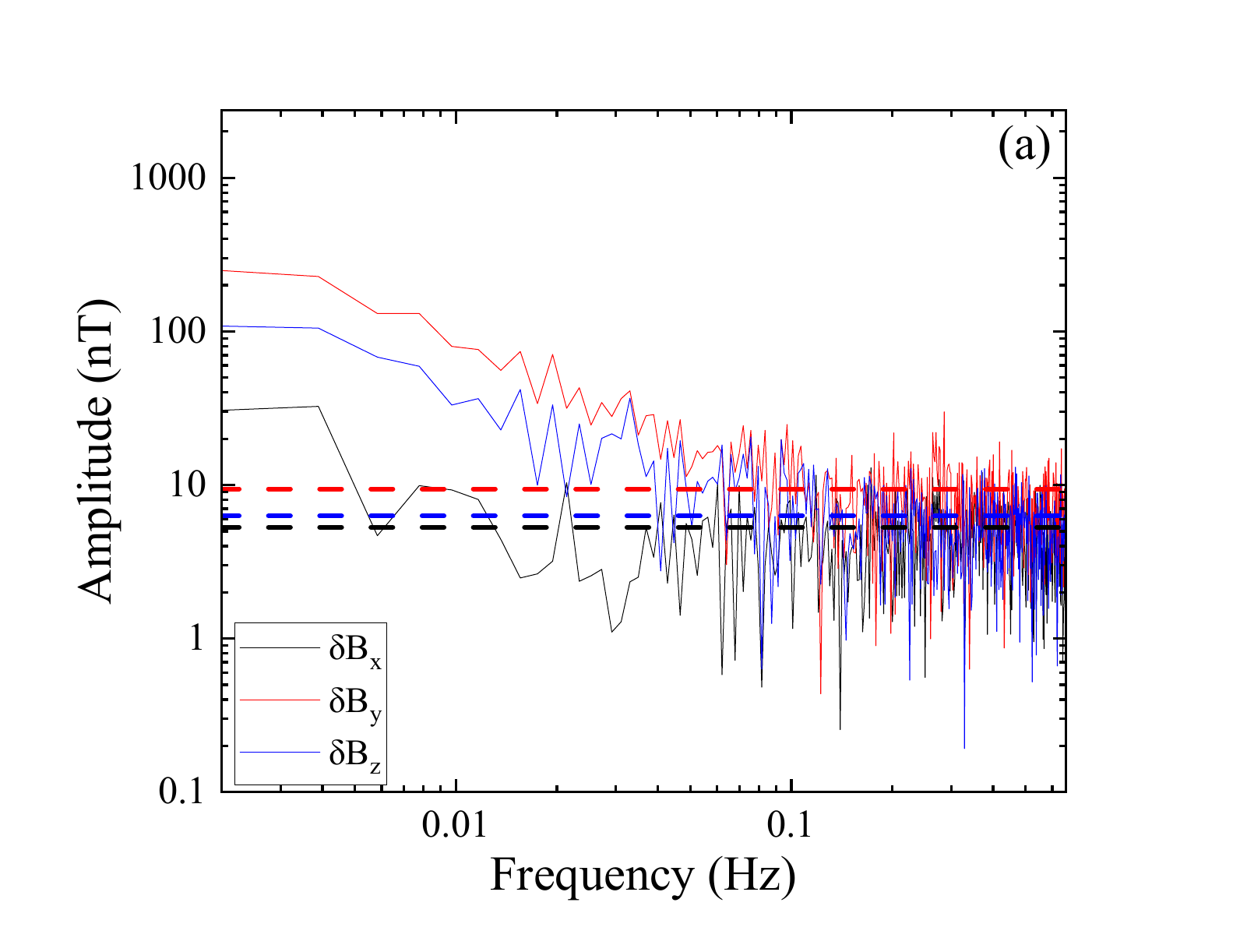}
\includegraphics[width=\columnwidth, trim={1.5cm 1.5cm 1.5cm 1.5cm}]{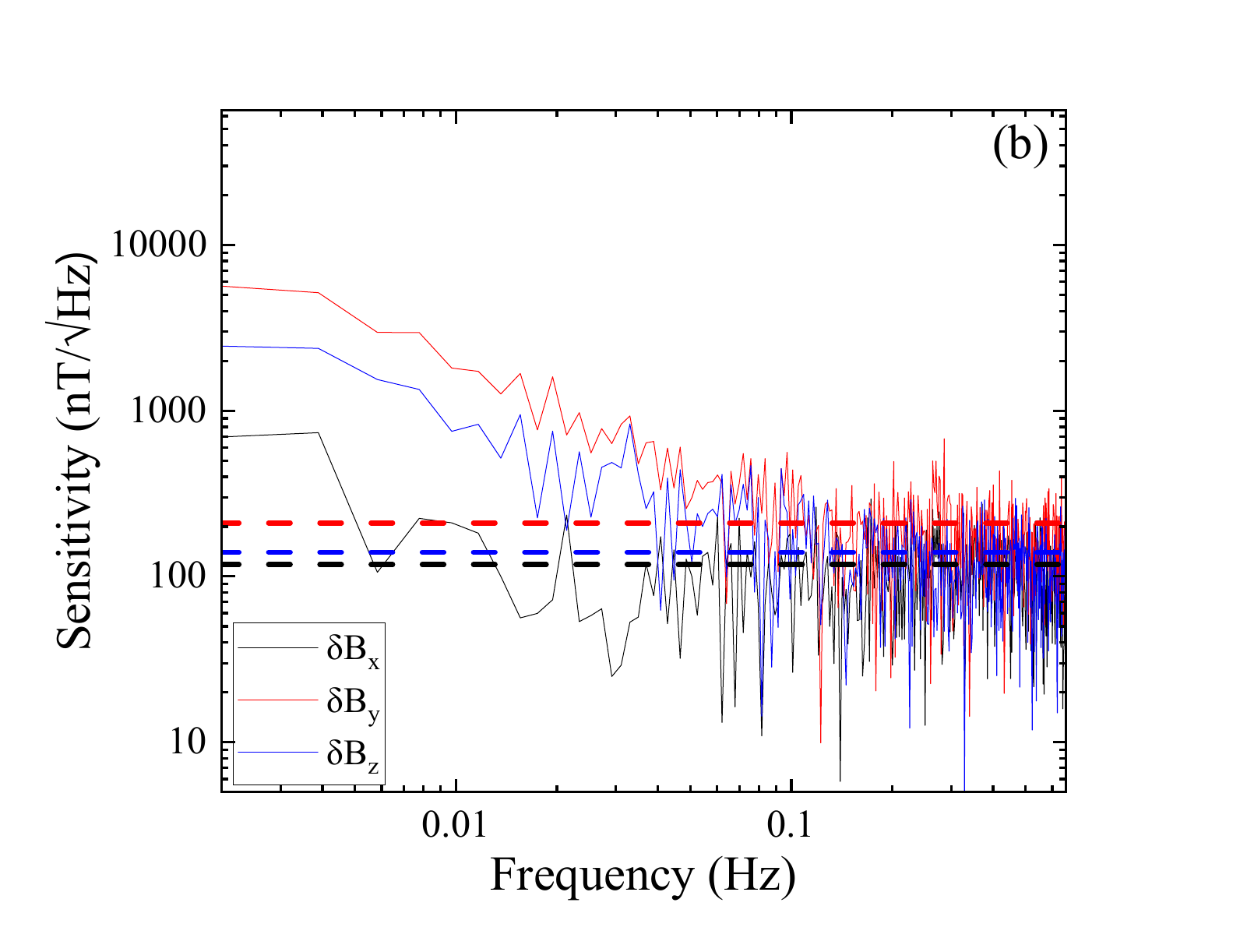}
\caption{\small (a) Amplitude spectra taken of the $\delta\textit{B}_{\textrm{x}}$, $\delta\textit{B}_{\textrm{y}}$, and $\delta\textit{B}_{\textrm{z}}$ vector feedback control time traces with the mean values of the noise floor determined in the 0.1 Hz to 0.66 Hz (white noise section) indicated by the dashed lines. The duration of the time traces are 500-s and the bin size is 0.002 Hz. (b) Sensitivity (ASD) spectra taken of the $\delta\textit{B}_{\textrm{x}}$, $\delta\textit{B}_{\textrm{y}}$, and $\delta\textit{B}_{\textrm{z}}$ vector feedback control time traces with the mean values of the sensitivity determined in the (0.1-0.66)-Hz frequency range indicated by the dashed lines. A rectangular window is used for the amplitude-spectral-densities.} 
\label{fig: vector1.5HzSensitivity}
\end{figure}

Figure \ref{fig: vector1.5HzSensitivity} shows amplitude and ASD measurements taken from the x, y, and z vector feedback controlled time traces (amplitude and ASD thus taken from a single acquisition), and with the limited 1.3 Hz sampling rate. These time traces are approximately 500-s in duration. Mean noise floors of (5.3 $\pm$ 0.3) nT, (9.4 $\pm$ 0.6) nT, and (6.3 $\pm$ 0.4) nT are determined in the (0.1-0.66)-Hz frequency range (neglecting the 1/f and random walk noise frequency band resulting from drift) for the $\delta\textit{B}_{\textrm{x}}$, $\delta\textit{B}_{\textrm{y}}$, and $\delta\textit{B}_{\textrm{z}}$ respectively. Mean sensitivities of (180 $\pm$ 7) nT/$\sqrt{\textrm{Hz}}$, (210 $\pm$ 13) nT/$\sqrt{\textrm{Hz}}$, and (140 $\pm$ 8) nT/$\sqrt{\textrm{Hz}}$ are determined in the (0.1-0.66)-Hz frequency range for the $\delta\textit{B}_{\textrm{x}}$, $\delta\textit{B}_{\textrm{y}}$, and $\delta\textit{B}_{\textrm{z}}$ respectively. The size of the dc magnetic field, relative to the bias field, is better represented by the amplitude spectra. 

In order to correctly obey Parseval's theorem the average of the 30 PSD plots is taken, as opposed to that of the ASD, and then the square root of this average results in the appropriate averaged ASD. Equivalently the root-mean square (rms) average of the 30 ASD plots could be taken \cite{barry2023sensitive}. To obtain the mean sensitivity in the selected frequency range the mean of the ASD plot with the 50 Hz and its harmonics masked is taken. The standard deviation of this mean is taken as the error in the sensitivity. For the non-averaged single acquisition measurements the mean of the selected frequency band is taken from the PSD and then the square root of this is taken: this is equivalent to taking the rms average of the ASD in the selected frequency band. For the non-averaged single-acquisition measurements the standard deviation is comparable or even larger than the mean. The standard error (= standard deviation/$\sqrt{\textrm{N}_{d}}$ where $\textrm{N}_{d}$ is the number of data points in the frequency band) is thus used as the error. 

\FloatBarrier

\section*{Appendix E: Allan Deviation}

AD measurements are taken in order to quantify the levels of noise over different timescales. This helps to demonstrate the stability of the magnetometer and its ability to operate in a wide range of environments. These measurements were made using the vector settings. 

\begin{figure}[h!]
\includegraphics[width=\columnwidth, trim={1.5cm 1.5cm 1.5cm 1.5cm}]{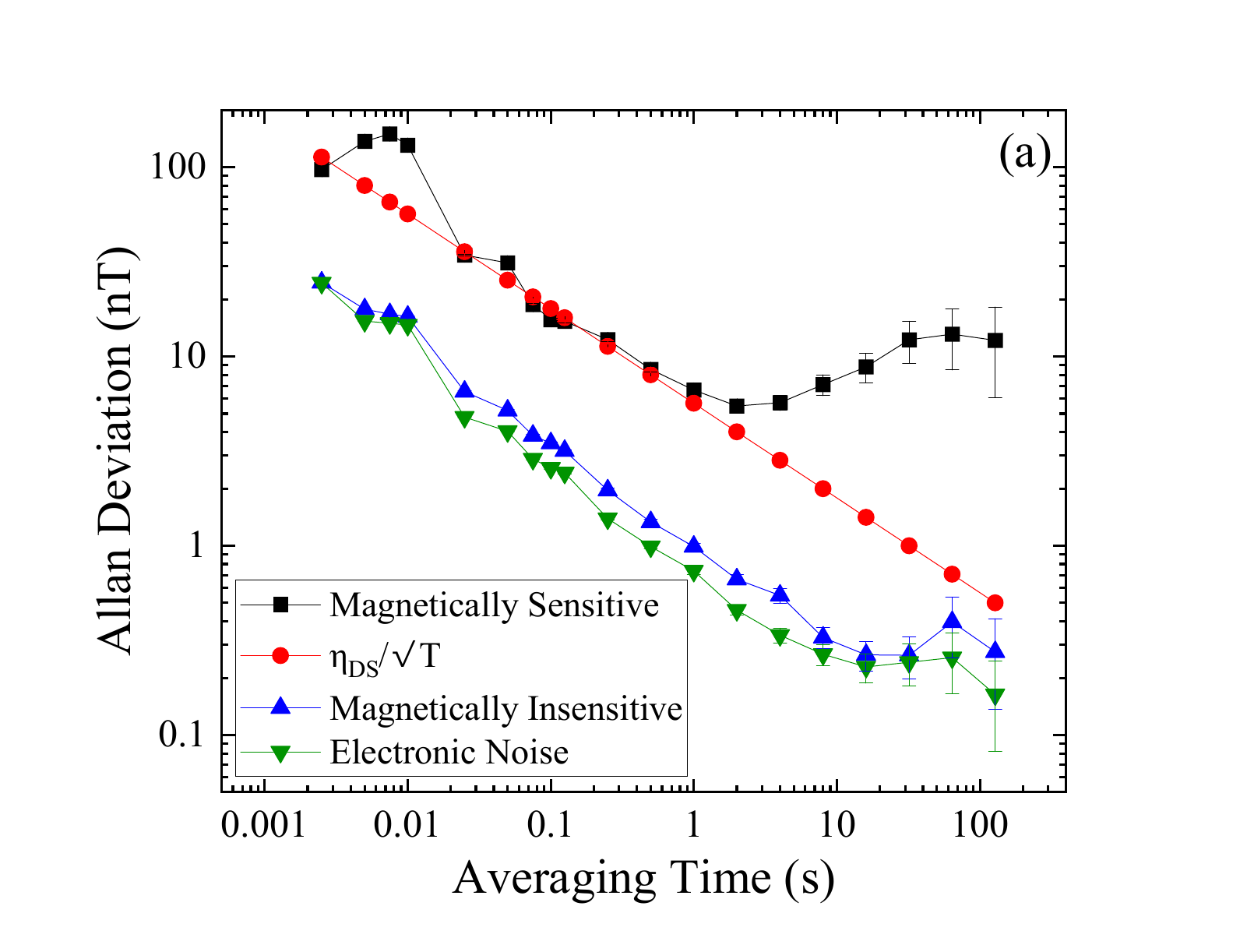}
\includegraphics[width=\columnwidth, trim={1.5cm 1.5cm 1.5cm 1.5cm}]{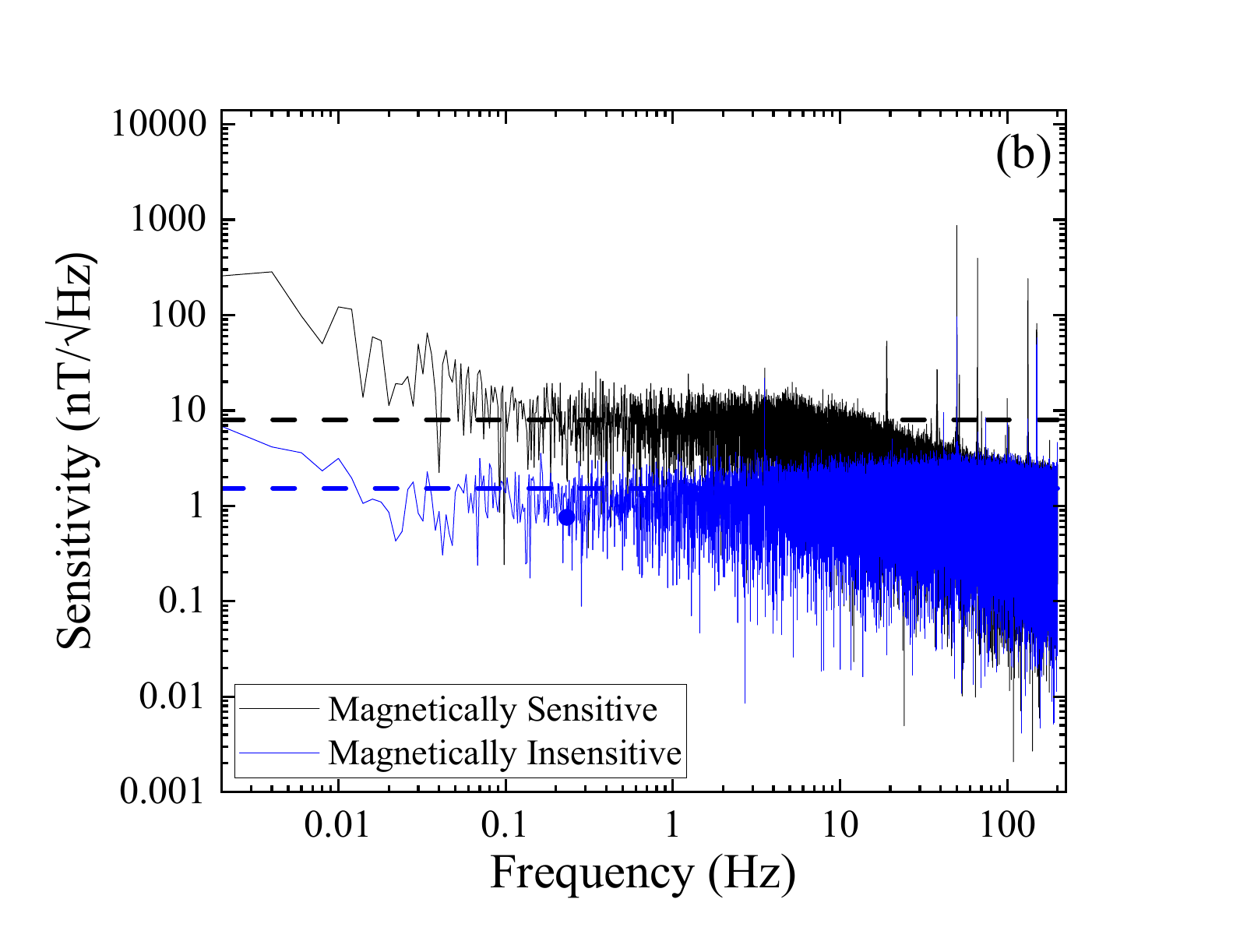}
\caption{\small (a) Allan deviation (AD) measurements of the fourth optically detected magnetic resonance (ODMR) magnetically sensitive time trace, alongside the ADs of time traces taken when magnetically insensitive (off-resonance at 3.5 GHz) and when looking at the electronic noise. The expected AD for each averaging time value, T, given the mean sensitivity (divided by $\sqrt{2}$ to be equivalent to double-sided sensitivity $\eta_{\textrm{DS}}$) determined from the white noise part of the amplitude-spectral-density sensitivity spectra (see Fig. \ref{fig: SubHz-SensitivityMeasurements}) is also shown. (b) Sensitivity spectra of the magnetically sensitive and insensitive case with the mean sensitivities in the 0.1 Hz to 5 Hz frequency range indicated by the dashed lines. The 3-dB point of the lock-in amplifier low-pass filter is set to 100 Hz for all of these measurements.} 
\label{fig: AllanDeviationF4}
\end{figure}

Figure \ref{fig: AllanDeviationF4}(a) shows the AD plots taken when magnetically sensitive (on-resonance), magnetically insensitive (off-resonance), with the laser off, and a theoretical line (AD = $\eta_{\textrm{DS}}$/$\sqrt{\textrm{T}}$ where $\eta_{DS}$ the sensitivity from a double-sided ASD spectrum and T the averaging time) with a gradient of -1/2 based on the mean sensitivity determined from the ASD measurements for the fourth of the labelled ODMR resonances. The time traces from which the AD is calculated are the same as those used for Fig. \ref{fig: SubHz-SensitivityMeasurements}, peak 4, and are 500-s in duration. The AD assumes the noise is frequency independent (i.e. white noise) though in fact sinusoidal noise such as the 50 Hz and its harmonics are present in the voltage trace. The slight oscillations observable in the magnetically sensitive AD at around 1 to 5 ms may be attributed to such sinusoidal noise and they are not observed in the off-resonance and electronic noise (laser off) data which shows -1/2 gradient white noise behaviour, before flattening out (indicating 1/f noise). The drop in the AD below 1.52 ms is because this is below the integration time of the LIA. For the on-resonance data, measurement drift (random walk noise) in the system causes the rise in the AD beyond 1.1-s: this drift is likely due to a combination of the variation in the laser power output (changing the ODMR ZCS) and the heating of the diamond causing changes in the ZFS. Neither of these effects would be apparent when off-resonance and thus the difference between the magnetically sensitive and magnetically insensitive cases cannot be attributed entirely to magnetic noise. Fig. \ref{fig: AllanDeviationF4}(b) shows the sensitivity spectra for the magnetically sensitive and insensitive cases and it can be seen that at higher frequencies the sensitivity is laser-noise limited (being approximately the same on and off-resonance) while sub-30 Hz the flat white noise and random walk noise of the magnetically sensitive case is higher. 

As can be seen the AD at an averaging time of 1-s, which is expected to be equal to the ASD sensitivity (normalised to a bin size of 1 Hz), are in approximate agreement with the mean noise floor of the ASD sensitivity plots. As noted in \cite{fescenko2020diamond} and \cite{barry2023sensitive} the definition of nT/$\sqrt{\textrm{Hz}}$ is inconsistent between the sensitivity calculated using

\begin{equation}
\label{eq:SensitivityStdEquation}
\eta = \textrm{AD}\sqrt{\textrm{T}},
\end{equation}

and that calculated from the amplitude spectral density. The ASD mean noise level in nT/$\sqrt{\textrm{Hz}}$ is a factor of $\sqrt{2}$ higher, with the value from Eq. (\ref{eq:SensitivityStdEquation}) being equivalent to the double-sided ASD noise level, $\eta_{DS}$. To account for this when plotting alongside the AD in Fig. \ref{fig: AllanDeviationF4}(a) the mean sensitivity from the single-sided ASD measurement is divided by $\sqrt{2}$ for consistency. Equation (\ref{eq:SensitivityStdEquation}) does also make the assumption you have frequency-independent white noise, whereas there is considerable 1/f, random walk, and sinusoidal noise in the data as can be clearly seen from the amplitude, ASD, and AD measurements. Therefore even with the appropriate correction factor some small difference between the sensitivity, in nT/$\sqrt{\textrm{Hz}}$, determined from a selected frequency band of the ASD and those calculated from the AD (or standard deviation for that matter) of the time trace is expected. The amplitude, ASD, and AD measurements all give the rms noise level and strictly the sensitivity plots thus have units of $nT_{\textrm{rms}}$/$\sqrt{\textrm{Hz}}$.

\begin{figure}[h!]
\includegraphics[width=\columnwidth, trim={1.5cm 1.5cm 1.5cm 1.5cm}]{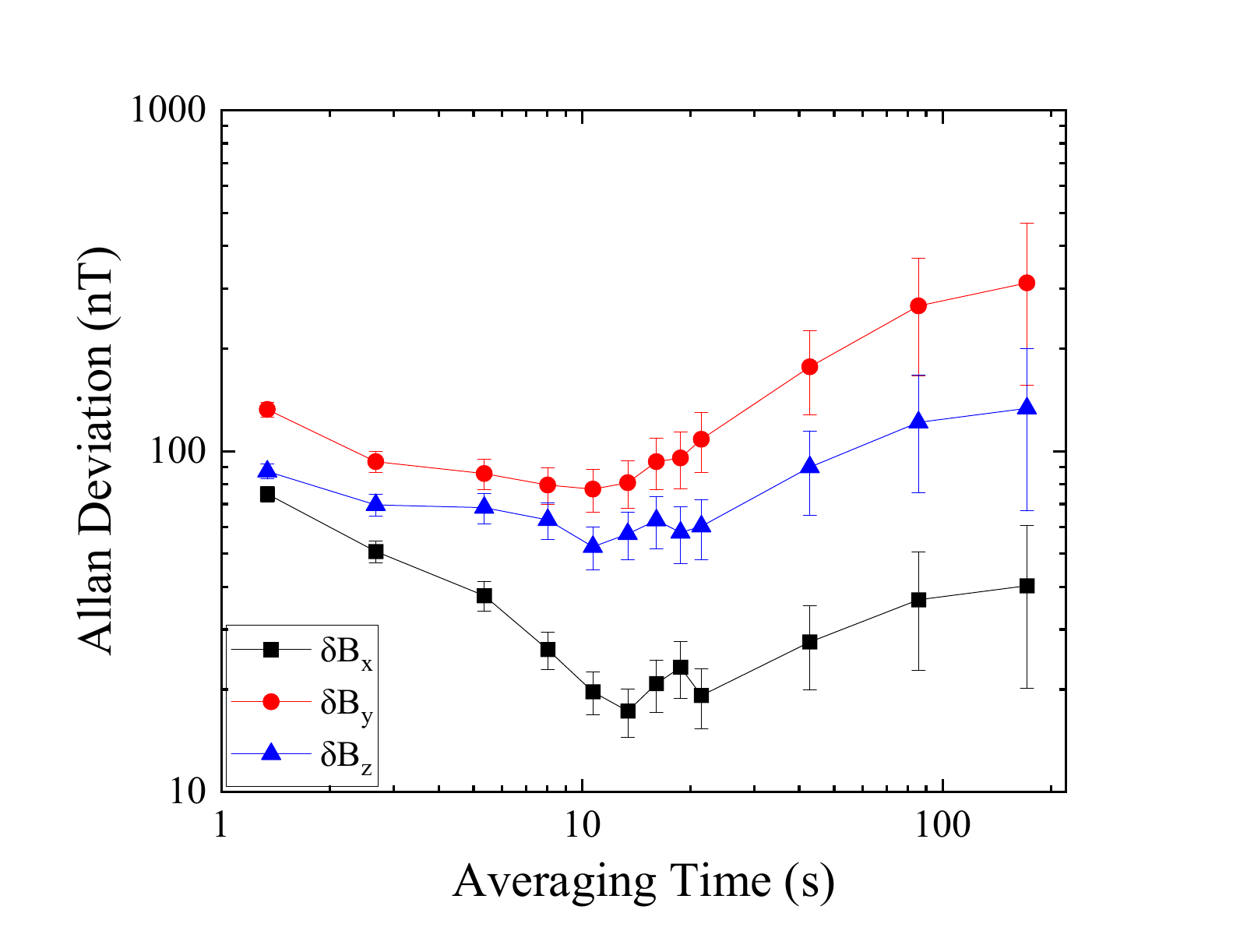}
\caption{\small Allan deviation (AD) measurements of the $\delta\textit{B}_{\textrm{x}}$, $\delta\textit{B}_{\textrm{y}}$, and $\delta\textit{B}_{\textrm{z}}$ vector feedback-controlled time traces assuming an averaging sampling rate of 1.33881 Hz.} 
\label{fig: AllanDeviationxyzVector}
\end{figure}

Figure \ref{fig: AllanDeviationxyzVector} shows the AD measurements taken on the vector x, y, and z time traces. The 1.3 Hz bandwidth means the minimum averaging time is approximately 0.75 s. These are consistent with the ASD measurements seen in Fig. \ref{fig: vector1.5HzSensitivity} above.  

With these measurements we are seeing the drift in the feedback control and vector over time starting from the initial frequency (magnetic field). This gives us some idea of the stability of the feedback and vector system. There is a noticeable drift in the y and z axes - this may be due to temperature and laser power induced fluctuations or potentially cross-talk between slightly overlapping ODMR resonances. Drift is also seen in the single projection AD measurements. It is possible that there are also slowly varying magnetic fields from magnetic mechanical objects on the trolley and from the bias field. All the measurements are taken without magnetic shielding. Looking at the single peak-tracking alone, this drift is not observed - suggesting it may be related to the sequential hopping. For instance, the microwave amplifier gain has a strong frequency dependence and thus the amount of power delivered to the diamond may be rapidly changing with the frequency - causing rapid changes in the temperature of the diamond. This issue could be solved by using a separate microwave source to address each symmetry axis \cite{schloss2018simultaneous}. In any case, clearly the long-term stability of the feedback control and vector could be significantly improved beyond its present performance. 

\FloatBarrier

\section*{Appendix F: Vector Method and Calibration}

The voltage data from the LIA is streamed directly into our Python vector and feedback control code with a sampling rate of 1.673 kHz. This data is unscaled, but is multiplied by $\times$250 in the code to match the ODMR ZCSs measured using the PicoScope oscilloscope. The four initial frequencies of the $m_s$ = 0 to $m_s$ = +1 resonances are set by the operator. During initialisation, the code measures the initial voltage at these zero-crossing points. When the magnetic fields at the diamond change, causing a shift in the Zeeman splitting for each resonance pair the voltage changes. This change in voltage relative to the initial value can then be used to calculate the frequency shift, using the ZCS of each of the four resonances. The frequency shifts can also be added onto the initial frequencies and sent to the Agilent N5172B microwave source for feedback control.  

To convert the four frequency shifts into $\delta\textit{B}_{\textrm{x}}$, $\delta\textit{B}_{\textrm{y}}$, and $\delta\textit{B}_{\textrm{z}}$, with x, y, and z being defined by carefully aligning a Helmholtz coil along three orthogonal axes, a vector $\mathbf{A}$-matrix made up of effective gyromagnetic ratios (showing the shift in frequency for each ODMR resonance for a known field applied along the respective axes) is obtained \cite{schloss2018simultaneous, newman2024tensor}. This matrix takes the form

\begin{equation}
\label{eq:AMatrix}
\mathbf{A} = \begin{pmatrix}
\frac{\partial{f_1}}{\partial{(\delta B_{\textrm{x}}})} & \frac{\partial{f_1}}{\partial{(\delta B_{\textrm{y}}})} & \frac{\partial{f_1}}{\partial{(\delta B_{\textrm{z}}})}\\
\frac{\partial{f_2}}{\partial{(\delta B_{\textrm{x}}})} & \frac{\partial{f_2}}{\partial{(\delta B_{\textrm{y}}})} & \frac{\partial{f_2}}{\partial{(\delta B_{\textrm{z}}})}\\
\frac{\partial{f_3}}{\partial{(\delta B_{\textrm{x}}})} & \frac{\partial{f_3}}{\partial{(\delta B_{\textrm{y}}})} & \frac{\partial{f_3}}{\partial{(\delta B_{\textrm{z}}})}\\
\frac{\partial{f_4}}{\partial{(\delta B_{\textrm{x}}})} & \frac{\partial{f_4}}{\partial{(\delta B_{\textrm{y}}})} & \frac{\partial{f_4}}{\partial{(\delta B_{\textrm{z}}})}\\
\end{pmatrix}
,
\end{equation}

and from our calibration procedure is determined to be

\begin{equation}
\label{eq:AMatrixExample}
\mathbf{A} \, (\textrm{MHz}/\textrm{mT}) = \begin{pmatrix}
-16.7 & -4.3 & 14.3\\
25.5 & -3.2 & 16.3\\
6.95 & -20.3 & -11.9\\
9.9 & -18.3 & 14.4\\
\end{pmatrix}
.
\end{equation}

Without feedback control the magnetometer's dynamic range is given by $\delta$B = $\delta$$\textit{f}$/$\gamma$, where $\delta$$\textit{f}$ is the linwidth of the ODMR resonance \cite{wang2022portable}. The use of a modulation amplitude larger than the approximately 1 MHz resonance linewidths reduces the potential sensitivity (by reducing the ZCS) but increases the effective linewidth of the LIA output and thus the dynamic range. However, the response is not entirely linear over the entire derivative lineshape resonance especially with the distortion caused by the large modulation amplitude. The dynamic range of the vector measurements with feedback control in principle would be limited by the frequency range of the microwave source, as well as the bias field alignment and strength. Shifts in the ODMR resonance position due to external magnetic fields can, depending on the separation of the resonances and the magnitude of the field, cause the ODMR resonances to be overlapped, as for an externally applied field the shifts are not necessarily either all positive or negative. This cross-talk means you can no longer take accurate measurements of the field along the defined x, y, and z axes. The minimum separation between the zero-crossing points of the $m_s$ = 0 to $m_s$ = +1 resonances seen in Fig. \ref{fig: ODMRVectorMeasurement} is approximately 26 MHz. From applying known test fields it is found that feedback control is typically lost when total fields of around 700 $\mu$T are applied. This makes sense given if a test field of this magnitude were applied directly along one of the four the NVC symmetry axes it would shift the resonance by 19.6 MHz, and it would produce equal frequency shifts of approximately 7 MHz for the remaining three resonances, and depending on the signs this could cause overlapping of the resonances. All this assumes you remain in the linear low-field regime (approximately $\leq$ 100 mT transverse field). Additionally, for larger fields the effect of transverse magnetic fields, as opposed to just the projection along the NVC symmetry axes become more significant \cite{rondin2014magnetometry}. For fields significantly larger than the anti-crossing point the NVC Hamiltonian is dominated by the Zeeman term, as opposed to the zero-field term \cite{rondin2014magnetometry, doherty2013nitrogen}. Rapid changes in the magnetic field can also cause a loss of feedback control.

\section*{Appendix G: Additional Vector Maps in Laboratory and Van(s)}

\begin{figure*}[t]
\hspace{-0.5cm}\vspace{0.1cm}\includegraphics[width=\columnwidth]{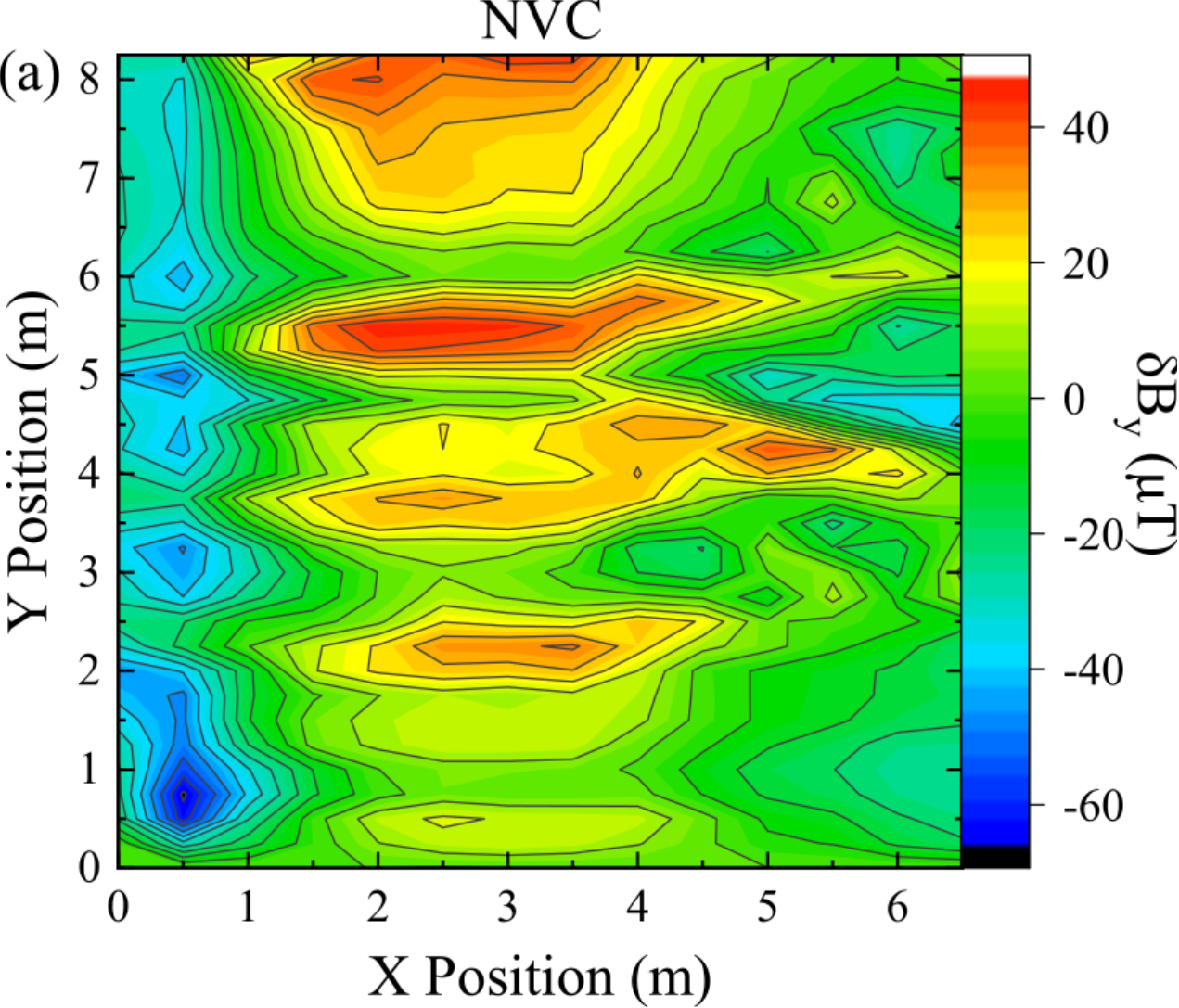} 
\hspace{0.5cm}\vspace{0.1cm}\includegraphics[width=\columnwidth]{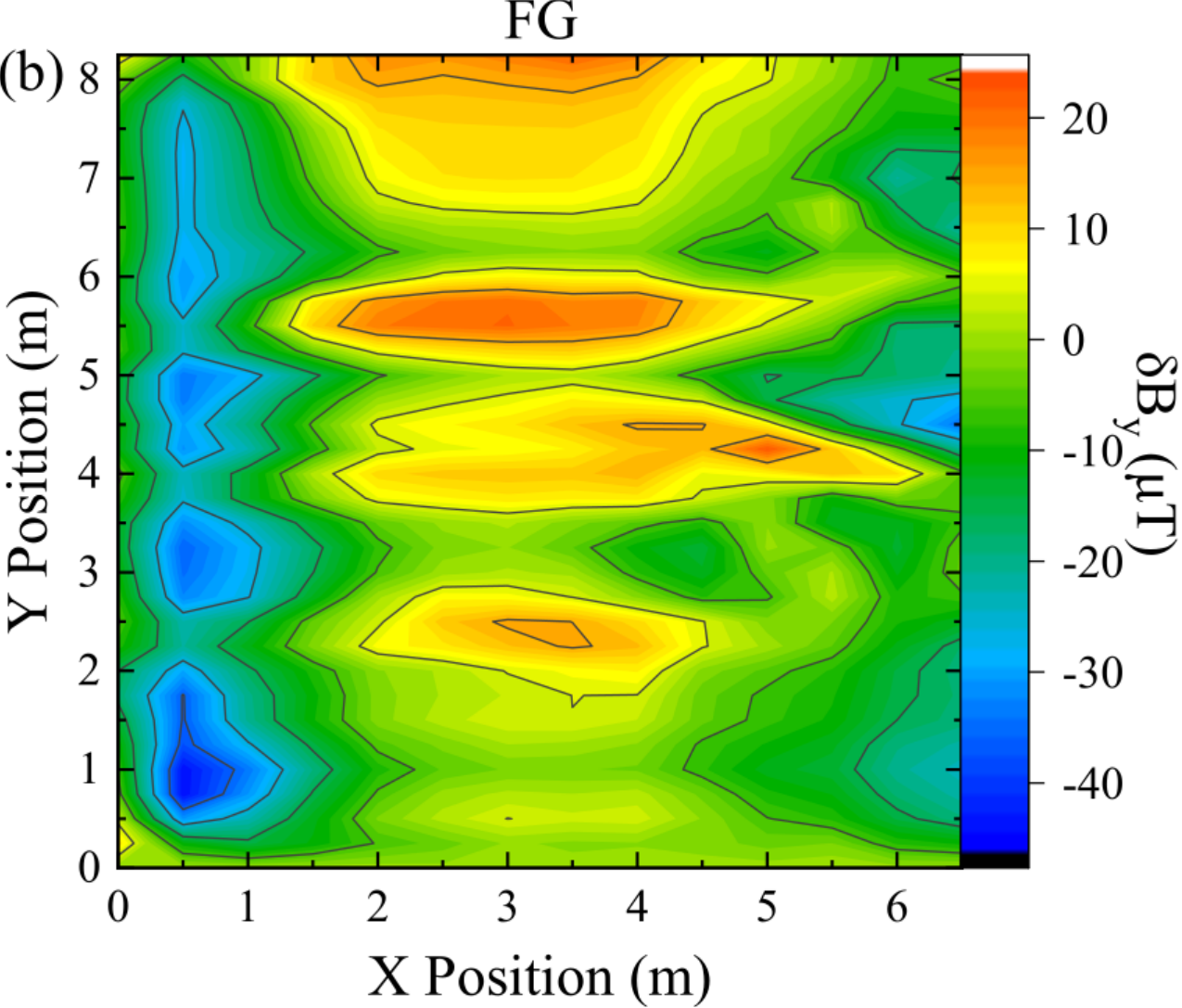} 
\hspace{-0.5cm}\includegraphics[width=\columnwidth]{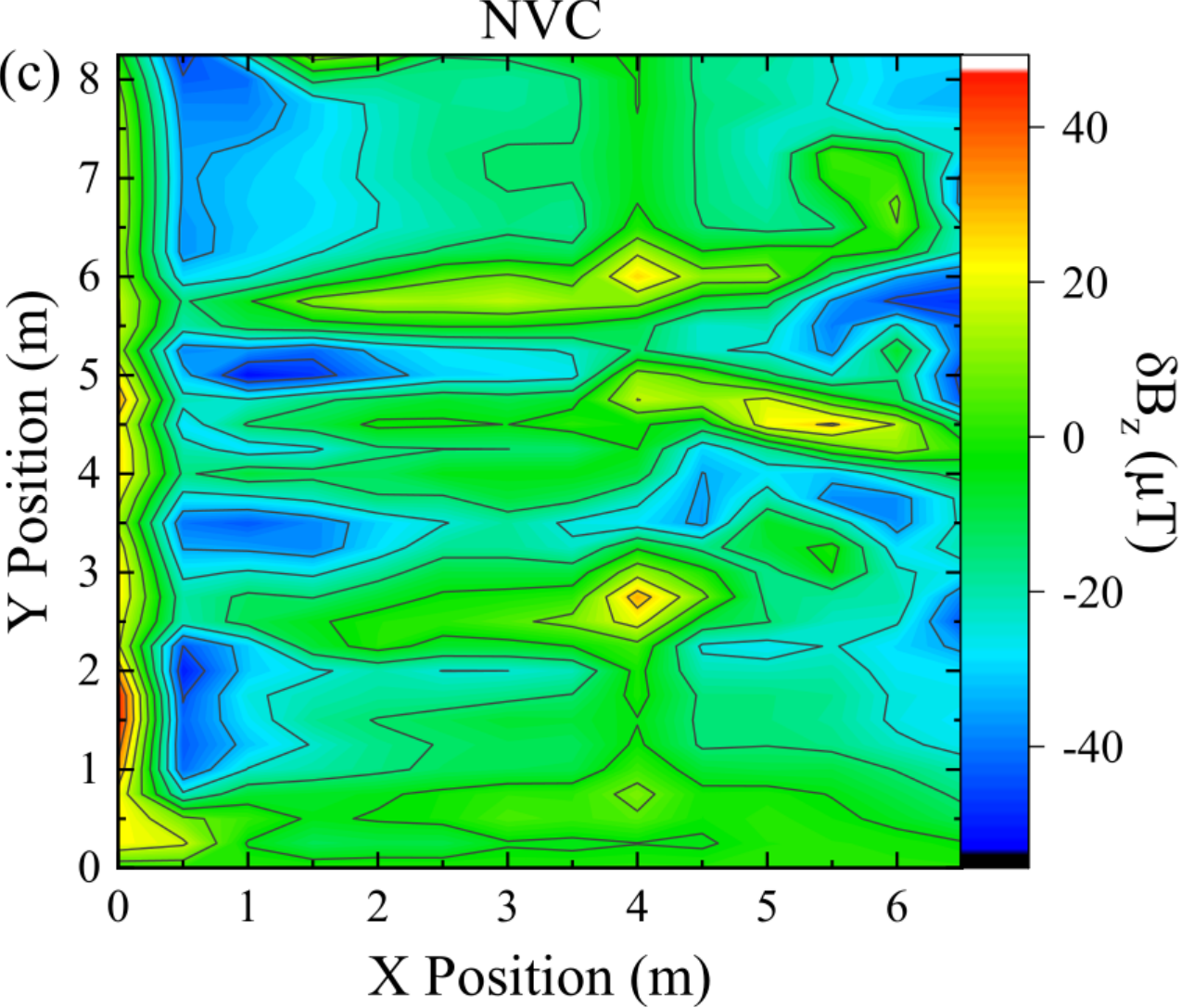} 
\hspace{0.5cm}\includegraphics[width=\columnwidth]{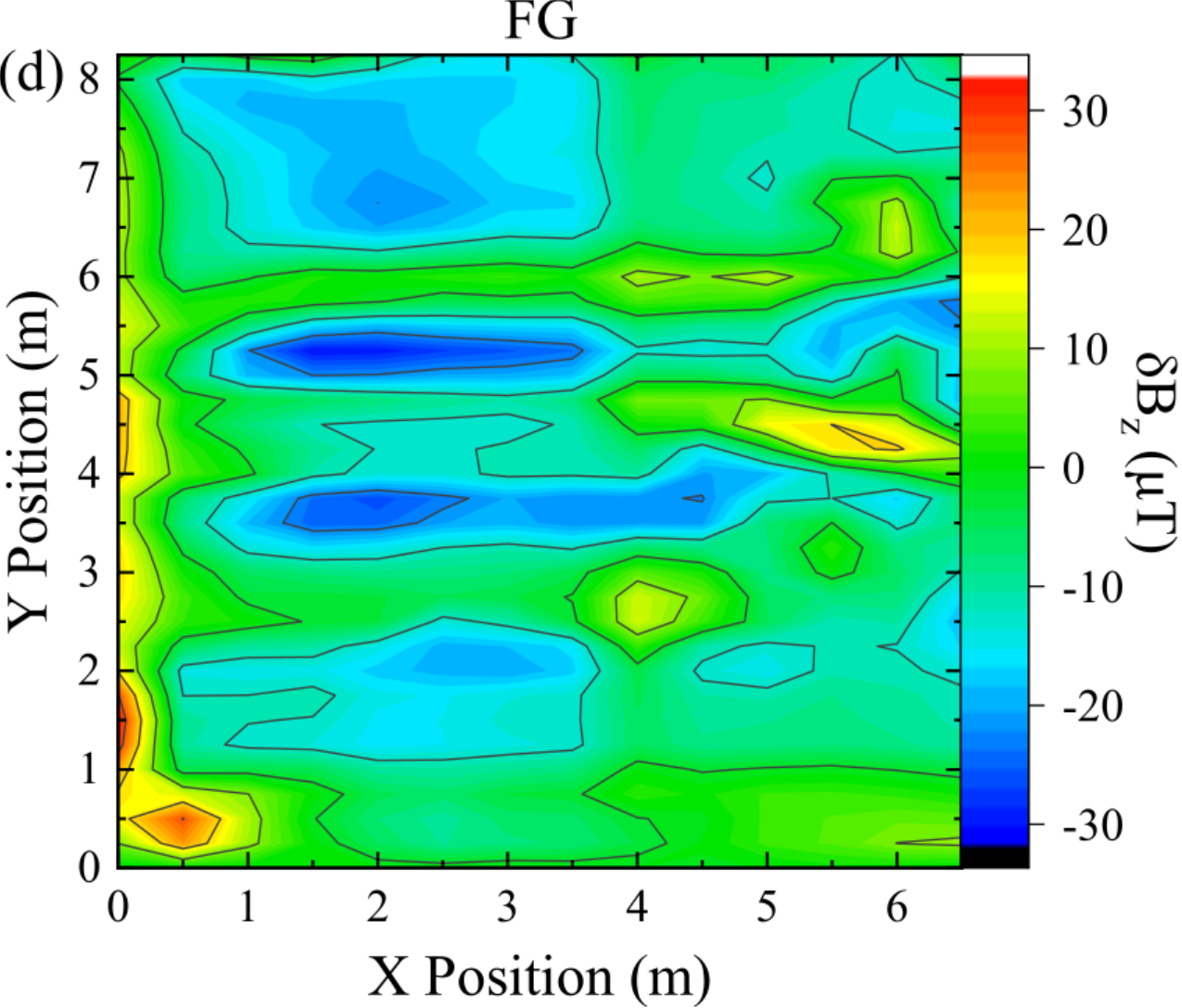} 
\caption{\small Laboratory room 2D point maps of the $\delta\textit{B}_{\textrm{y}}$ ((a) and (b)), and $\delta\textit{B}_{\textrm{z}}$, ((c) and (d)) with 0.5 m and 0.25 m resolutions along the X and Y axes respectively taken with the NVC and fluxgate (FG) magnetometers as labelled.} 
\label{fig: 2DMapByBz}
\end{figure*}

In the main paper Fig. \ref{fig: Lab2DPointMap} shows 2D point maps of the $\delta\textit{B}_{\textrm{x}}$ component of the magnetic field measured with the FG and NVC. Figure \ref{fig: 2DMapByBz} shows the 2D point maps taken with the FG and NVC for the $\delta\textit{B}_{\textrm{y}}$ and $\delta\textit{B}_{\textrm{z}}$ components. As can be seen these are also in qualitative agreement, with the differences in magnitude being at least in part attributable to the difference in position of the FG and NVC magnetometers. 

\begin{figure*}[t]
\hspace{-0.5cm}\includegraphics[width=\columnwidth]{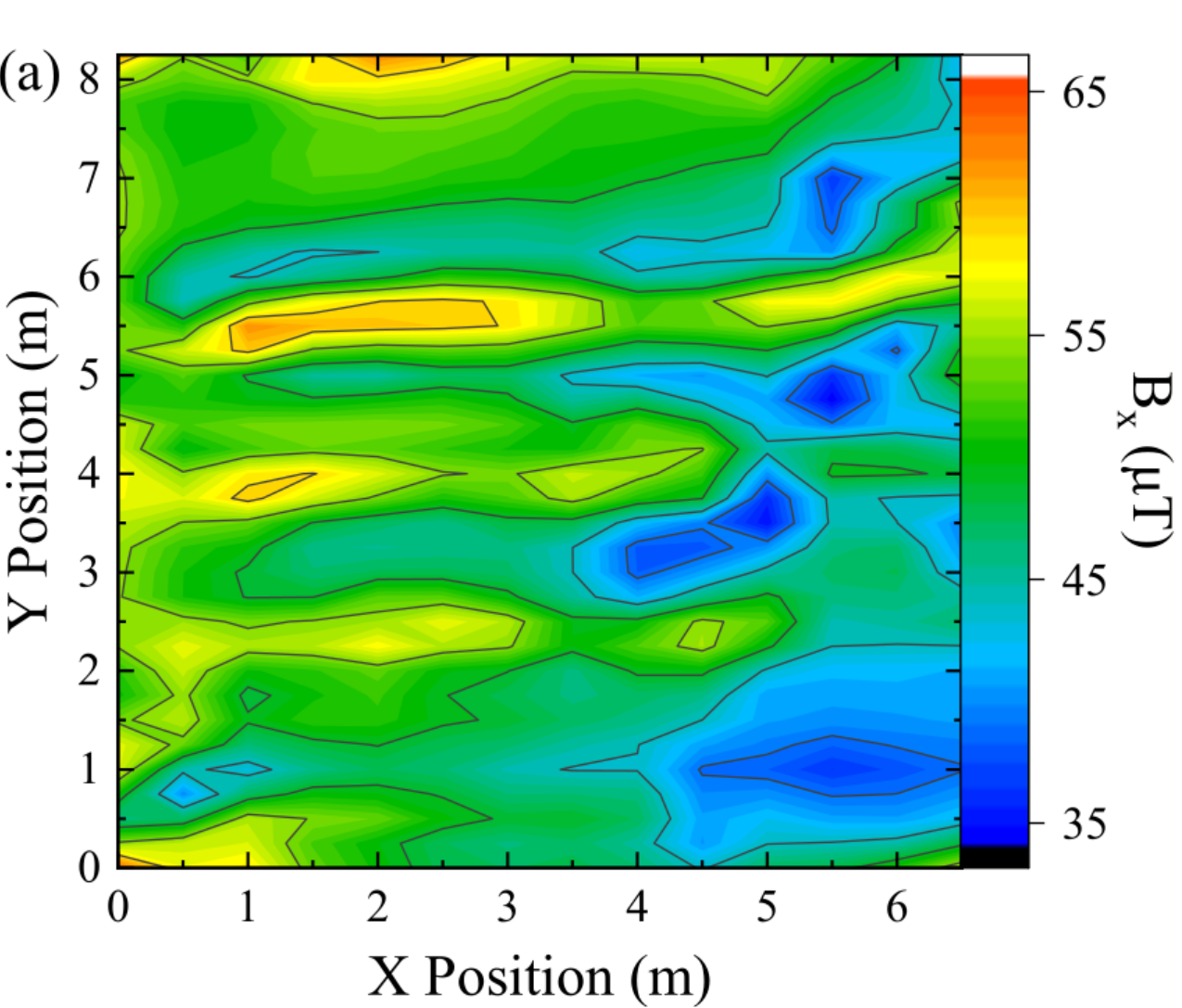} 
\hspace{0.5cm}\includegraphics[width=\columnwidth]{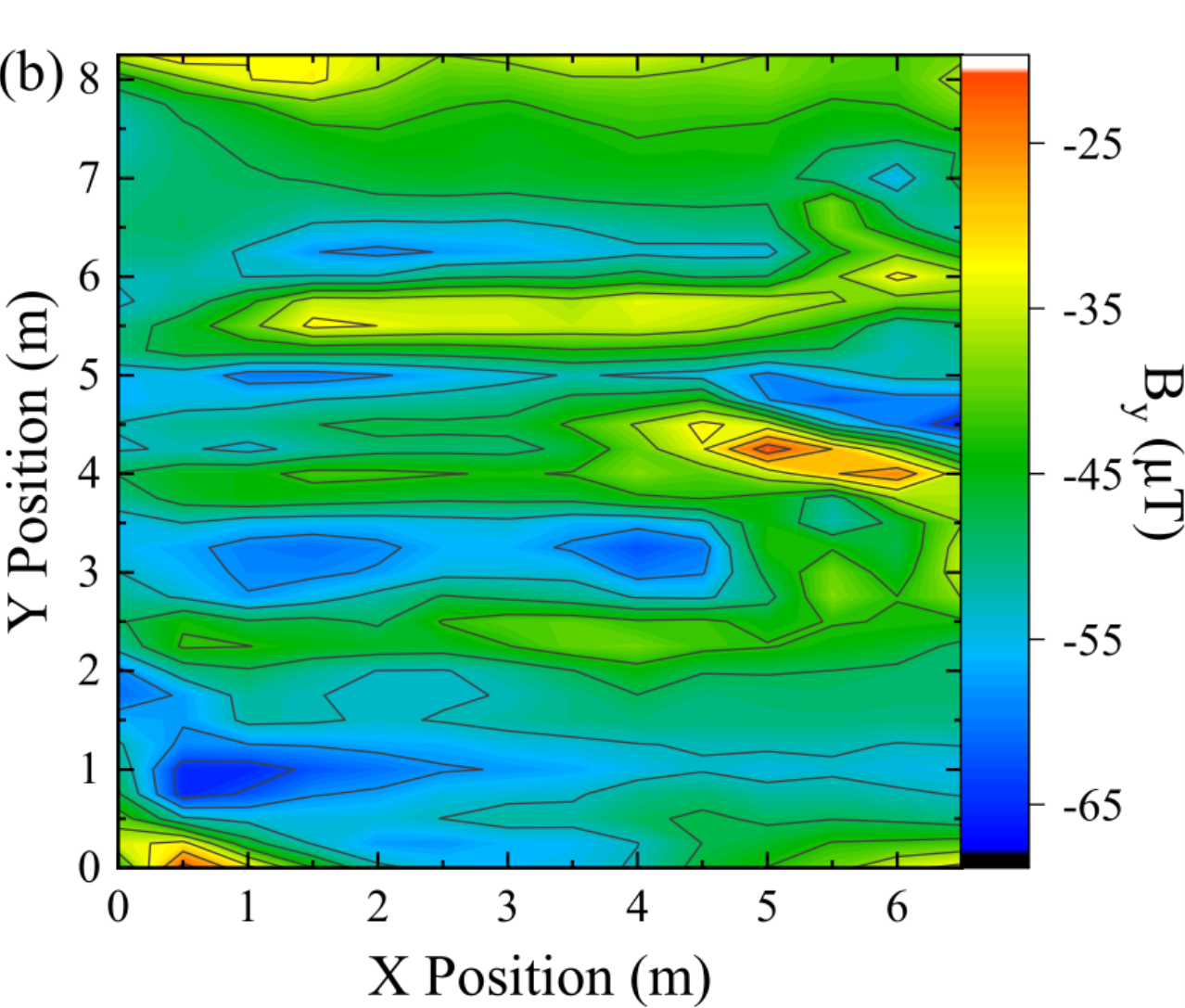} 
\hspace{-0.5cm}\includegraphics[width=\columnwidth]{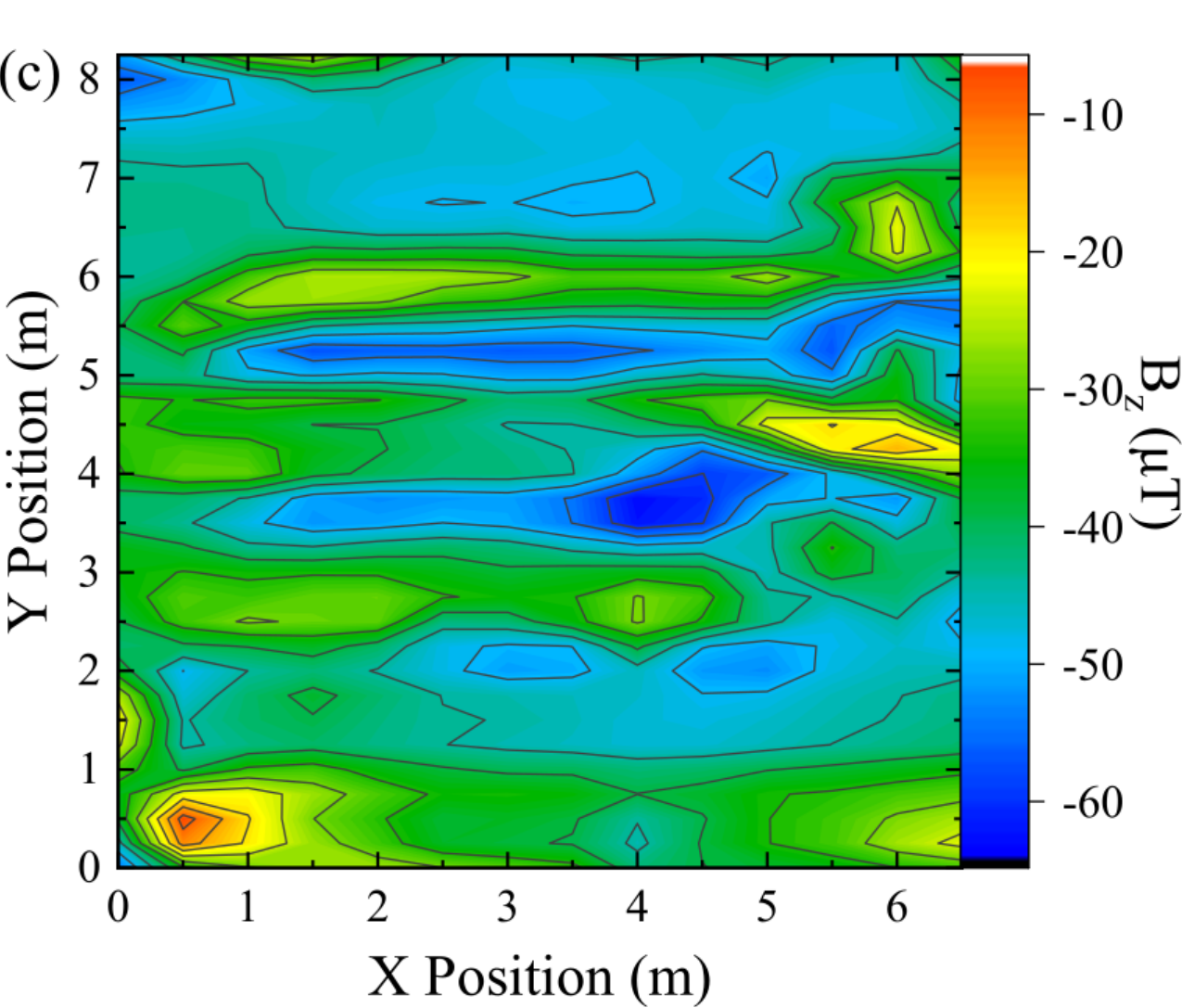} 
\hspace{0.5cm}\includegraphics[width=\columnwidth]{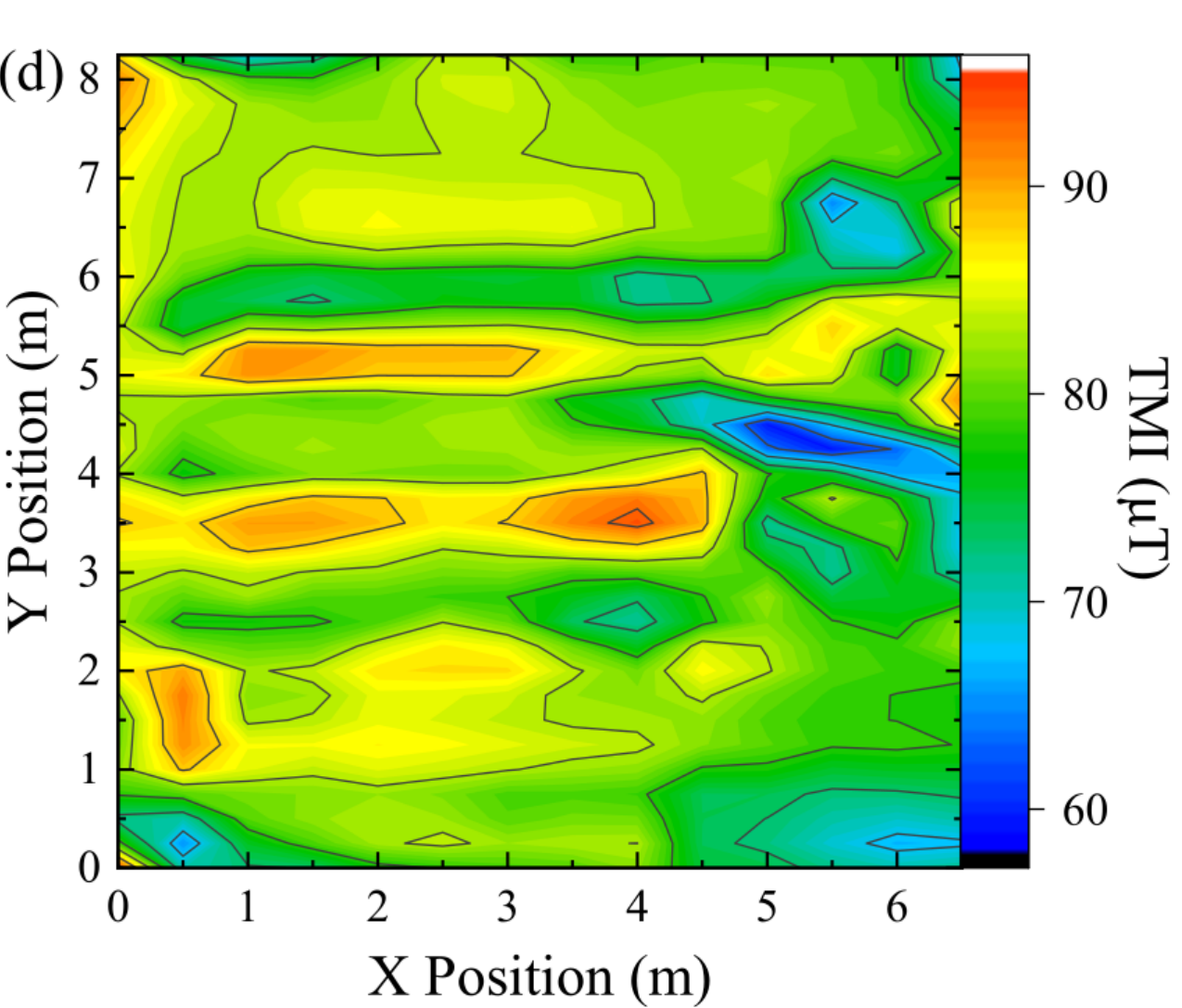} 
\caption{\small Laboratory room 2D point maps of the $\textit{B}_{\textrm{x}}$ (a), $\textit{B}_{\textrm{y}}$ (b), $\textit{B}_{\textrm{z}}$ (c), and total magnetic intensity (TMI) (d) with 0.5 m and 0.25 m resolutions along the X and Y axes respectively. All measurements taken with the fluxgate (FG) magnetometer.} 
\label{fig: 2DMapAbsFGBxByBzTMI}
\end{figure*}

Unlike the NVC magnetometer, with which we are measuring the shift in magnetic field relative to some initial value, the FG is capable of measuring the absolute magnetic field at a position. As the vector sum of the permanent fields from the trolley, a notable contribution of which to the FG is the bias magnet of the NVC magnetometer, are fixed in the reference frame of the FG the shift seen in the absolute magnetic field may be largely attributed to the changing contribution of the Earth's magnetic field. Figure \ref{fig: 2DMapAbsFGBxByBzTMI} shows the absolute magnetic field 2D map of the laboratory measured with the FG for the x, y, and z axes as well as the TMI. It is not possible to take the TMI from the NVC magnetic field shift measurements as TMI = $\sqrt{\textit{B}_{\textrm{x}}^2+\textit{B}_{\textrm{y}}^2+\textit{B}_{\textrm{z}}^2}$ is not equivalent to $\sqrt{\delta\textit{B}_{\textrm{x}}^2+\delta\textit{B}_{\textrm{y}}^2+\delta\textit{B}_{\textrm{z}}^2}$.

\begin{figure*}[h!]
\hspace{-0.5cm}\vspace{0.05cm}\includegraphics[width=\columnwidth]{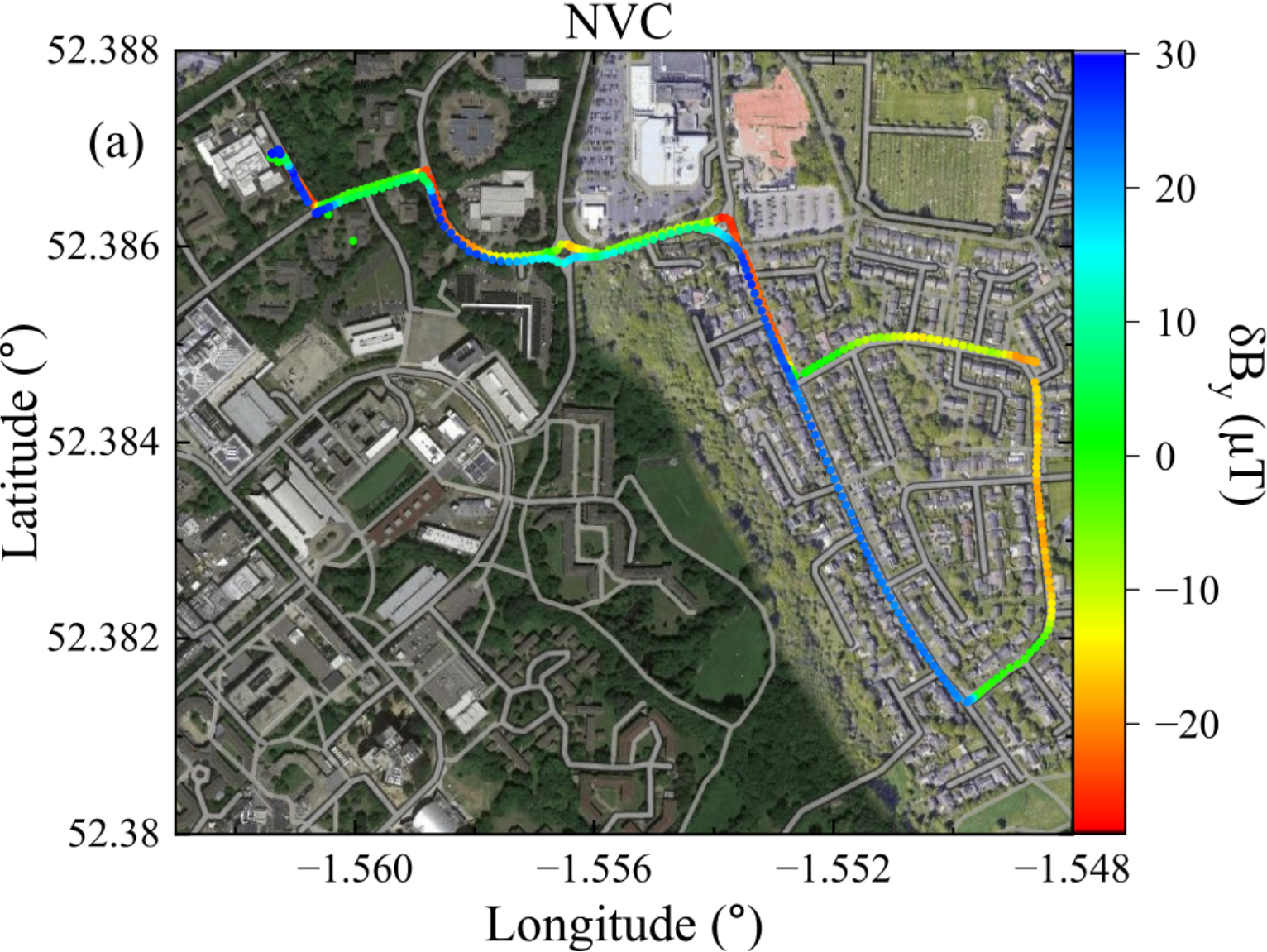} 
\hspace{0.5cm}\vspace{0.05cm}\includegraphics[width=\columnwidth]{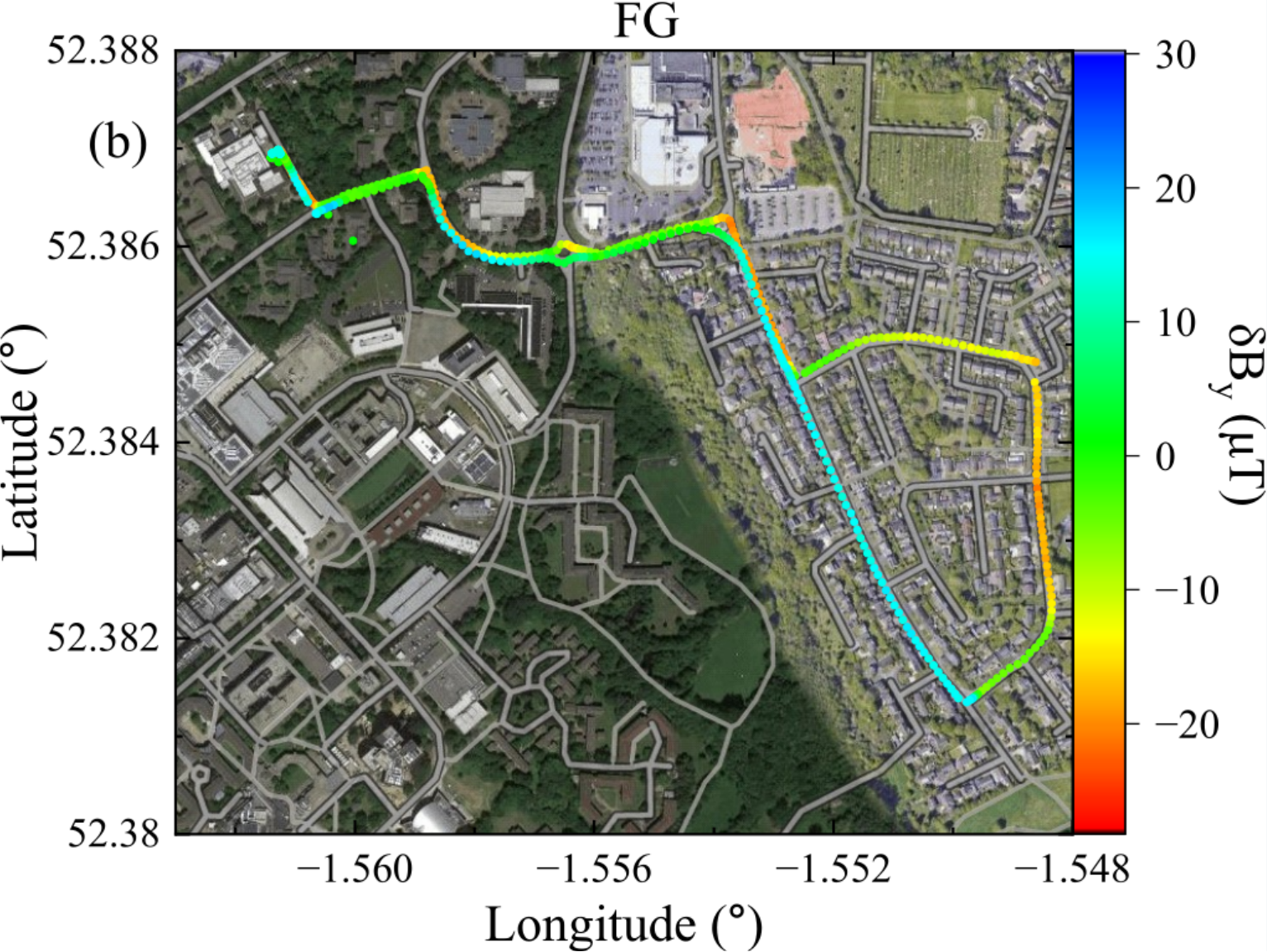} 
\hspace{-0.5cm}\includegraphics[width=\columnwidth]{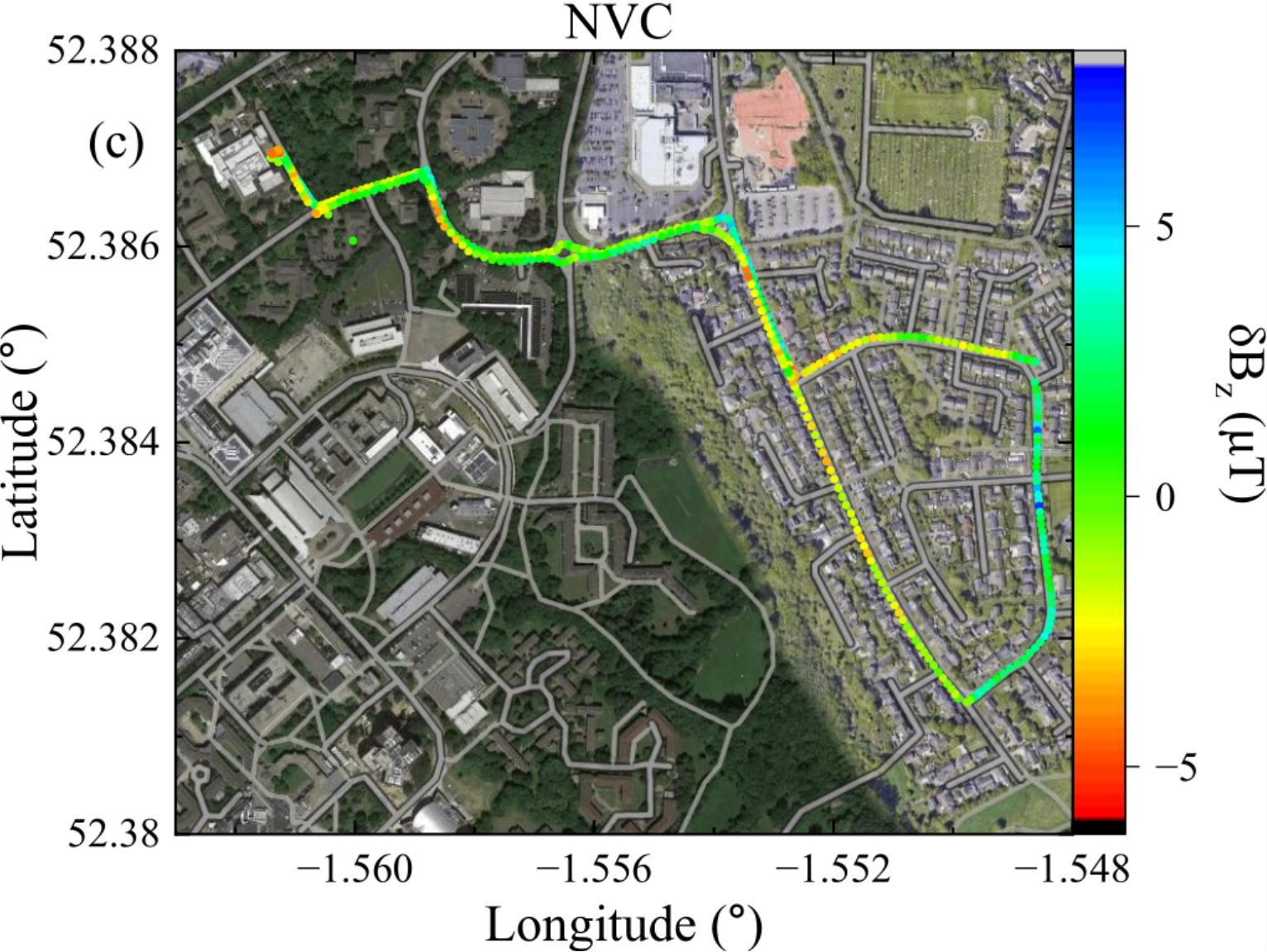} 
\hspace{0.5cm}\includegraphics[width=\columnwidth]{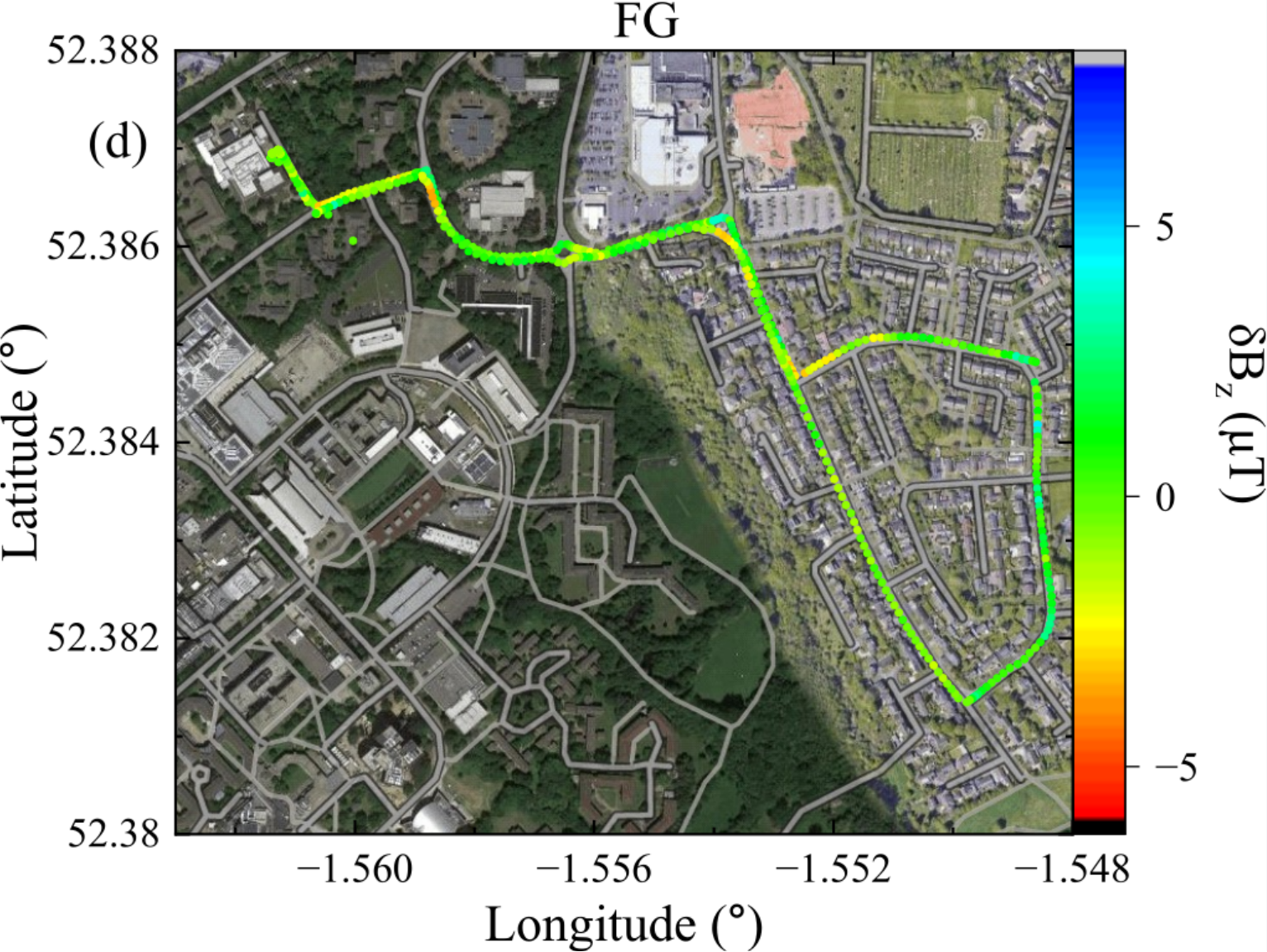} 
\caption{\small Magnetic field shifts of; (a) NVC magnetometer $\delta\textit{B}_{\textrm{y}}$, (b) fluxgate (FG) magnetometer $\delta\textit{B}_{\textrm{y}}$, (c) NVC magnetometer $\delta\textit{B}_{\textrm{z}}$, and (d) FG magnetometer $\delta\textit{B}_{\textrm{z}}$. All plotted as a function of latitude and longitude on a Google hybrid map. These measurements were taken in the electric van.} 
\label{fig: GPSMap-ElectricVanByBzFGvsNVC}
\end{figure*}

Figure \ref{fig: GPSMap-ElectricVanByBzFGvsNVC} shows the GPS tagged magnetic field shift measurements taken in the electric van for the $\delta\textit{B}_{\textrm{y}}$ and $\delta\textit{B}_{\textrm{z}}$ components. As can be seen there is relatively little variation in the z-shift component. This is unsurprising as the roads have relatively few significant magnetic anomalies. Furthermore, there is limited variation in height or orientation relative to the z-direction on the path we travelled and so you would not expect to see large changes in the Earth's core field along the z-direction. If we were measuring the absolute field of the Earth, however, at the latitude of the measurements the z-component of the Earth's core field is the largest. 

\begin{figure*}[h!]
\hspace{-0.5cm}\includegraphics[width=\columnwidth]{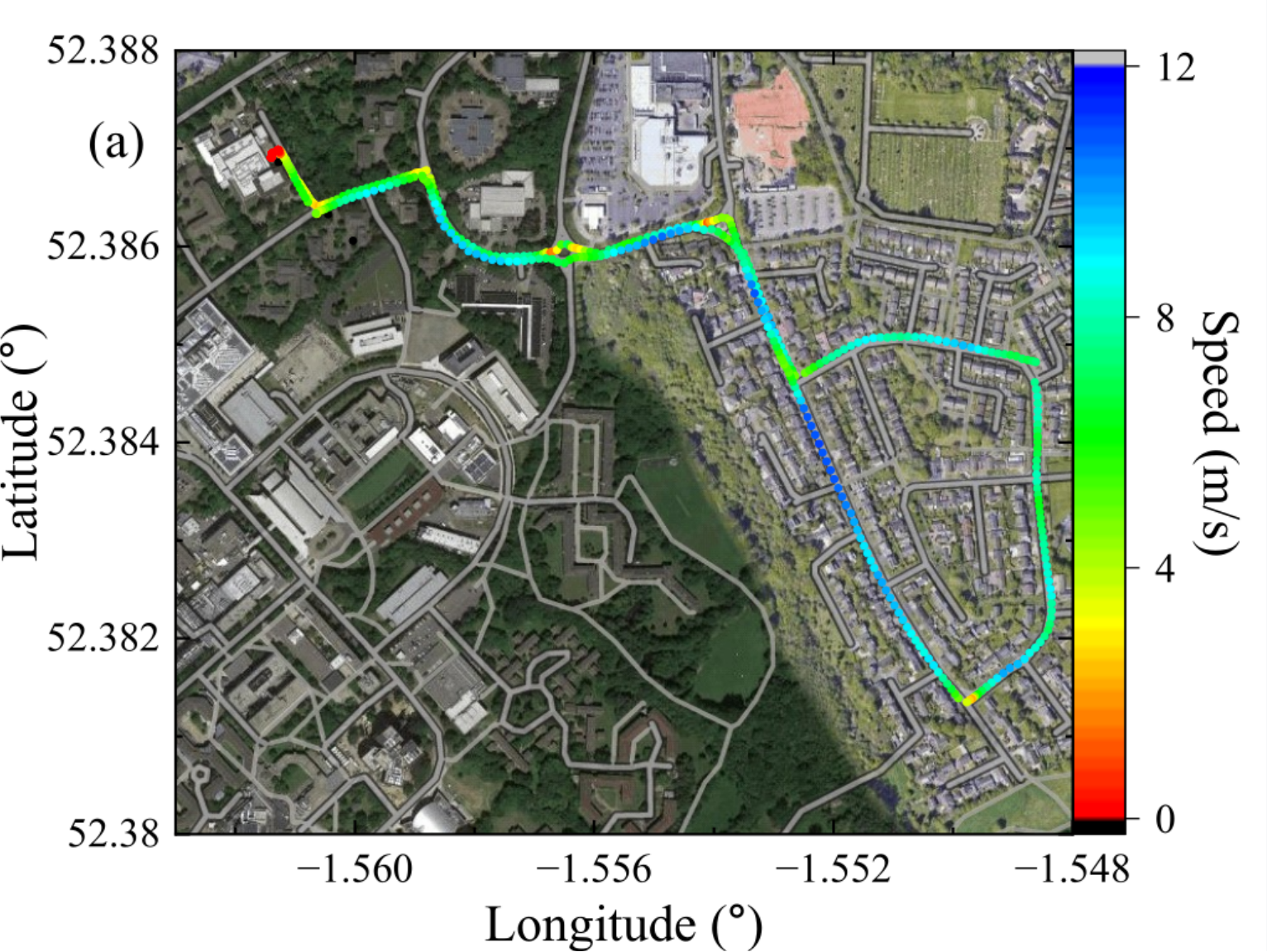} 
\hspace{0.5cm}\includegraphics[width=\columnwidth]{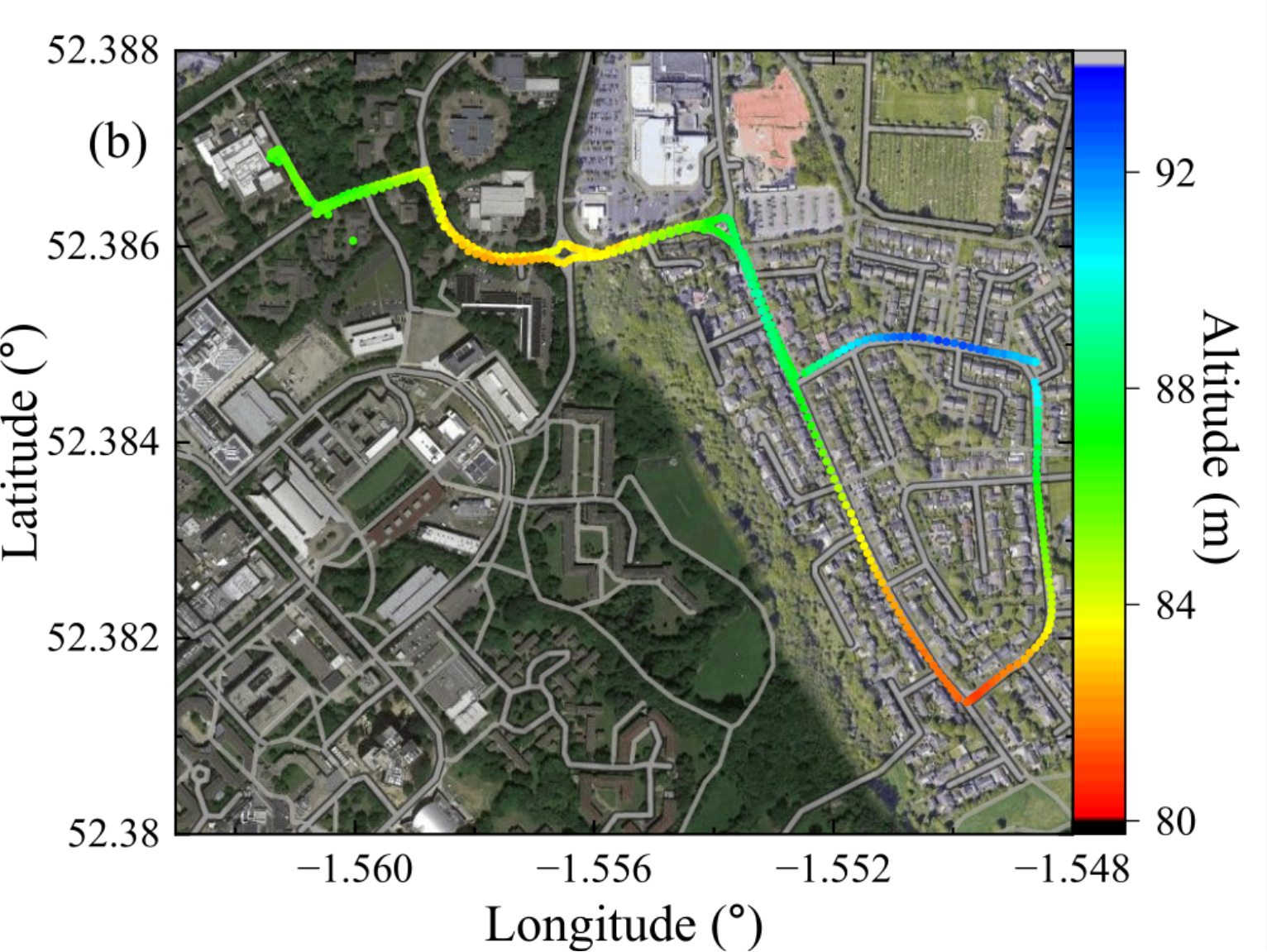} 
\caption{\small (a) The speed measured from GPS data-logger as a function of latitude and longitude for the electric van drive. (b) The altitude above mean sea level measured from GPS data-logger as a function of latitude and longitude for the electric van drive.} 
\label{fig: GPSSpeed+Altitude}
\end{figure*}

Figure \ref{fig: GPSSpeed+Altitude}(a) shows the speed of the electric van plotted as a function of the latitude and longitude coordinates. The speed is measured from the GPS measurements of the smartphone data-logger. The average speed is approximately 7 m/s (15.7 mi/h or 25.2 km/h). Given our sampling rate of 1.3 Hz, it takes approximately 0.8 s to go through the sequential vector code. This does mean that the frequency shift measured for each of the four resonances will have been recorded at different spatial positions as the van moves. This means that the spatial resolution achievable with the NVC magnetometer in the van at this speed would be approximately 5 to 7 m, and would vary with the actual speed of the van. This is comparable to the accuracy of the GPS measurements, which is consistently between 4 and 6 m. This accuracy depends on the availability of the signals from the satellites. Figure \ref{fig: GPSSpeed+Altitude}(b) shows the altitude above mean sea level (and thus the spatial changes along the z-axis of the magnetometers) as a function of latitude and longitude. As noted previously, the path travelled was relatively flat, and these measurements were also taken from the GPS data-logger. 

\FloatBarrier

\section*{Appendix H: Diesel vs Electric Van Measurements}

\begin{figure}[h!]
\includegraphics[width=\columnwidth, trim={1cm 1cm 0.5cm 1cm}]{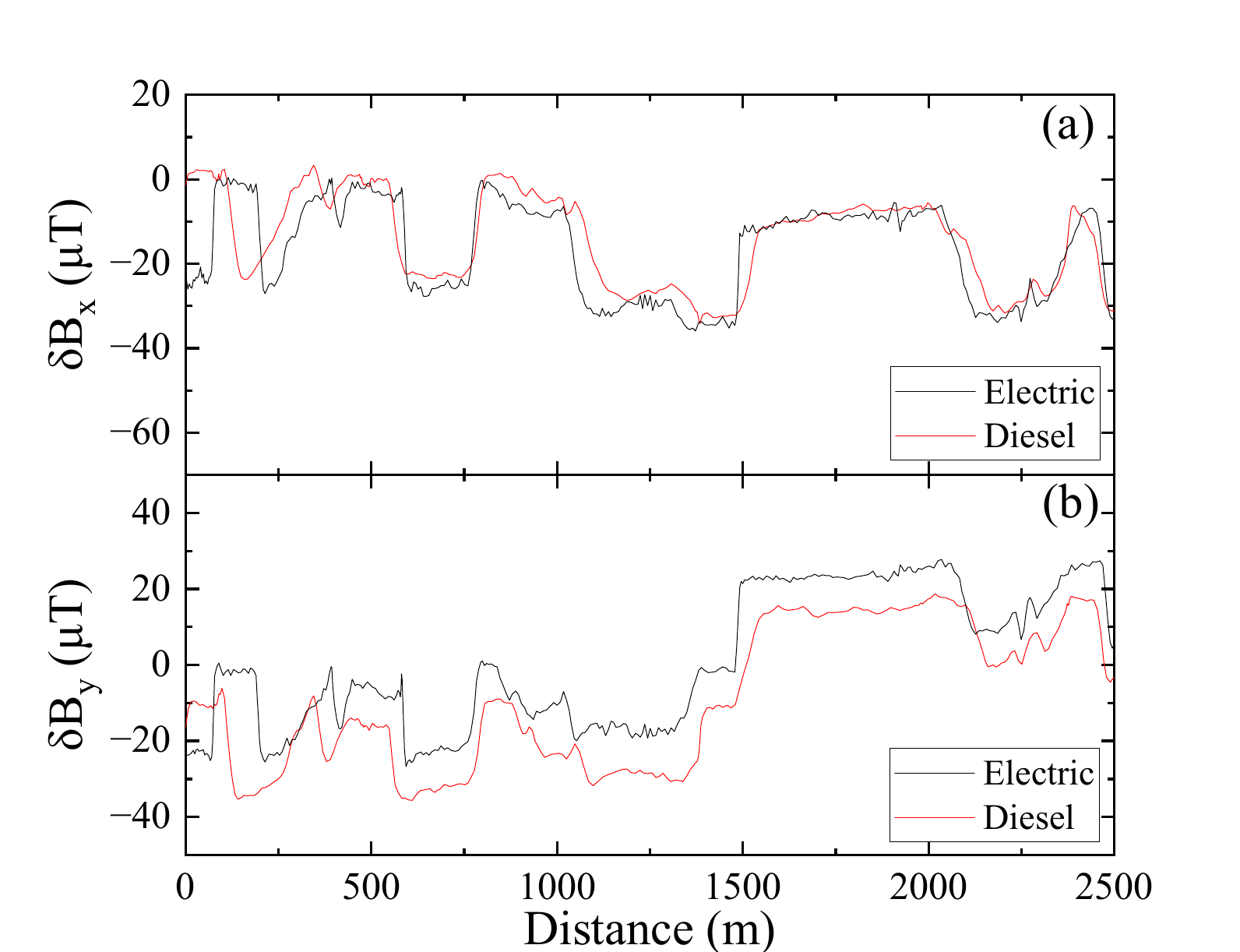} 
\caption{\small Line profiles comparing the magnetic field shifts; (a) $\delta\textit{B}_{\textrm{x}}$  and (b) $\delta\textit{B}_{\textrm{y}}$, all taken with the NVC magnetometer in the electric and diesel vans. The distance is calculated by integrating the speed as measured with the GPS smartphone data-logger, the distance travelled in the diesel and electric vans differed so the x-axis is shifted for easier comparison.} 
\label{fig: LineProfile-DieselvsElectric}
\end{figure}

Given the argument that the magnetic field shifts observed by the NVC and FG magnetometers may be principally attributed to the change in the magnitude of the components of the Earth's core field along the x, y, and z axes with the permanent magnetic field of the van and trolley equipment being constant in the frame of the reference of the magnetometers, and thus subtracted off automatically when looking at the magnetic field shifts, it would be expected that the same field shifts would be observed regardless of the vehicle. You would also expect such fields to be constant over long periods of time \cite{canciani2016absolute}. Figure \ref{fig: LineProfile-DieselvsElectric}(a) shows the line profile of the x and y magnetic field shifts measured with the NVC magnetometer when going on approximately the same drive path with the magnetometers in the two different vans (the electric eNV200 and diesel Trafic SL30). The differences observed could be due to changes in the calibration (both in terms of the ZCSs and the $\mathbf{A}$-matrix), different vehicles and objects on the roads, and slight variations in the path travelled by the van. They could also be due to differences in the induced and eddy current magnetic fields in the electric and diesel vans.

\begin{figure*}[h!]
\hspace*{-0.5cm}\includegraphics[width=\columnwidth]{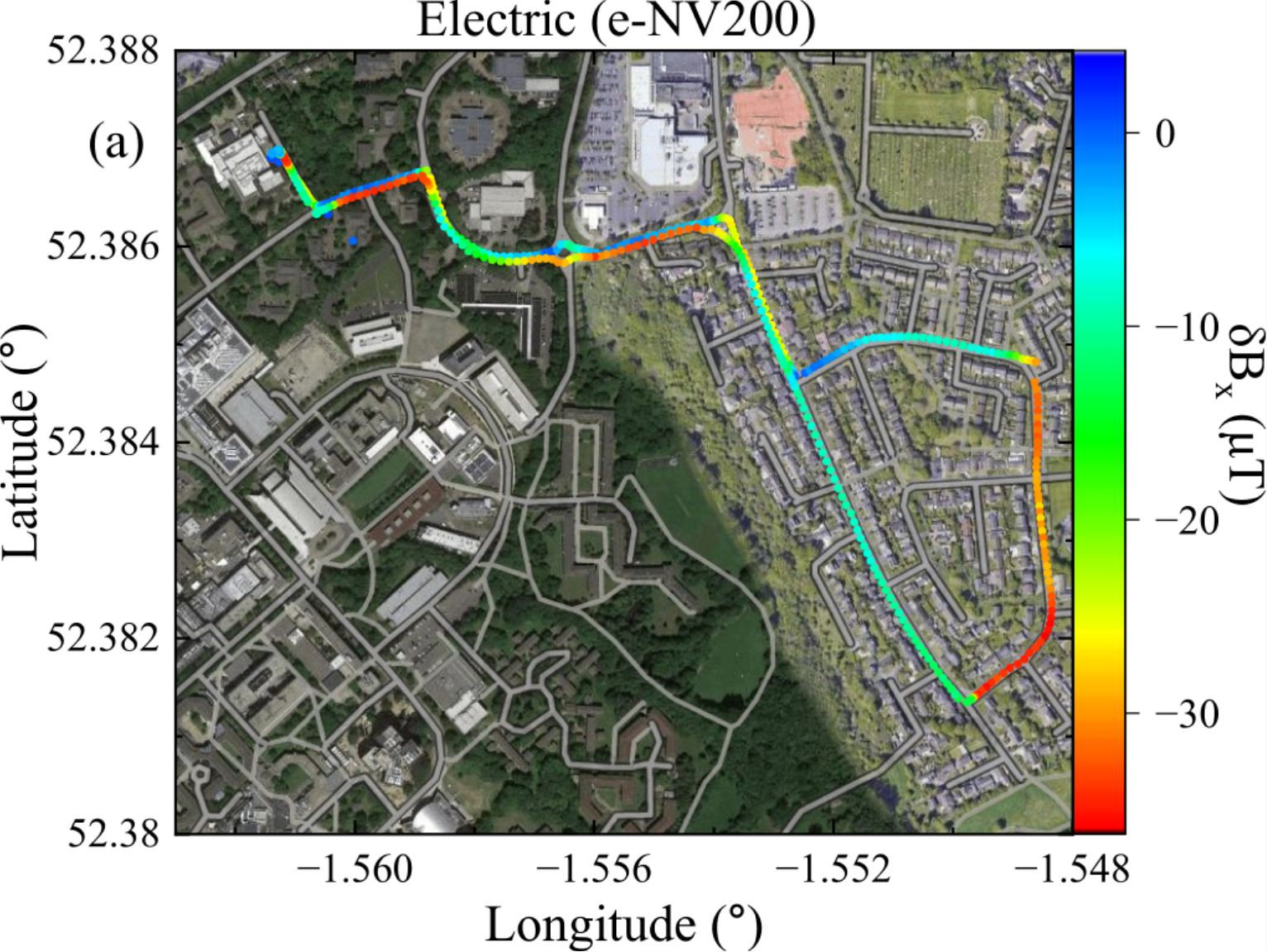} 
\hspace*{0.5cm}\includegraphics[width=\columnwidth]{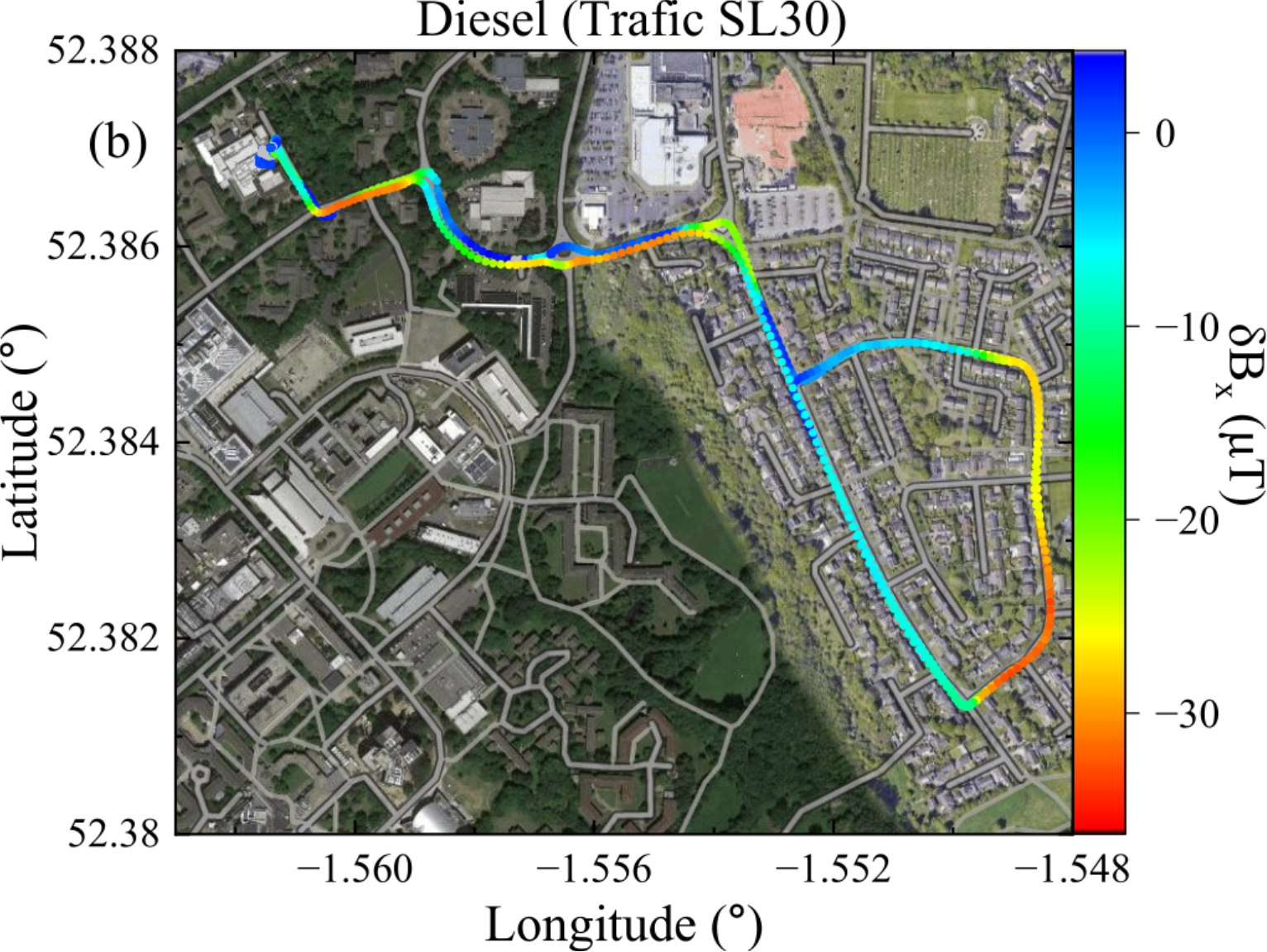} 
\caption{\small Magnetic field shifts of; (a) NVC magnetometer $\delta\textit{B}_{\textrm{x}}$ in electric van, and (b) NVC magnetometer $\delta\textit{B}_{\textrm{x}}$ in diesel van. All plotted as a function of latitude and longitude on a Google hybrid map.} 
\label{fig: GPSMap-DieselvsElectric}
\end{figure*}

Figure \ref{fig: GPSMap-DieselvsElectric} shows the GPS maps of the magnetic field shifts corresponding to the line profiles for the $\delta\textit{B}_{\textrm{x}}$ component. 

\begin{figure}[h!]
\includegraphics[width=\columnwidth, trim={1.5cm 1.5cm 1.5cm 1.5cm}]{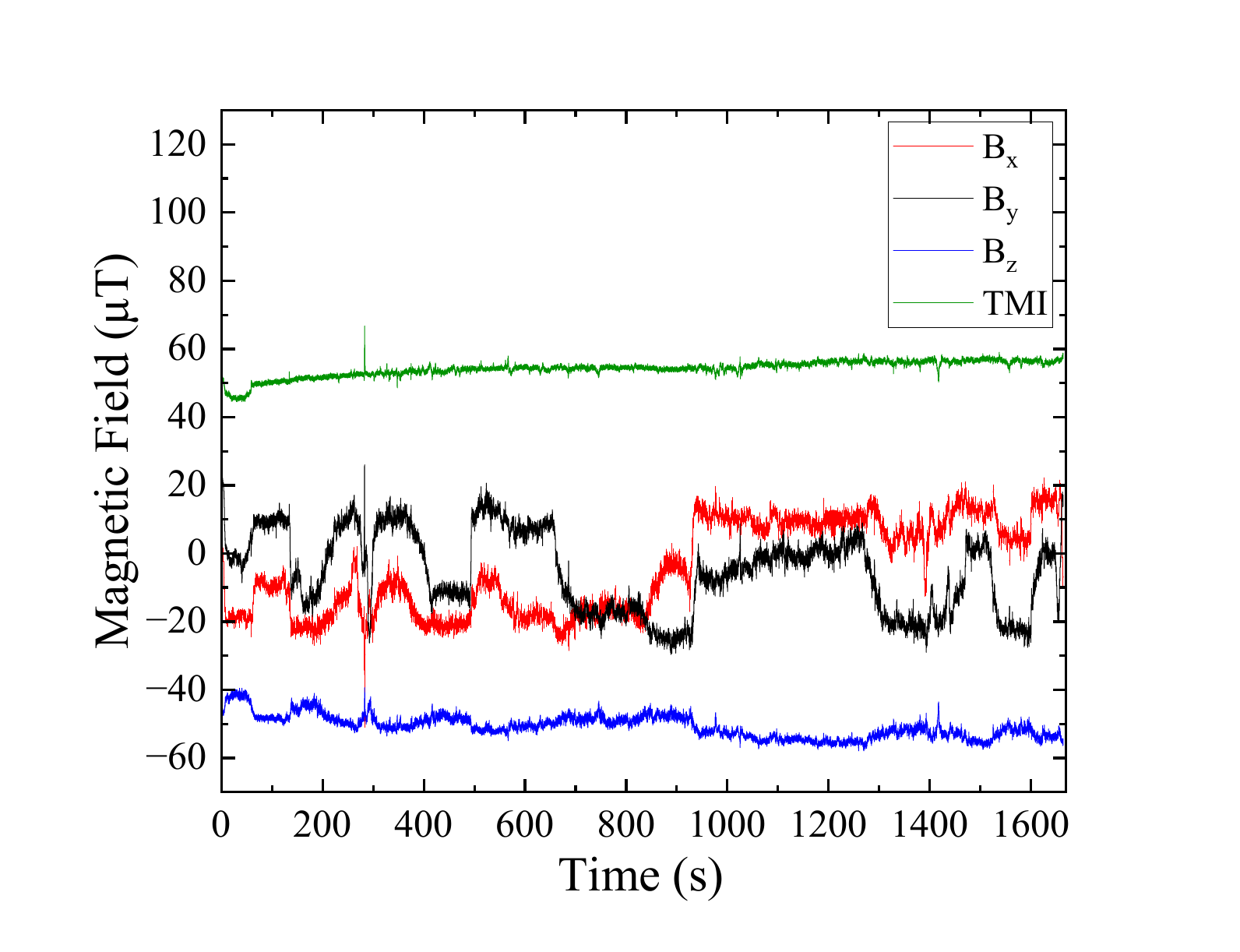} 
\caption{\small Line profiles of the magnetic field components as well as the total magnetic intensity (TMI) measured with a smartphone magnetometer, orientated in approximately the same direction as the fluxgate (FG) and NVC magnetometers. Approximately the same path as the vans was followed but on foot without disturbance magnetic fields from the van.} 
\label{fig: LineProfilesFoot}
\end{figure}

\begin{figure*}[h!]
\hspace*{-0.5cm}\includegraphics[width=\columnwidth]{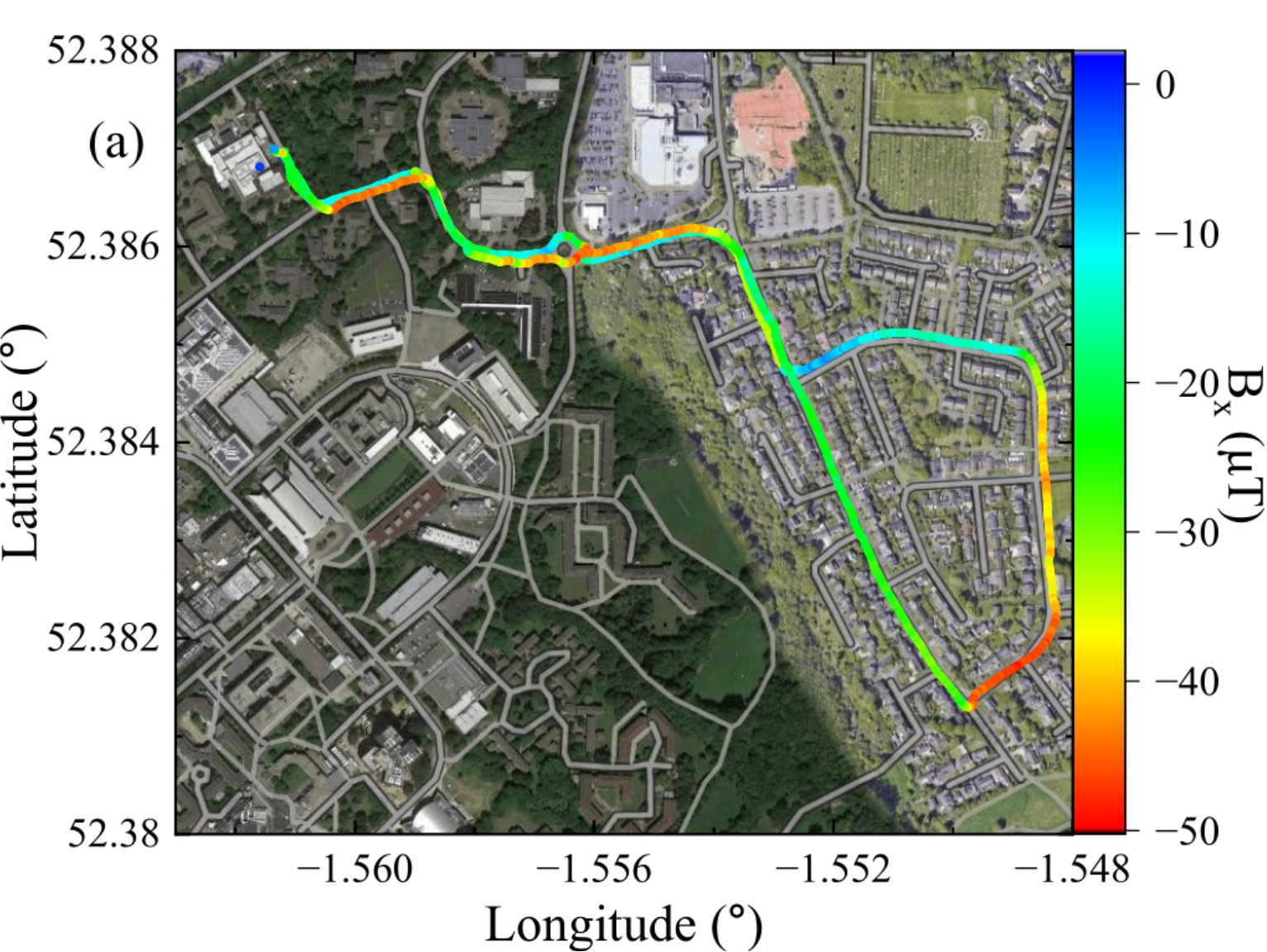}
\hspace*{0.5cm}\includegraphics[width=\columnwidth]{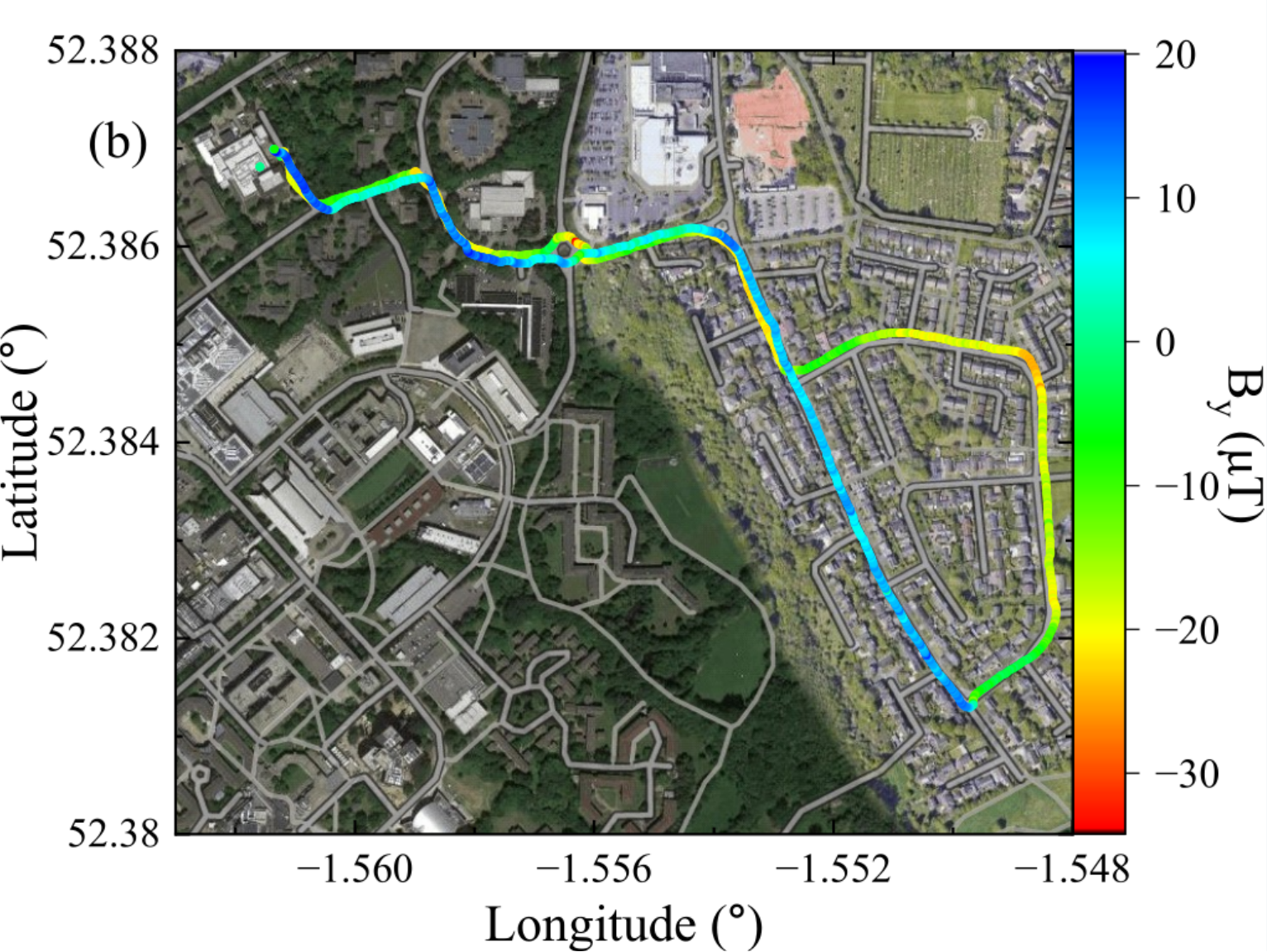} 
\hspace*{-0.5cm}\includegraphics[width=\columnwidth]{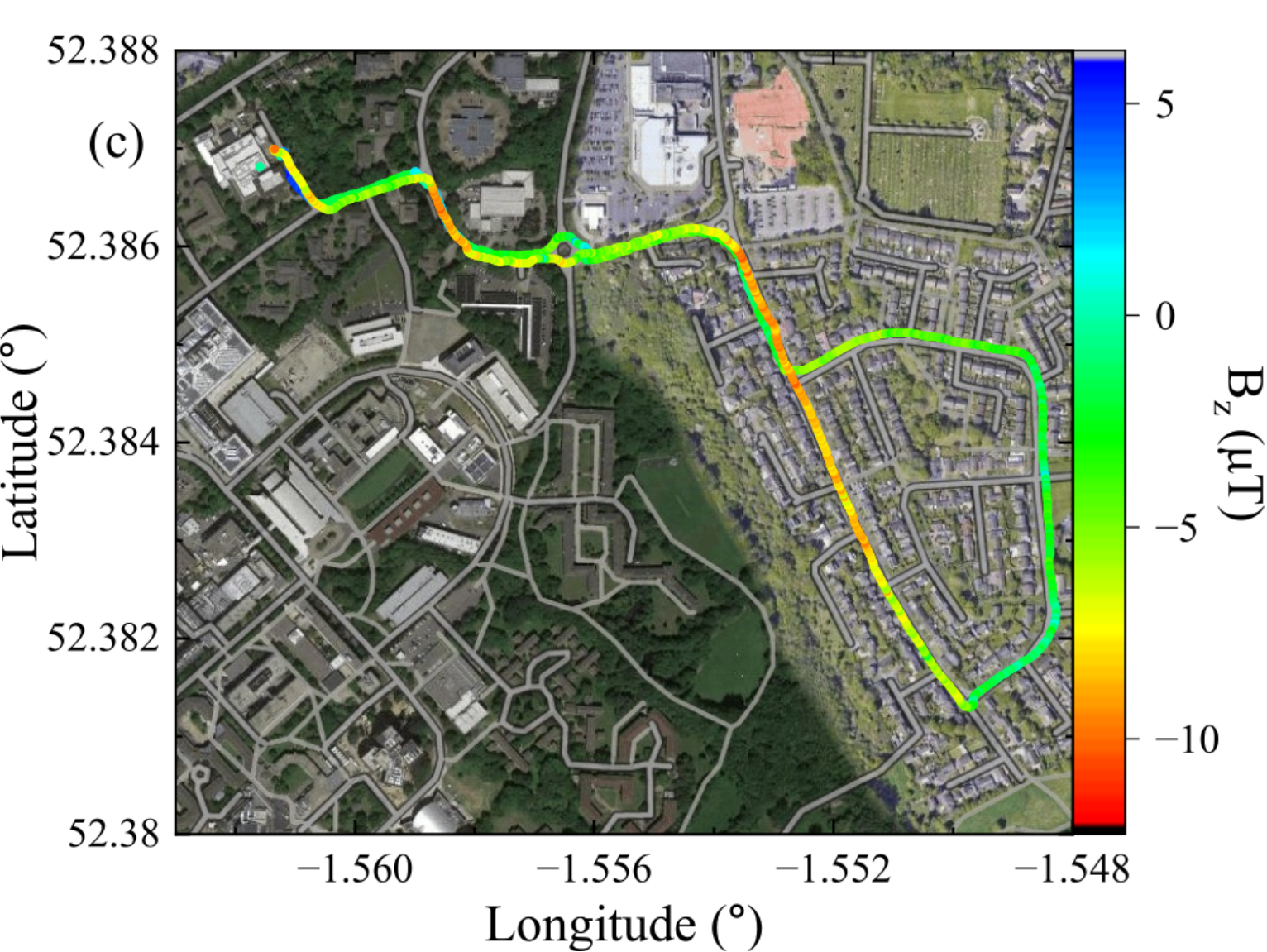} 
\hspace*{0.5cm}\includegraphics[width=\columnwidth]{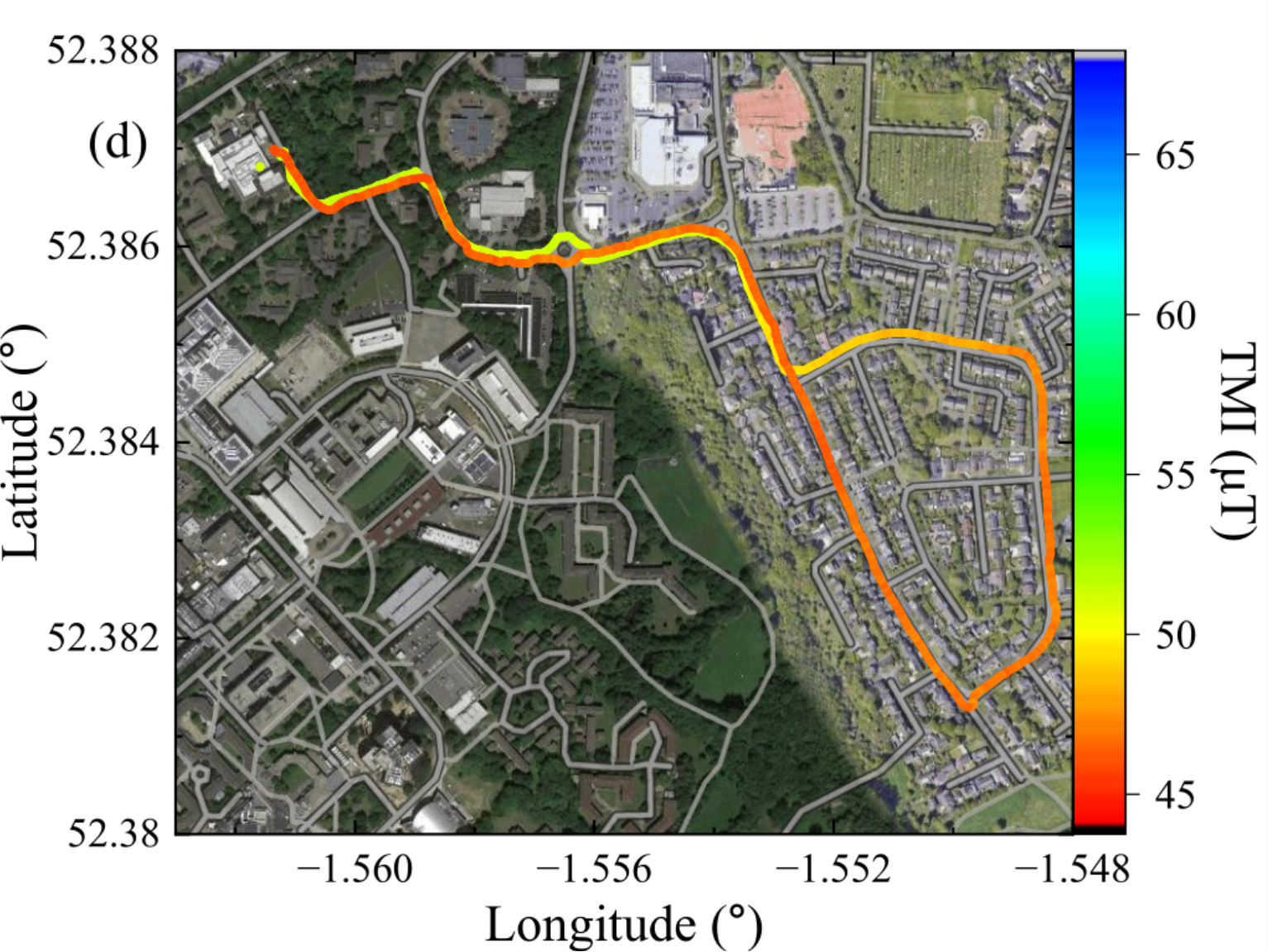} 
\caption{\small Magnetic field shifts measured using smartphone magnetometer; (a) $\textit{B}_{\textrm{x}}$, (b) $\textit{B}_{\textrm{y}}$, (c) $\textit{B}_{\textrm{z}}$, and (d) total magnetic intensity (TMI). All these measurements are taken on foot and plotted as a function of latitude and longitude on a Google hybrid map.} 
\label{fig: GPSMap-Foot}
\end{figure*}

Additionally, the approximate path taken by the van containing the magnetometers was also completed on foot using a smartphone magnetometer (approximately orientated in the same manner as the NVC and FG magnetometers in the van). Figure \ref{fig: LineProfilesFoot} shows the line profiles for the absolute fields $\textit{B}_{\textrm{x}}$, $\textit{B}_{\textrm{y}}$, and $\textit{B}_{\textrm{z}}$ along with the TMI. The GPS maps of $\delta\textit{B}_{\textrm{x}}$ and $\delta\textit{B}_{\textrm{y}}$ are shown in Fig. \ref{fig: GPSMap-Foot}. The TMI calculated from the absolute magnetic fields is also shown. It is not possible to follow entirely the same path on foot, but the changes in magnetic field observed are in overall agreement with the FG and NVC magnetometers.

\FloatBarrier

\section*{Appendix I: Van Magnetic Signals}

\begin{figure}[h!]
\includegraphics[width=\columnwidth, trim={1cm 1cm 0.6cm 1cm}]{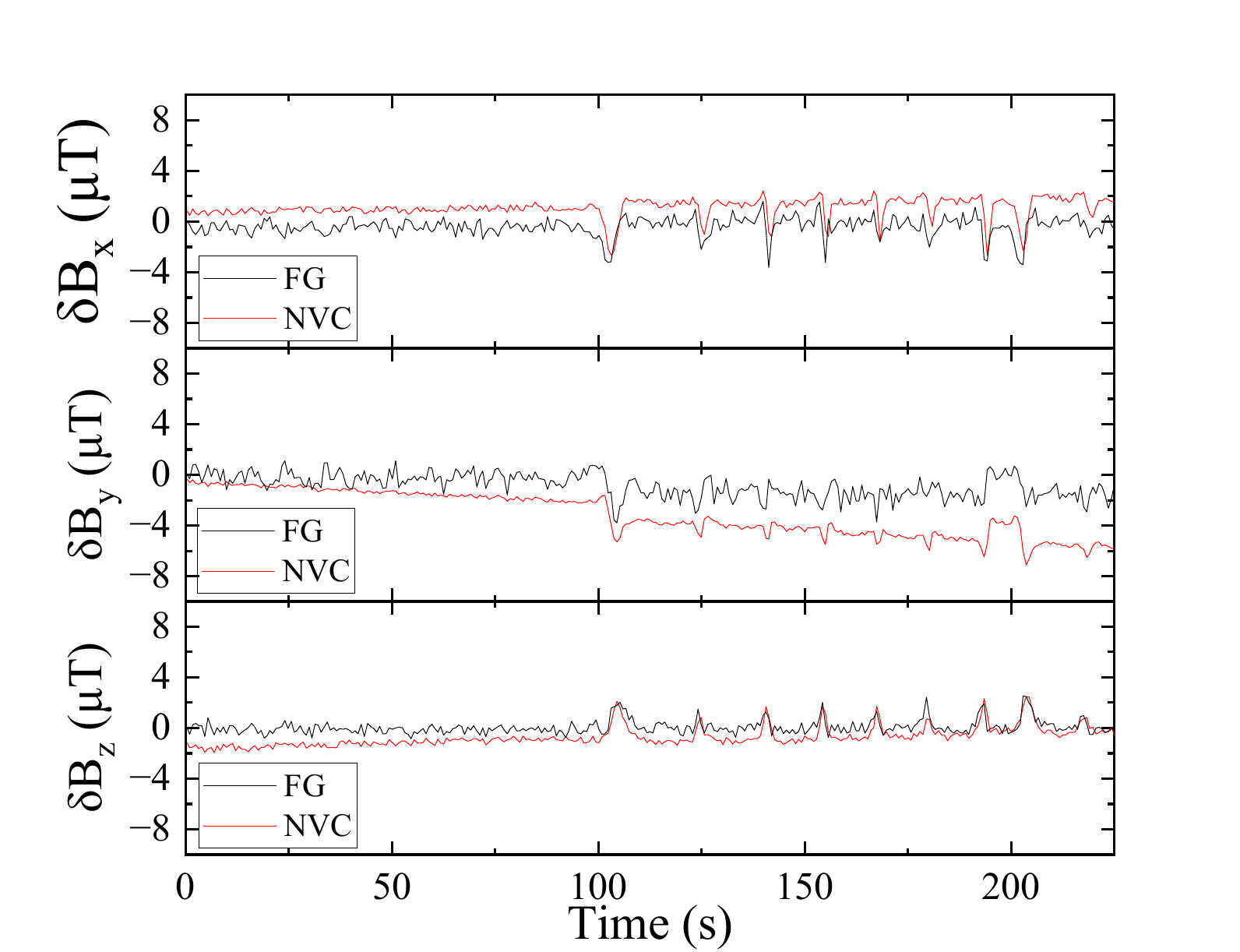} 
\caption{\small Line profiles comparing the magnetic field shifts; (a) $\delta\textit{B}_{\textrm{x}}$, (b) $\delta\textit{B}_{\textrm{y}}$, and (c)$\delta\textit{B}_{\textrm{z}}$, taken with the fluxgate (FG) and NVC magnetometers as the electric van was driven past repeatedly.} 
\label{fig: DrivingVanPastMagneticSignals}
\end{figure}

Additionally, with the trolley located outside of the electric van, the electric van was driven past the trolley multiple times at a distance of approximately 2 m. Figure \ref{fig: DrivingVanPastMagneticSignals} shows the magnetic signals that are observed from the van as it drove past with both the NVC and FG.

\FloatBarrier

\section*{Appendix H: Power Saturation}

\begin{figure}[h!]
\includegraphics[width=\columnwidth, trim={1.5cm 1.5cm 1.5cm 1.5cm}]{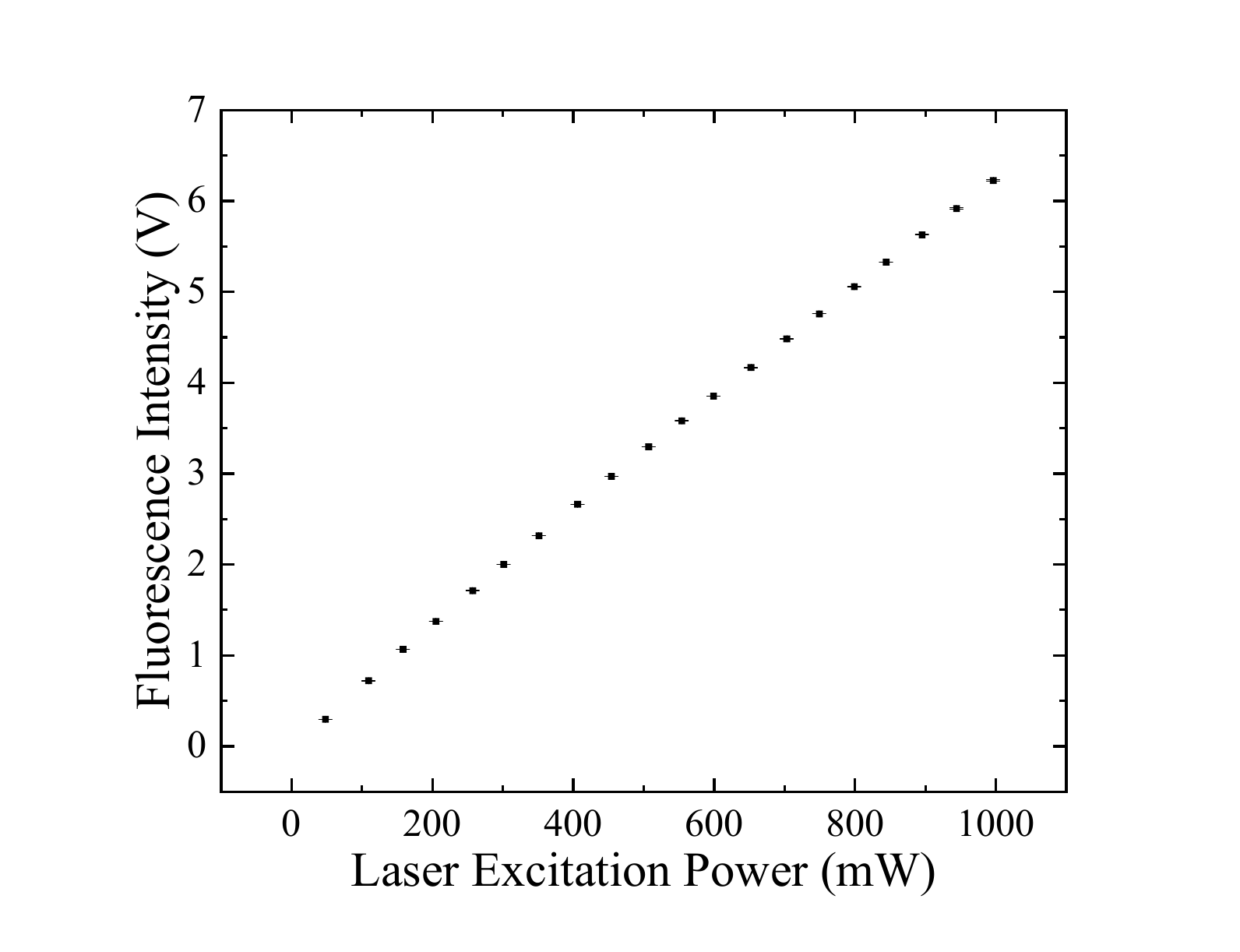} 
\caption{\small Power saturation curve displaying the fluorescence intensity as a function of laser excitation power at one of the photodiodes of the balanced detector.} 
\label{fig: PSat}
\end{figure}

Figure \ref{fig: PSat} shows a power-saturation measurement for the portable magnetometer. It is clear from this that at a laser power of 0.5 W we are not saturating our NVC ensemble. The data was fitted with

\begin{equation}
\label{eq:PowerSaturation}
I = I_{\textrm{sat}}\frac{P}{P+P_{\textrm{sat}}}
,
\end{equation}

where I is the fluorescence intensity in V, $I_{\textrm{sat}}$ is the saturation fluorescence intensity in V, P is the laser excitation power, and $P_{\textrm{sat}}$ is the saturation laser excitation power \cite{aslam2013photo}. The values of $I_{\textrm{sat}}$ and $P_{\textrm{sat}}$ are found to be (71 $\pm$ 4) V and (10.4 $\pm$ 0.6) W respectively.

\FloatBarrier

\section*{Appendix J: Laser Noise}

\begin{figure}[h!]
\includegraphics[width=\columnwidth, trim={1.5cm 1.5cm 1.5cm 1.5cm}]{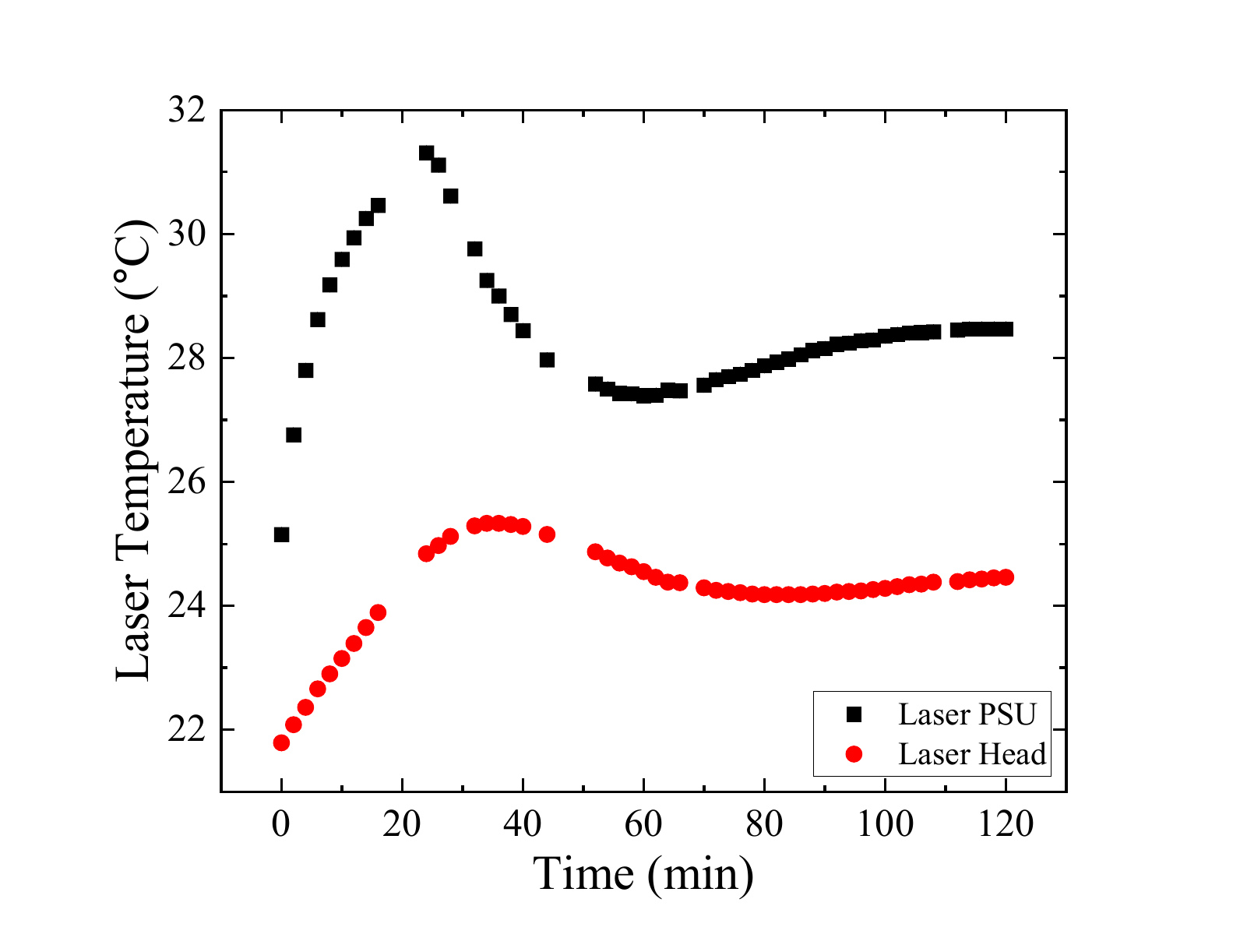}
\caption{\small The laser power supply unit (PSU) and head temperature (as recorded on the Laser Quantum control software) between 7:38 am and 9:38 am. The error bars are too small to be seen.} 
\label{fig: LaserTempInside}
\end{figure}

An issue with operating the diamond magnetometer in real-world situations is the reliance on a laser. The laser is sensitive to changes in the temperature of its environment. Generally, when operating the magnetometer within the laboratory we wait for approximately 30 minutes for the laser head and PSU to heat up and reach thermal equilibrium. Prior to this the laser is found to be unstable, leading to high levels of noise. Figure \ref{fig: LaserTempInside} shows the temperature of the laser head and PSU as a function of time from first turning the laser on to 0.5 W output power with the trolley in the laboratory. When operating the laser outside of the laboratory the temperature of the air surrounding the laser is more changeable and this means the temperatures of the laser head and PSU generally take longer to stabilise, are readily knocked out of thermal equilibrium and thus the laser is generally noisier. This could impact not only our sensitivity, but also our accuracy - as the magnetic field measurements depend upon the calibration constant obtained from the ZCS of the ODMR spectra, and thus on the laser power. This dependence on laser power could be eliminated using a scheme such as that employed in Ref. \cite{clevenson2018robust}. 

\begin{figure}[h!]
\includegraphics[width=\columnwidth, trim={1.5cm 1.5cm 1.5cm 1.5cm}]{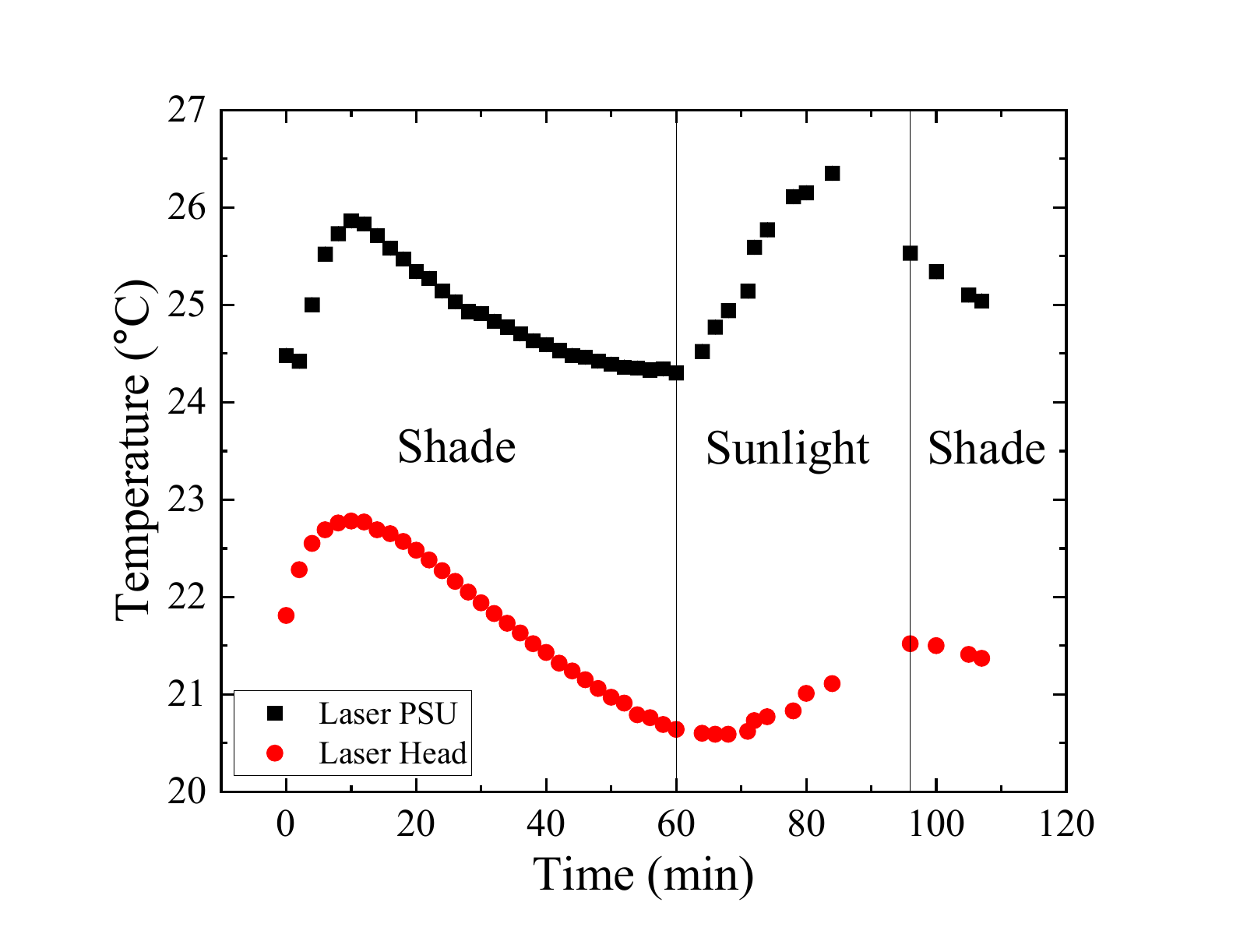}
\caption{\small The laser power supply unit (PSU) and head temperature (as recorded on the Laser Quantum control software) between 13:19 pm and 15:07 pm separated into time segments in which the trolley is under the shade of the laboratory building and trees and that in which it was in direct sunlight. The error bars are too small to be visible.} 
\label{fig: LaserTempOutside}
\end{figure}

Figure \ref{fig: LaserTempOutside} shows the temperature of the laser head and PSU as a function of time from first turning the laser on to 0.5 W output power with the trolley outside of the laboratory on the afternoon of a clear day with a temperature between $10^{\circ}$C and $14^{\circ}$C in the shade as measured with the temperature data-logger.

As the diamond, laser head and laser PSU are all contained within the trolley - which is sealed with blinds to ensure light tightness - the temperature within the trolley is relatively stable. However, a fan is connected to the front of the trolley for cooling purposes. A pipe directs the air pulled in from this fan to the side of the portable magnetometer box units intake fan. As can be seen from Fig. \ref{fig: LaserTempOutside} after approximately 40 to 60 minutes the laser PSU and head reach a relatively stable temperature, the rise in temperature between 60 minutes and 96 minutes (indicated with the vertical lines) as labelled is due to the trolley being in direct sunlight for this period. These measurements show the sensitivity of the laser temperature, and thus at least to some degree performance, upon changing external conditions. 

\FloatBarrier

\section*{Appendix K: Portable Configuration}

\begin{figure}[h!]
\includegraphics[width=0.9\columnwidth]{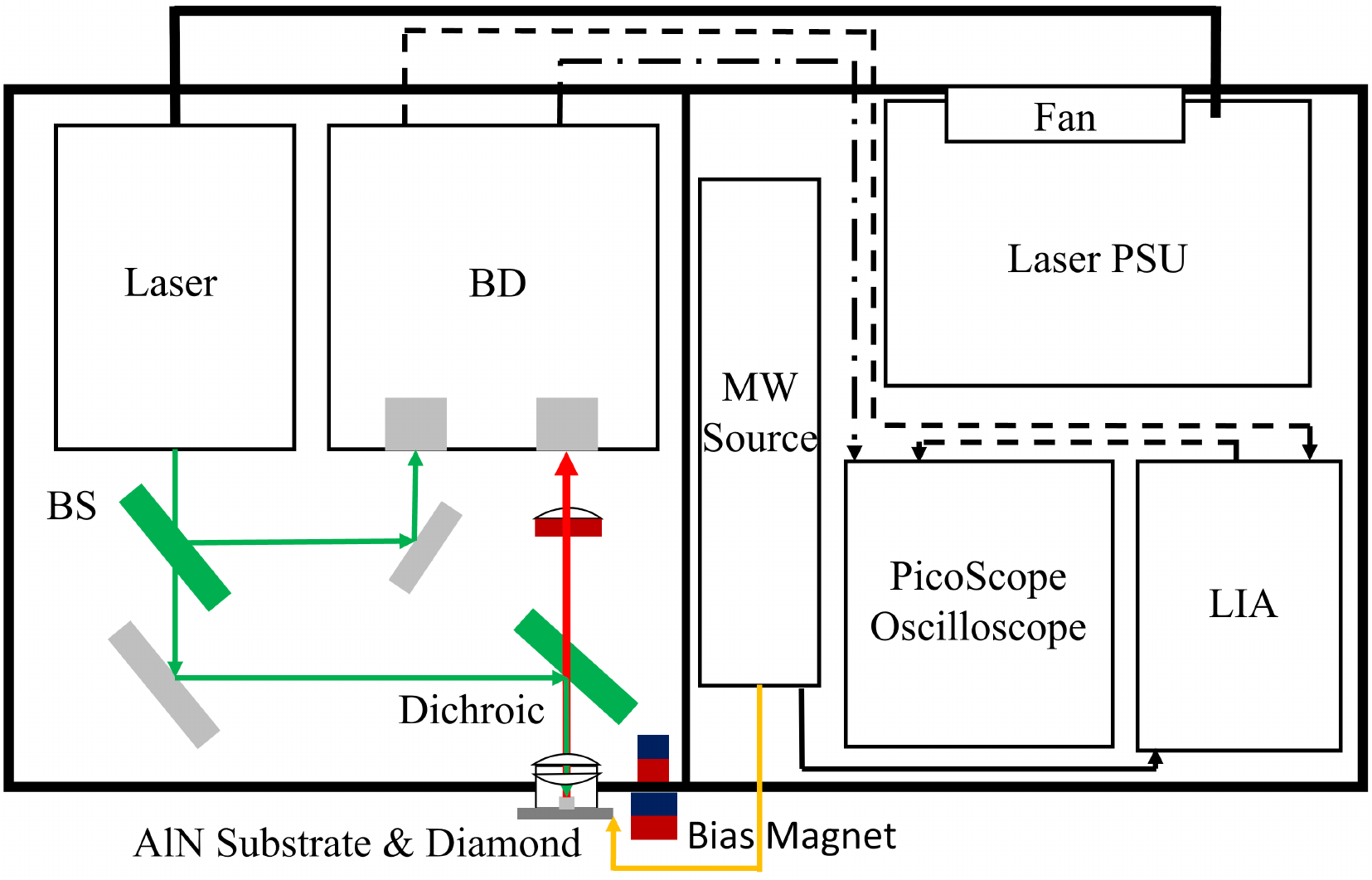}
\caption{\small A schematic of the lower SWaP (size, weight, and power) experimental setup: BS, 100:1 beam sampler; BD, balanced detector; PSU, power supply unit; MW, microwave source; LIA, lock-in amplifier.} 
\label{fig: ExperimentalSetupPortableConfiguration}
\end{figure}

Figure \ref{fig: ExperimentalSetupPortableConfiguration} shows an alternative single-box configuration of the magnetometer. This employs a RF-Consultant TPI-1005-B microwave source, microwave amplifier, and Red-Pitaya LIA all of which fit into the electronics section of the box. 

Using these items of equipment greatly reduces the SWaP (size, weight, and power) of the magnetometer at the expense of sensitivity and vector capability. The approximate power draw for this configuration is only 34 W, compared to the 403 W required with the external microwave source, microwave amplifier and LIA. Vector is not possible as the TPI-1005-B cannot change frequency rapidly enough for sequential hopping for vector and feedback control. Additionally, at the modulation frequencies we employ the Red Pitaya has a high input noise level and this limits our sensitivity to approximately 20 nT/$\sqrt{\textrm{Hz}}$ in a (10-100)-Hz frequency range. 

The sensor head itself could also be operated in two configurations - one fixed and the other employing a portable head. The first of these involves screwing the AlN sensor head to the side of the portable magnetometer optics box as in the main text. A lens is employed to focus the laser light down onto the diamond, and also to collimate the red fluorescence emitted from the diamond. This is then directly focused down onto one of the photodiodes of the balanced detector with a second aspheric lens. The optical setup was initially designed to be employed with a separate sensor head and it is clear that significant improvements in the photon collection efficiency could be made with some simple optimisation of the lens type and the positioning of the lenses. A potential source of noise with this configuration is the fact that the diamond and mount are mechanically attached to a flexible aluminium wall - which could thus move in relation to the laser beam - potentially generating significant amounts of noise. This is not found to be a major limiting factor for our mobile measurements, however, as the roof of the portable magnetometer box securely holds the wall in place.

In the second configuration a fiber adaptor can be attached where the AlN substrate with Al holder is usually fixed. This would be useful for applications requiring a lower magnetic noise level, as it would be possible to place the fiber-coupled sensor head away from the electronics of the box and trolley equipment. In principle, a central hub box containing optics and electronics could have multiple diamond-containing fiber-coupled sensor heads attached for gradiometry and tensor gradiometry measurements \cite{newman2024tensor, blakley2015room, blakley2016fiber, zhang2021diamond, masuyama2021gradiometer}. The sensor heads could be secured at different positions within a vehicle and the sensitive electronics, laser and optics kept in a temperature controlled area. Movement of the optical fiber(s) could lead to modal noise, however, which could negatively impact performance \cite{epworth1978phenomenon}.

\FloatBarrier

%

\end{document}